\documentclass[11pt]{article}

\usepackage[dvips]{graphicx}
\usepackage{amsmath}
\usepackage{amsfonts}
\usepackage{amssymb}
\usepackage{epsfig}
\usepackage{times}
\usepackage{cite}
\usepackage{calc}
\usepackage{version}
\usepackage[english]{babel}
\usepackage{epsf}
\usepackage{graphics}
\usepackage{ulem}
\usepackage{hyperref}
\usepackage{upgreek}

\usepackage[cmtip,arrow]{xy}
\usepackage{pst-plot, pst-coil, pstricks}

\newcommand{\alphas}{\alpha_{\rm s}}
\newcommand{\alphasmZ}{\alphas(\rm m^2_{_{\rm Z}})}
\newcommand{\sqrts}{\sqrt{\rm s}}

\newcommand{\dd}{{\rm d}}

\def\cO#1{{{\cal{O}}}\left(#1\right)}

\newcommand{\lqcd}{\Lambda_{_{\rm QCD}}}

\newcommand{\MSbar}{\overline{\rm MS}}

\newcommand{\epem}{e^+e^-}
\newcommand{\meff}{m{_{\rm eff}}}
\newcommand{\ximax}{\xi{_{\rm max}}}

\providecommand{\mean}[1]{\ensuremath{\left<#1\right>}}

\begin{document}
%\hfill {\sf CERN-PH-/2013-iii}
%%\begin{flushright}
%%CERN-PH-TH/2011-302\\
%%\end{flushright}
%\vspace{0.5cm}

\begin{center}
{\Large\bf Energy evolution of the moments of the hadron distribution in QCD jets \\\vspace{0.2cm} 
%jet fragmentation function\\\vspace{0.2cm} 
%next-to-leading-log resummations and %\\\vspace{0.2cm}
including NNLL resummation and %\\\vspace{0.2cm} 
NLO running-coupling corrections}
%$\text{Next-to-MLLA}+\text{NLO}$ energy evolution of the energy-momentum distribution of hadrons
\end{center}
\vspace{0.1cm}

\begin{center}
Redamy P\'erez-Ramos$^{1,}$\footnote{e-mail: redamy.perez@uv.es}\, and David~d'Enterria$^{2,}$\footnote{e-mail: dde@cern.ch}
\vspace{0.15cm}

{\it $^1$ Department of Physics, University of Jyv\"askyl\"a, P.O. Box 35 (YFL), F-40014 Jyv\"askyl\"a, Finland\\}
{\it $^2$ CERN, PH Department, CH-1211 Geneva 23, Switzerland\\}
\end{center}

\vspace{0.4cm}

\begin{center}
{\bf \large Abstract}
\end{center}
The moments of the single inclusive momentum distribution of hadrons %(``hump backed plateau'') 
in QCD jets, are studied in the next-to-modified-leading-log approximation (NMLLA)  
including next-to-leading-order (NLO) corrections to the $\alphas$ strong coupling.
%, a framework that we call NMLLA$+$NLO$^*$. 
The evolution equations are solved using a distorted Gaussian %(DG) 
parametrisation, which successfully reproduces the spectrum of charged hadrons of jets measured 
in $\epem$ %, deep-inelastic, and hadronic 
collisions.
%in MLLA single energy-momentum spectrum of charged hadrons
%$\epem$ annihilation, hadron-hadron collisions and deep inestalic scattering (DIS) 
%supported by the local parton hadron duality hypothesis (LPHD). In the NMLLA$+$NLO$^*$ approach, 
%we recover the energy evolution of the mean particle multiplicity
%and its ratio for gluon and quark jets \cite{Dremin:2000ep}.
The energy dependencies of the maximum peak, multiplicity, width, kurtosis and skewness of the jet hadron
distribution are computed analytically. 
%in this NMLLA$+$NLO$^*$ framework. 
Comparisons of all the existing jet data measured in $\epem$ collisions in the range
$\sqrts\approx$~2--200~GeV to the NMLLA$+$NLO$^*$ predictions allow one to extract 
%The complete set of NMLLA$+$NLO$^*$ corrections to the energy evolution of the DG mean peak position is found, that 
%allows to extract an approximated
a value of the QCD parameter $\lqcd$, and associated two-loop coupling constant at the Z resonance
$\alphasmZ$~=~0.1195~$\pm$~0.0022, in excellent numerical agreement with the current world average obtained
using other methods.
%$\alphas(\rm m_{{\rm Z}}^2)$ the strong coupling.
%We compare our NMLLA$+$NLO$^*$ single energy-momentum spectrum with the experimental data in the $\epem$ 
%annihilation.

%\clearpage

\tableofcontents

%%%%%%%%%%%%%%%%%%%%%%%%%%%%%%%%%%%%%%%%%%%%%%%%%%%%%%%%%%%%%%%%%%%%%%%%%%%%%%%%%%%%%%%%

%\scrollmode

\section{Introduction}

One of the most ubiquitous manifestations of the fundamental degrees of freedom of Quantum Chromodynamics
(QCD), quark and gluons, are the collimated bunches of hadrons (``jets'') produced in high-energy particle
collisions. The evolution of a parton into a final distribution of hadrons is driven by perturbative dynamics 
dominated by soft and collinear gluon bremsstrahlung~\cite{Dokshitzer:1982ia,Azimov:1984np} followed by
the final conversion of the radiated partons into hadrons at non-perturbative scales approaching
$\lqcd\approx$~0.2~GeV. The quantitative description of the distribution of hadrons of type $h$ in a jet is
encoded in a (dimensionless) fragmentation function (FF) which can be experimentally obtained, e.g. in $\epem$
collisions at c.m. energy $\sqrts$, via 
$$
D^{\rm h}(\ln(1/x),\rm s) = \frac{d\sigma(ee\to h X)}{\sigma_{\rm tot} \,d\ln(1/x)},\label{eq:1}
$$ 
where $x=2\,p_h/\sqrts$ is the scaled momentum of hadron $h$, and $\sigma_{\rm tot}$ the total $\epem$
hadronic cross section. Its integral over $x$ gives the average hadron multiplicity in jets. %the collision. 
Writing the FF as a function of the (log of the) inverse of $x$, $\xi=\ln(1/x)$, emphasises the region of
relatively low momenta that dominates the spectrum of hadrons inside a jet. 
%As a consequence of QCD colour coherence~\cite{Dokshitzer:1982ia}, 
%Effectively, 
Indeed, the emission of successive gluons inside a jet follows a parton cascade where the emission
angles decrease as the jet evolves towards the hadronisation stage, the so-called ``angular
ordering''~\cite{Ermolaev:1981cm,Dokshitzer:1982ia,Dokshitzer:1991wu}.
Thus, due to QCD colour coherence and interference of gluon radiation,
%(resulting, on average, in {\it angular ordering}\footnote{Angular ordering (or coherence) implies
%$\theta_{p_1p_2} \gg \theta_{k_1p_1} \gg \theta_{k_2k_1} \gg \theta_{k_3k_2} \gg ...$, where $\theta_{k_1p_1}$ 
%is the emission angle of the primary soft gluon from the direction of the hard parton, 
%$\theta_{k_2k_1}$ is that of the softer secondary gluon from the direction of the primary gluon, etc.}
%of the sequential branching), 
not the softest partons but those with intermediate energies ($E_h\propto E_{\rm jet}^{0.3}$) multiply most
effectively in QCD cascades~\cite{Dokshitzer:1991wu}. 
As a result, the energy spectrum of hadrons as a function of $\xi$ 
%as a function of the (logarithm of the) inverse of the fraction of the parton momentum $x=E_h/E_{jet}$ 
%carried out by them, $\ln(1/x)$, 
takes  a typical ``hump-backed plateau'' (HBP) shape~\cite{Dokshitzer:1991wu,Azimov:1985by},
confirmed by jet measurements at LEP~\cite{Akrawy:1990ha} and Tevatron~\cite{:2008ec} colliders,
that can be written to first approximation in a Gaussian form of peak $\bar\xi$ and width $\sigma$:
%Soft gluon coherence suppresses the showering of soft gluons in such a way that, only particles with
%intermediate energies multiply with the highest probability. Consequently, the energy spectrum 
%of hadrons takes a hump-backed shape that can be written as 
\begin{equation}\label{eq:dlapeak}
D^{\rm ch}(\ln(1/x),Q)\simeq\exp\left[-\frac1{2\sigma^2}(\xi-\bar\xi)^2\right],\quad
\bar\xi=\ln(1/x_{\rm max})\to\frac12\ln\left(\frac{Q}{Q_0}\right),
\end{equation}
where $Q_0$ is the collinear cut-off parameter of the perturbative expansion which can be pushed
down to the value of $\lqcd$ (the so-called ``limiting spectrum''). Both the HBP peak and width
evolve approximately logarithmically with the energy of the jet: the hadron distribution peaks at
$\bar\xi~\approx$~2~(5)~GeV with a dispersion of $\sigma~\approx$~0.7~(1.4)~GeV, for a parton with
$E_{\rm jet}$~=~10~GeV~(1~TeV).\\  

%$\bar\xi(Q)$ is 
%the DLA energy evolution of the mean peak position, that in this case coincides with 
%the maximum reached by the distribution, $\sigma$ is the DLA width of the distribution
%that grows like $\sigma\propto(\ln Q)^{3/4}$ asymptotically~\cite{Dokshitzer:1991wu,Azimov:1985by}.

\begin{figure}[htpb]
\centering
\epsfig{file=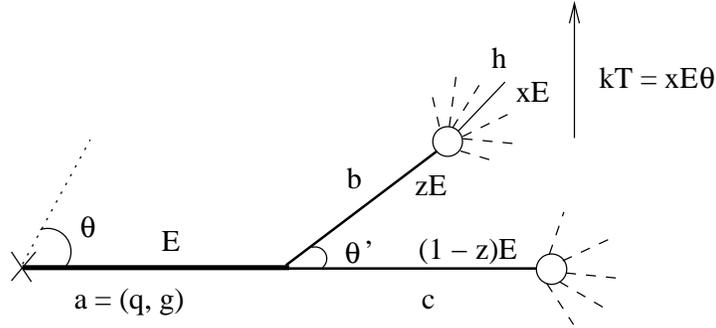,width=9.5truecm}
%\psset{unit=0.7cm}
%\begin{pspicture}(-2,-1)(8,3)
%%\psaxes[ticksize=2pt,tickstyle=bottom]{->}(0,0)(-3,0)(6,6)
%\psline[linestyle=dashed,linecolor=gray](-2,0)(0,0)
%\put(-1.8,0.5){$Q$}
%\put(-1.8,-0.5){$E_{jet}$}
%\psline[linestyle=dashed](0,0)(2.5,1.9)
%\psline[linestyle=dashed](0,0)(2.5,-1.9)
%\put(0,0){\line(1,0){7}}
%\put(1.5,0){\vector(1,0){0.8}}
%\pscoil[coilheight=1.2,coilwidth=0.2,coilarm=0](3,0)(6,1.2)
%\psarc(3,0){1.5}{0}{22}
%\put(5,0.2){$\theta$}
%\put(6.2,0.9){$E_g$}
%\end{pspicture}
\vskip 0.15cm
\caption{Relevant kinematical variables in the parton splitting process $a\to bc$: 
%of gluon radiation from a parent parton $a$ in the splitting $a\to bc$:
%$k_\perp=z(1-z)E\theta$ is the relative transverse momentum of the offsprings $b$ and $c$ 
%with the condition $Q_0\leq k_\perp\leq Q$, 
$E$ is the energy of the leading quark or gluon of virtuality $Q=E\theta$; $z$ and ($1-z$) are the energy
fractions of the intermediate offsprings $b$ and $c$ which finally fragment (at virtualities $Q_0$) into
hadrons carrying a fraction $x$ of the parent parton momentum.}
\label{fig:1}
\end{figure}

The measured fragmentation function (\ref{eq:1}) corresponds to
the sum of contributions from the fragmentation $D_{i}^{h}$ of different primary partons $i=u,d,\cdots,g$:
$$
D^{\rm h}(\ln(1/x),{\rm s})=\sum_i\int_{0}^{1}\frac{dz}{z}C_i(s;z,\alphas)\,D_{i}^{h}(x/z,{\rm s}),
$$
%where $C_i(s;z,\alphas)$ are 
and, although one cannot compute from perturbation theory the final parton-to-hadron transition encoded in
$D_{i}^{h}$, the evolution of the ``intermediate'' functions $D_{a}^{bc}$ describing the branching of a parton
of type $a$ into partons of type $b$,$c$ %(with the kinematics shown in Fig.~\ref{fig:1}) 
can be indeed theoretically predicted. The relevant kinematical variables in the parton splitting process are
shown in Fig.~\ref{fig:1} for the splitting $a(k)\to b(k_1)+c(k_2)$, such that 
$b$ and $c$ carry the energy-momentum fractions $z$ and $(1-z)$ of $a$ respectively. The 
Sudakov parametrisation for $k_1$ and $k_2$, the four-momentum of partons $b$ and $c$, can be written as
\begin{eqnarray}
k_1^{\mu}=zk^{\mu}-k_\perp^{\mu}+\frac{\vec{k}^2+k_1^2}{z}\frac{n^{\mu}}{n\cdot k},\quad
k_2^{\mu}=(1-z)k^{\mu}+k_\perp^{\mu}+\frac{\vec{k}^2+k_2^2}{1-z}\frac{n^{\mu}}{2n\cdot k},
\end{eqnarray}
with the light-like vector $n^2=0$, and time-like transverse momentum $k_\perp^2>0$ such that, 
$k\cdot k_\perp=n\cdot k_\perp=0$. Then, the scalar product $k_1\cdot k_2$ reads:
\begin{equation}\label{eq:k1k2}
k_\perp^2=2z(1-z)k_1\cdot k_2.
\end{equation}
Writing now the 4-momenta $k=\left(E,\vec{k}\right)$, $k_1=\left(zE,\vec{k}_1\right)$, 
$k_2=\left((1-z)E,\vec{k}_2\right)$ one has, 
$\mid\!\vec{k}_1\!\mid=zE$, $\mid\!\vec{k}_2\!\mid=(1-z)E$ for on-shell and massless partons $k_i^2\approx0$. 
From energy-momentum conservation: %and the previous choices: 
\begin{equation}\label{eq:ksq}
k^2=2k_1\cdot k_2=2z(1-z)E^2(1-\cos\theta)
\end{equation}
such that, replacing Eq.~(\ref{eq:ksq}) in (\ref{eq:k1k2}), one finally obtains:
\begin{equation}\label{eq:kperp}
k_\perp=2z(1-z)E\sin\frac{\theta}2.
\end{equation}
In the collinear limit, one is left with $k_\perp\approx z(1-z)Q$, where $Q=E\theta$ is the 
jet virtuality, or transverse momentum of the jet. %introduced in Eq.~(\ref{eq:dlapeak}).

The calculation of the evolution of $D_{a}^{bc}$ %radiation pattern 
inside a jet suffers from two types of singularities at each order in the strong coupling
$\alphas$: collinear $\ln \theta$-singularities when the gluon emission angle is very small
($\theta\to 0$), and infrared $\ln(1/z)$-singularities when the emitted gluon takes a very small fraction $z$ of
the energy of the parent parton. Various perturbative resummation schemes have been developed to deal with such
singularities: (i) the Leading Logarithmic Approximation (LLA) resums single logs of the type
$\left[\alphas\ln\left(k_\perp^2/\mu^2\right)\right]^n$ where $k_\perp$ is the transverse momentum of the emitted gluon 
with respect to the parent parton, (ii) the Double Logarithmic Approximation (DLA) resums soft-collinear and
infrared gluons, $g\to gg$ and $q(\bar q)\to gq(\bar q)$, for small values of $x$ and $\theta$
%where $x$ is the energy fraction of the jet taken away by one parton before the hadronisation into hadrons occurs. 
%the Double Logarithmic Approximation (DLA) resums collinear and infrared singularities
$[\alphas\ln(1/z)\ln\theta]^n\sim {\cal O}(1)$~\cite{Dokshitzer:1982xr,Fadin:1983aw}, 
(iii) Single Logarithms (SL)~\cite{Dokshitzer:1991wu,Khoze:1996dn} %take into
account for the emission of hard collinear gluons ($\theta\to 0$), $[\alphas\ln\theta]^n\sim{\cal
  O}(\sqrt{\alphas})$, and (iv) the Modified Leading Logarithmic Approximation (MLLA) provides a SL
correction to the DLA, resumming terms of order 
$\left[\alphas\ln(1/z)\ln\theta+\alphas\ln\theta\right]^n\sim [{\cal O}(1)+({\cal 
O}(\sqrt{\alphas})]$~\cite{Dokshitzer:1991wu}.
%The standard radiation pattern in QCD \cite{basics} has been extensively tested by measurements in high energy
%$e^+ e^-$ and $pp (p\bar{p})$ collisions. 
While the DLA resummation scheme~\cite{Khoze:1996dn} is known to overestimate the cascading process, 
as it neglects the recoil of the parent parton with respect to its offspring after radiation~\cite{Fadin:1983aw},
%as it ignores theenergy-momentum conservation since the radiating particle does not loose any energy after
%soft gluon radiation, 
the MLLA approximation reproduces very well the $\epem$ data, 
%from OPAL and %TASSO and the MLLA approximation, while 
although Tevatron jet results require further (next-to-MLLA, or NMLLA) refinements~\cite{PerezRamos:2007cr,Arleo:2007wn}. 
%Perturbative resummation schemes have been implemented in order to cast the single inclusive distributions of final-state hadrons in the 
%$\epem$-annihilation, hadron-hadron ($p\bar p$, $pp$) high energy collisions and DIS. Such is the case 
%of the double logarithmic approximation (DLA) and the modified leading logarithmic approximation (MLLA)~\cite{Dokshitzer:1991wu}. 
%The DLA resums soft-collinear gluons ($g\to gg$ and $q(\bar q)\to gq(\bar q)$) at small $x$ and small emission angles 
%$\theta_i\ll1$~\cite{Dokshitzer:1982xr,Fadin:1983aw}, where $x$ is the energy fraction 
%of the jet taken away by one parton before the hadronisation into hadrons occurs. 
%In the DLA approach, Feynman ladder diagrams were demonstrated to provide the largest soft-collinear 
%logarithmic contributions to the single inclusive energy-momentum distributions of final state hadrons at small $x$ 
%with a condition of strict energy ordering $\omega_{i+1}\ll\omega_i$ and strict AO: $\theta_{i+1}\ll\theta_{i}$ 
%over successive emissions of gluons that made possible the computation 
%of single inclusive observables and particle-correlations (for a review see~\cite{Khoze:1996dn}). 
%In this approximation, energy conservation is violated by neglecting the 
%recoil of the parent parton with respect to its offpring in the propagator of the emitting parton
%\cite{Fadin:1983aw}. 
The MLLA~\cite{Dokshitzer:1991wu}, partially restores 
the energy-momentum balance by including SL corrections of order $\cO{\sqrt{\alphas}}$ 
coming from the emission of hard-collinear gluons and quarks at large $x\sim1$ and small $\theta_i$ 
($g\to gg$, $q(\bar q)\to gq(\bar q)$ and $g\to q\bar q$).
%the so-called single logarithmic (SL) contributions~\cite{Dokshitzer:1991wu,Khoze:1996dn}. 
Such corrections are included %after the integration over the regular part of the 
in the standard Dokshitzer-Gribov-Lipatov-Altarelli-Parisi
(DGLAP)~\cite{Gribov:1972ri,Altarelli:1977zs,Dokshitzer:1977sg} 
splitting functions which describe the parton evolution %of %hard-collinear 
at intermediate and large $x$ in the  
%($e^+e^-$, $pp(\bar p)$\ldots) 
(time-like) FFs and (space-like) parton distribution functions (PDFs).
%in deep-inelastic scattering (DIS) $e^-p$ collisions 
The first comparison of the MLLA analytical results to the inclusive particle spectra in jets, determining the
energy evolution of the HBP peak position was performed in~\cite{Dokshitzer:1992jv}.\\

The solution of the evolution equations for the gluon and quark jets is usually obtained
%will be written in the NMLLA$+$NLO$^*$ scheme following the symbolic form,
writing the FF in the form
$$
D\simeq C(\alphas(t))\exp\left[\int^t \gamma(\alphas(t')) dt\right],\quad t=\ln Q \label{eq:Dexponentiated}
$$
where $C(\alphas(t))=1+\sqrt\alphas+\alphas\ldots$ are the coefficient functions, and
$\gamma=1+\sqrt\alphas+\alphas\ldots$ is the so-called anomalous dimension, 
%integrated over $t$, the evolution ``time" parameter chosen as a consequence of AO: $t\sim\ln\theta$ and 
%the parton shower picture. 
%In both schemes however, the diagonalisation of the matrix elements leads to the so-called
%anomalous dimension that is embedded in the fragmentation functions and structure functions, 
which in Mellin space at LLA reads,
$$
\gamma^{\rm LLA}(\omega,\alphas)=\frac{1}{4}\left(-\omega+\sqrt{\omega^2 +8N_c\alphas/\pi}\right).
$$
where $\omega$ is the energy of the radiated gluon and $N_c$ the number of colours.
At small $\omega$ or $x$, the expansion of the FF expression leads to a series of half-powers of
$\alphas$, $\gamma\simeq\sqrt{\alphas}+\alphas+\alphas^{3/2}+\ldots$, while at  
larger $\omega$ or $x$ in DGLAP, the expansion yields to a series of integer powers of $\alphas$,
$\gamma\simeq\alphas+\alphas^2+\alphas^3+\ldots$ for FFs and PDFs. In the present work 
we are mostly concerned with series of half-powers of $\sqrt{\alphas}$ generated at small $\omega$, which can be
truncated beyond $\cO\alphas$ in the high-energy limit.\\

In this paper, %as a further ingredient to cast the single inclusive distribution, 
the set of next-to-MLLA corrections of order $\cO\alphas$ for the single inclusive
hadron distribution in jets, which further improve energy conservation~\cite{Capella:1999ms,Cuypers:1991hm},
including in addition the running of the coupling constant $\alphas$ at two-loop or next-to-leading order
(NLO)~\cite{Caswell:1974gg}, are computed for the first time.
Corrections beyond MLLA were considered first in~\cite{Dremin:2000ep}, and more recently
in~\cite{Bolzoni:2013rsa}, for the calculation of the jet mean multiplicity ${\cal N}$ and the ratio
$r={\cal N}_g/{\cal N}_q$ in gluon and quark jets. We will follow the resummation
scheme presented in~\cite{Dremin:2000ep} and apply it not just to the jet multiplicities 
but extend it to the full properties of parton fragmentation functions using the distorted Gaussian (DG)
parametrisation~\cite{Fong:1990nt} for the HBP which was only used so far to compute the evolution of FFs at MLLA. 
%for the HBP that was introduced by Fong and Webber in 1991~\cite{Fong:1990nt}. 
%Thus, some results as the NMLLA$+$NLO$^*$ jet mean multiplicities, the ratio given
%in~\cite{Dremin:2000ep} as well as the MLLA DG of \cite{Fong:1990nt} should be recovered. 
The approach followed consists in writing the exponential of Eq.~(\ref{eq:Dexponentiated})
as a DG with mean peak $\bar\xi$ and width $\sigma$, including higher moments (skewness and kurtosis) that
provide an improved shape of the quasi-Gaussian behaviour of the final distribution of hadrons, and compute the
energy evolution of all its (normalised) moments at NMLLA+NLO$^{*}$ accuracy, which just depend on $\lqcd$
as a single free parameter.\\

Since the evolution of each moment is independent of the ansatz for the initial conditions assumed for
the jet hadron spectrum, and since each moment evolves independently of one another, we can obtain five different
constraints on $\lqcd$.
By fitting all the %energy evolution of the different moments of the DG measured 
measured $\epem$ jet distributions in the range of collision energies
$\sqrts\approx$~2--200~GeV~\cite{Dunwoodie:2003xt,Braunschweig:1988qm,Braunschweig:1990yd,Aihara:1988su,Itoh:1994kb,Barate:1996fi,Adeva:1991it,Akrawy:1990ha,Ackerstaff:1998hz,Buskulic:1996tt,Abreu:1996mk,Alexander:1996kh,Heister:2003aj,Heister:2003aj,Ackerstaff:1997kk,Abbiendi:1999sx,Abbiendi:2002mj}
%an approximated 
a value of $\lqcd$ can be extracted which agrees very well with that obtained from the NLO
coupling constant evaluated at the $Z$ resonance, $\alphasmZ$, 
in the minimal subtraction ($\overline{\rm MS}$) 
factorisation scheme~\cite{Altarelli:1979kv,Furmanski:1981cw,Curci:1980uw}.
Similar studies --at (N)MLLA+LO accuracy under different approximations, and with a more reduced experimental
data-set-- were done previously for various parametrizations of the input fragmentation
function~\cite{Albino:2004xa,Albino:2004yg,Albino:2005gg,Albino:2005gd} but only with
a relatively modest data-theory agreement, and an extracted LO value of $\lqcd$ with large uncertainties.\\ 

The paper is organised as follows. In Sect.~\ref{sec:eveqnmllanlo} we write the evolution equations and
provide the generic solution including the set of $\cO{\alphas}$ terms from the splitting functions in Mellin
space. In subsection~\ref{subsec:approxeveq}, the new NMLLA$+$NLO$^*$ anomalous dimension,
$\gamma_\omega^{_{\rm NMLLA+NLO^*}}$, is obtained from the evolution equations in Mellin space, being the main
theoretical result of this paper. 
In subsection~\ref{subsec:moftheDG} the Fong and Webber DG parametrisation~\cite{Fong:1990nt} for the single-inclusive 
hadron distribution is used and the energy evolution of its moments (mean multiplicity, peak position, width,
skewness and kurtosis) is computed making use of $\gamma_\omega^{_{\rm NMLLA+NLO^*}}$.
% the NMLLA$+$NLO$^*$ anomalous dimension. The corresponding 
%are then determined accordingly.
%\item 
In subsection~\ref{subsec:CFlambda}, 
%the coefficient functions will be extended from 
%the MLLA to our approach and the final distribution 
the results of our approach are compared for the quark and gluon multiplicities, %with previous results, 
recovering the NMLLA multiplicity ratio first obtained in~\cite{Capella:1999ms}.
The energy-evolution  for all the moments in the limiting spectrum 
case ($Q_0\to\lqcd$) are derived in subsection~\ref{subsec:lphdandlimspec},
%and the LPHD will be discussed in the framework of this approximation;
%\item{
%the exact MLLA solution of the evolution equations, written in terms of confluent
%hypergeometric functions was 
%provided and matched with the MLLA components (multiplicity, mean peak position, dispersion, 
%skewness and kurtosis) of the Fong-Webber distorted Gaussian. 
and the role of higher-order corrections contributing to the resummed components of the DG
which improve the overall behaviour of the perturbative series, are discussed in
subsection~\ref{sec:beyondNMLLA}, %and Appendix~\ref{section:hocorrtomoments},
and the final analytical formul{\ae} %in the DG limiting spectrum parametrisation 
are provided.
%in subsection~\ref{subsec:higherordercorrecs}.
%the median peak position, mean peak position and maximum peak position will be given in the 
%present approach;
%Subsection \ref{subsec:meanmedian} deals with 
Subsection \ref{subsec:powcorrecs} discusses our treatment of finite-mass effects and heavy-quark thresholds,
as well as other subleading corrections.
%the (limited) role of mass and power corrections in our analysis.
The phenomenological comparison of our analytical results to the world $\epem$ jet data is carried out in
Sect.~\ref{section:extracting}, 
%the maximum peak position will be compared with $\epem$ data and the 
from which a value of $\lqcd$ can be extracted from the fits.
Our results are summarised in Sect.~\ref{sec:summ} and the appendices provide more details on various
ingredients used in the calculations.

%%%%%%%%%%%%%%%%%%%%%%%%%%%%%%%%%%%%%%%%%%%%%%%%%%%%%%%%%%%%%%%%%%%%%%%%%%%%%%%%%%%%%%%%

\section{Evolution equations for the low-$x$ parton fragmentation functions} %exact solution}
\label{sec:eveqnmllanlo}

%Consider a parton of energy E and virtuality $Q\approx E\theta$
The fragmentation function of a parton $a$ splitting into partons $b$ and $c$ satisfies the
following system of evolution equations~\cite{Dokshitzer:1991wu,Azimov:1985by} as a function of the variables
defined in Fig.~\ref{fig:1}:
%As a consequence of AO, the parton shower picture and 
%the probabilistic interpretation of cascading processes, the parton fragmentation inside 
%jets can be described from the first splitting $a\to bc$ occurring inside the shower in 
%the scale $Q\approx E\theta$ (the jet virtuality). Thus, the MLLA system of evolution equations 
%for particle spectra that follows from the parton shower picture 
%reads:
\begin{equation}\label{eq:mllaeveqgen}
\frac{\partial}{\partial\ln\theta}x{D}_a^b(x,\ln E\theta)=\sum_c\int_0^1dz
\frac{\alphas(k_\perp^2)}{2\pi}P_{ac}(z)\left[\frac{x}{z}{D}_c^b\left(\frac{x}z,\ln zE\theta\right)\right],
\end{equation}
%where $k_\perp=z(1-z)E\theta$ is the relative transverse momentum of the offsprings $b$ and $c$ in
%the splitting $a\to bc$ with the condition $Q_0\leq k_\perp\leq Q$, $E$ is the energy of 
%the jet leading parton, $z$ and ($1-z$) are the energy fractions of the offsprings $b$ and $c$, 
%$Q=E\theta$ is the total virtuality of the jet, $Q_0$, as mentioned above is the parton minimum transverse
%momentum and $\lqcd$ is the QCD mass scale. 
where $P_{ac}(z)$ are the regularised DGLAP splitting functions%that can be found in the literature
~\cite{Gribov:1972ri,Altarelli:1977zs,Dokshitzer:1977sg}, which at LO are given by
\begin{eqnarray}
P_{qg}(z)\!\!&\!\!=\!\!&\!\!4C_F\left(\frac1z+\frac{z}2-1\right),
\quad P_{qq}(z)=2C_F\left(\left[\frac1{1-z}\right]_+-\frac{z}2-\frac12\right),\label{eq:pqgqq}\\
P_{gg}(z)\!\!&\!\!=\!\!&\!\!2C_A\left(\frac1z+\left[\frac1{1-z}\right]_++z(1-z)-2\right)\label{eq:pgggq},\quad 
P_{gq}(z)=n_fT_R[z^2+(1-z)^2],
\end{eqnarray} 
with $C_F=(N_c^2-1)/2N_c$ and $N_c$ respectively the Casimirs of the fundamental and adjoint 
representation of the QCD colour group $SU(3)_c$, $T_R=1/2$, and $n_f$ is the number 
of active (anti)quark flavours.
The regularisation of the splitting functions in Eq.~(\ref{eq:mllaeveqgen}) is 
performed through the $+$ distribution\footnote{The plus distribution applied to a function $F(x)$, written
  $[F(x)]_+$, is defined as $\int_0^1dx [F(x)]_+g(x)=\int_0^1dx [F(x)](g(x)-g(1))$ for any function $g(x)$.}
in Eqs.~(\ref{eq:pqgqq}) and (\ref{eq:pgggq}). 
The $\alphas$ is the strong coupling which at the two-loop level reads~\cite{Caswell:1974gg}
\begin{equation}\label{eq:twoloop}
\alphas(q^2)=\frac{4\pi}{\beta_0\ln q^2}
\left[1-\frac{2\beta_1}{\beta_0^2}\frac{\ln\ln q^2}
{\ln q^2}\right], \quad \mbox{ for }\quad q^2=\frac{k_\perp^2}{\lqcd^2},
\end{equation}
with
$$
\beta_0=\frac{11}{3}N_c-\frac{4n_fT_R}{3},\quad \beta_1=\frac{51}{3}N_c-\frac{38n_fT_R}{3}\,,
$$
being the first two coefficients involved in the perturbative expansion of the $\beta$-function through
the renormalisation group equation:
$$
\beta(\alphas)=-\beta_0\frac{\alphas^2}{2\pi}-\beta_1\frac{\alphas^3}{4\pi^2}+{\cal O}(\alphas^4).
$$

The initial condition for the system of evolution equations (\ref{eq:mllaeveqgen}) is given by a delta function 
$$
x{D}_a^b(x,\ln E\theta)\mid_{(\ln E\theta=\ln Q_0)}=\delta_a^b\cdot\delta(1-x)
$$
running ``backwards'' from the end of the parton branching process, with a clear physical interpretation: when
the transverse momentum of the leading parton is low enough, it can not fragment ($x=1$) and hadronises into a single hadron. 
The equations (\ref{eq:mllaeveqgen}) are identical
to the DGLAP evolution equations but for one detail, the shift in $\ln z$ in the second argument of the
fragmentation function $\frac{x}{z}{D}_c^b\left(\frac{x}z,\ln z+\ln E\theta\right)$, %in the r.h.s., 
that for hard partons is set to zero, $\ln z\sim0$, in the LLA. 
It corresponds to the so-called scaling violation of DGLAP
FFs in time-like evolution, and that of space-like evolution of PDFs in 
in DIS. In our framework, however, this term is responsible for the double soft-collinear contributions
that are resummed at all orders as $(\alphas\ln^2)^n$, justifying the fact that
the approach is said to be modified (MLLA) with respect to the LLA.\\

%The following changes of variables can be made for the sake of simplicity,
The evolution equations are commonly expressed as a function of two variables:
\begin{equation}
\label{eq:Ylambda}
Y=\ln\frac{E\theta}{Q_0}, \quad \lambda=\ln\frac{Q_0}{\lqcd},
\end{equation}
where $Y$ provides the parton-energy dependence of the fragmentation process, and the %hadronisation parameter
%guarantees the convergence of the perturbative approach and 
$\lambda$ specifies, in units of $\lqcd$, the value of the hadronisation scale $Q_0$ down to which the parton
shower is evolved. Standard parton showers Monte Carlo codes, such as {\sc pythia}~\cite{Sjostrand:2006za}, use $Q_0$ 
values of the order of $\cO{\rm 1~GeV}$ whereas in the limiting spectrum%approximation%where the HBP has been evaluated 
~\cite{Dokshitzer:1991wu}, that will be used here, it can be taken as low as $\lambda\to 0$, i.e. $Q_0~\to~\lqcd$.
Applying the Mellin transform to the single inclusive 
distribution in Eq.~(\ref{eq:mllaeveqgen}) %with the following variables,
\begin{equation}
{\cal D}(\omega,Y)=
\int_0^{\infty}d\xi e^{-\omega\xi}D(\xi,Y),\; %\xi=\ln\frac1x,\;y=\ln\frac{xE\theta}{Q_0},\;\xi+y=Y,
\end{equation}
and introducing 
\begin{equation}
\label{eq:kinem}
\hat\xi=\ln\frac1z,\quad \hat{y}=\ln\frac{k_\perp}{Q_0},\quad \hat\xi+\hat{y}
=\ln\frac{E\theta}{Q_0}\equiv Y, 
\end{equation}
with $k_\perp\approx zE\theta$ in the soft approximation ($z\ll1$), one is left with 
the integro-differential system of evolution equations for the non-singlet distributions
\begin{equation}\label{eq:mllaeveqgenbis}
\frac{\partial}{\partial Y}{\cal D}(\omega,Y)
=\int_0^{\infty}d\hat\xi e^{-\omega\hat\xi}P(\hat\xi)
\frac{\alphas(Y-\hat\xi)}{2\pi}{\cal D}(\omega,Y-\hat\xi),
\end{equation}
where
\begin{equation}\label{eq:dglap}
P(\hat\xi)=
\begin{pmatrix}
P_{qq}(\hat\xi)&P_{q g}(\hat\xi)\\
P_{gq}(\hat\xi)&P_{gg}(\hat\xi) 
\end{pmatrix},
\quad
{\cal D}(\omega,\hat{y})=\begin{pmatrix}
{\cal D}_{{q}}(\omega,\hat{y}) \\
{\cal D}_{g}(\omega,\hat{y})
\end{pmatrix}
\end{equation}
and the lower and upper indices have been omitted for the sake of simplicity. The NLO strong coupling 
(\ref{eq:twoloop}) can be rewritten as a function of the new variables (\ref{eq:kinem}), such that
\begin{equation}\label{eq:alphaparm}	
\alphas(\hat{y})=\frac{2\pi}{\beta_0(\hat{y}+\lambda)}\left[1-\frac{\beta_1}{\beta_0^2}
\frac{\ln2(\hat{y}+\lambda)}{\hat{y}+\lambda}\right],\quad \hat{y}=Y-\hat{\xi}.
\end{equation}
%where $\lambda$ was defined in (\ref{eq:Ylambda}).
The parton density $xD(x,Y)$ is then obtained through the inverse Mellin transform:
\begin{equation}\label{eq:inversemellin}
D(\hat\xi,Y)=\int_C\frac{d\omega}{2\pi i}e^{\omega\xi}{\cal D}(\omega,Y)
\end{equation}
where the contour $C$ lies to the right of all singularities in the $\omega$-complex plane.
%With the variables introduced in (\ref{eq:kinem}) the coupling constant (\ref{eq:twoloop}) could 
%be rewritten in (\ref{eq:alphaparm}), such that
In the high-energy limit ($Q\gg Q_0$) and hard fragmentation region ($Y\gg\hat\xi$ or $x\sim1$), 
one can replace in the r.h.s. of Eq.~(\ref{eq:mllaeveqgenbis}) the following expansion\footnote{Note 
that the MLLA solution~\cite{Dokshitzer:1991wu} to the evolution equations corresponds to the 
replacement $\alphas(Y-\bar\xi){\cal D}(\omega,Y-\bar\xi)\approx\alphas(Y){\cal D}(\omega,Y)$ 
accounting for the single logarithmic corrections of relative order $\cO{\sqrt{\alphas}}$.}:
\begin{equation}\label{eq:kernelexpansion}
\alphas(Y-\bar\xi){\cal D}(\omega,Y-\bar\xi)=
e^{-\bar\xi\frac{\partial}{\partial Y}}\alphas(Y){\cal D}(\omega,Y),
\quad e^{-\bar\xi\frac{\partial}{\partial Y}}=\sum_{n=0}^{\infty}\frac{(-1)^n}{n!}\frac{\partial^n}{\partial Y^n}.
\end{equation}
Thus, replacing Eq.~(\ref{eq:kernelexpansion}) into (\ref{eq:mllaeveqgenbis}) one obtains
\begin{equation}\label{eq:mllaeveqgen4}
\frac{\partial}{\partial Y}{\cal D}(\omega,Y)
=\left(\int_0^{\infty}d\hat\xi e^{-\Omega\hat\xi}P(\hat\xi)\right)\frac{\alphas(Y)}{2\pi}{\cal D}(\omega,Y),
\end{equation} 
which allows for the factorisation of $\alphas(Y){\cal D}(\omega,Y)$, and leads to the equation
\begin{equation}
\label{eq:mllaeveqgenter}
\frac{\partial}{\partial Y}{\cal D}(\omega,Y)=P(\Omega)
\frac{\alphas(Y)}{2\pi}{\cal D}(\omega,Y),\quad P(\Omega)=\int_0^{\infty}d\hat\xi e^{-\Omega\hat\xi}P(\hat\xi),
\end{equation}
more suitable for analytical solutions. 
Truncating the series at higher orders translates into including corrections
$\cO\alphas$ which better account for energy conservation, particularly at large $x$. 
In Mellin space, the expansion can be made in terms of the differential operator
$\Omega\equiv\omega+\partial/\partial Y$ 
such that, up to the second term in $\Omega$, one is left with NMLLA corrections
of order $\cO{\alphas}$~\cite{PerezRamos:2007cr}. Explicitly, the inclusion of higher-order 
corrections from the second term of $\alphas(Y-\bar\xi){\cal D}(\omega,Y-\bar\xi)\approx\alphas{\cal D}-
\bar\xi\partial(\alphas{\cal D})/\partial Y$, followed by the integration over the splitting functions 
(\ref{eq:pqgqq})--(\ref{eq:pgggq}) in $x$ space in the r.h.s. of Eq.~(\ref{eq:mllaeveqgenbis}), 
is equivalent to the expansion $P(\Omega)=P^{(0)}+P^{(1)}\Omega$ in Mellin space in the r.h.s. of
(\ref{eq:mllaeveqgenter}), where $P^{(0)}$ and $P^{(1)}$ are constants. 
The expansion of the matrix elements $P(\Omega)$ in $\Omega$ can be obtained 
from the original expressions of the Mellin transformed splitting functions~\cite{Dokshitzer:1978hw}, as given
in Eqs.~(\ref{eq:expomega3})--(\ref{eq:expomega4}) in Appendix~\ref{sec:mellinsplittings}, 
which leads to the following expressions:
\begin{subequations}
\begin{eqnarray}\label{eq:expomega1}
P_{gg}(\Omega)\!&\!=\!&\!\frac{4N_c}{\Omega}-\frac{11}3N_c-\frac43n_fT_R+4N_c\left(\frac{67}{36}-
\frac{\pi^2}{6}\right)\Omega+{\cal O}(\Omega^2),\\
P_{gq}(\Omega)\!&\!=\!&\!\frac{8n_fT_R}{3}-\frac{26n_fT_R}{9}\Omega+{\cal O}(\Omega^2),\\
P_{qg}(\Omega)\!&\!=\!&\!\frac{4C_F}{\Omega}-3C_F+\frac72C_F\Omega+{\cal O}(\Omega^2),\\
P_{qq}(\Omega)\!&\!=\!&\!4C_F\left(\frac58-\frac{\pi^2}6\right)\Omega+{\cal O}(\Omega^2).
\label{eq:expomega2}
\end{eqnarray}
\end{subequations}
where the finite terms for $\Omega\to0$ constitute the new subset
to be computed for the first time in this work. The solution of the evolution 
equations in the MLLA were considered in~\cite{Dokshitzer:1991wu}
up to the regular terms with $\delta P_{qq}(\Omega)\Omega=0$.
By including those proportional to $\Omega$, one is in addition considering the set of higher-order 
corrections $\cO\alphas$ known as NMLLA that improve energy conservation~\cite{Dremin:2000ep}. 
The diagonalisation of the matrix (\ref{eq:dglap}) in order to solve
(\ref{eq:mllaeveqgenter}) results into two trajectories (eigenvalues), 
which can be written as~\cite{Dokshitzer:1991wu,Dokshitzer:1978hw}
\begin{equation}\label{eq:extrajectories}
P_{\pm\pm}(\Omega)=\frac12\left[P_{gg}(\Omega)+P_{qq}(\Omega)
\pm\sqrt{\left(P_{gg}(\Omega)-P_{qq}(\Omega)\right)^2+4P_{gq}(\Omega)P_{qg}(\Omega)}\right].
\end{equation}
Substituting Eqs.~(\ref{eq:expomega1})--(\ref{eq:expomega2}) into (\ref{eq:extrajectories}) and
performing the expansion again up to terms $\cO\Omega$, yields:
\begin{subequations}
\begin{eqnarray}
\label{eq:postraj}
P_{++}(\Omega)\!&\!=\!&\!\frac{4N_c}{\Omega}-a_1+4N_ca_2\Omega+{\cal O}(\Omega^2),\\
P_{--}(\Omega)\!&\!=\!&\!-b_1+4C_Fb_2\Omega+{\cal O}(\Omega^2),
\label{eq:postraj1}
\end{eqnarray}
\end{subequations}
where the terms proportional to $\Omega$ are new in this framework.
The set of constants involved in Eqs.~(\ref{eq:postraj}) and (\ref{eq:postraj1}) reads:
\begin{subequations}
\begin{eqnarray}
a_1\!&\!=\!&\!\frac{11}{3}N_c+\frac43n_fT_R\left(1-2\frac{C_F}{N_c}\right),\\
a_2\!&\!=\!&\!\frac{67}{36}-\frac{\pi^2}{6}
-\frac{n_fT_RC_F}{18N_c^2}\left[\frac{11}{3}N_c-4\frac{n_fT_R}{N_c}\left(1-2\frac{C_F}{N_c}\right)\right],\\
b_1\!&\!=\!&\!\frac{8n_fT_RC_F}{3N_c},\\
b_2\!&\!=\!&\!\frac58-\frac{\pi^2}{6}+\frac{n_fT_R}{18N_c}\left[\frac{11}{3}N_c-4\frac{n_fT_R}{N_c}
\left(1-2\frac{C_F}{N_c}\right)\right].
\end{eqnarray}
\end{subequations}
Therefore, the diagonalisation of Eq.~(\ref{eq:mllaeveqgenter}) leads to two equations:
\begin{equation}\label{eq:trajectories}
\frac{\partial}{\partial Y}{\cal D}^{\pm}(\omega,Y,\lambda)=P_{\pm\pm}(\Omega)
\frac{\alphas(Y)}{2\pi}{\cal D}^{\pm}(\omega,Y,\lambda),
\end{equation}
such that in the new $D^{\pm}$-basis the respective solutions read:
\begin{subequations}
\begin{eqnarray}\label{eq:DqDgdistr}
{\cal D}_q(\omega,Y,\lambda)\!&\!=\!&\!\frac{P_{qg}(\Omega)}
{P_{++}(\Omega)-P_{--}(\Omega)}\left[{\cal D}^+(\omega,Y,\lambda)
-{\cal D}^-(\omega,Y,\lambda)\right],\\
{\cal D}_g(\omega,Y,\lambda)\!&\!=\!&\!\frac{P_{++}
(\Omega)-P_{qq}(\Omega)}{P_{++}(\Omega)-P_{--}(\Omega)}{\cal D}^+(\omega,Y,\lambda)-
\frac{P_{--}(\Omega)-P_{qq}(\Omega)}{P_{++}(\Omega)
-P_{--}(\Omega)}{\cal D}^-(\omega,Y,\lambda).
\end{eqnarray} 
\end{subequations}
where the ratios in front of ${\cal D}^{\pm}$ are the coefficient functions that 
will be evaluated hereafter.
Notice that in the ${\cal D}^{\pm}$ basis, the off-diagonal terms $P_{+-}(\Omega)=0$ and 
$P_{-+}(\Omega)=0$ vanish for LO splitting functions, while this is no longer true for time-like splitting
functions obtained from the $\MSbar$ factorisation scheme 
beyond LO~\cite{Vogt:2011jv}, as explained in~\cite{Bolzoni:2013rsa} for multiparticle production. Following
this logic, ${\cal D}^{\pm}$ should first be determined in order 
to obtain the gluon and quark jets single inclusive distributions. 

%%%%%%%%%%%%%%%%%%%%%%%%%%%%%%%%%%%%%%%%%%%%%%%%%%%%%%%%%%%%%%%%%%%%%%%%%%%%%%%%%%%%%%%%

\section{Evolution of the parton fragmentation functions at NMLLA +NLO$^*$}
\label{sec:NMLLANLO}

%\subsection{Approximated evolution equations}
\subsection{Anomalous dimension at NMLLA +NLO$^*$}
\label{subsec:approxeveq}
%
%The approach of the DG to solve the evolution equations
%in~\cite{Fong:1990nt} was a good approximation which in 
Our NMLLA$+$NLO$^*$ scheme involves adding further corrections
$\cO\alphas$ from contributions proportional to $\Omega$ in the Mellin representation 
of the expanded splitting functions, %, Eqs.~(\ref{eq:}), 
and considering the two-loop strong coupling,
Eq.~(\ref{eq:alphaparm}). We label our approach as NLO$^*$ to indicate that the full set of NLO corrections
are only approximately included, as the two-loop splitting functions (discussed e.g. in~\cite{Bolzoni:2013rsa})
are not incorporated. After diagonalisation of the original evolution equations (\ref{eq:mllaeveqgen}), 
the Eqs.~(\ref{eq:trajectories}) for ${\cal D}^{\pm}$ result in the following expressions for ${\cal D}^{+}$
and ${\cal D}^{-}$:
\begin{eqnarray}
\left(\omega+\frac{\partial}{\partial Y}\right)\frac{\partial}{\partial Y}{\cal D}^+(\omega,Y,\lambda)=\left[
1-\frac{a_1}{4N_c}\left(\omega+\frac{\partial}{\partial Y}\right)+a_2\left(\omega
+\frac{\partial}{\partial Y}\right)^2\right]
4N_c\frac{\alphas}{2\pi}{\cal D}^+(\omega,Y,\lambda)
\label{eq:gluonDg}
\end{eqnarray}
\begin{equation}
\frac{\partial}{\partial Y}{\cal D}^-(\omega,Y,\lambda)=-b_1\frac{\alphas}{2\pi}{\cal D}^-(\omega,Y,\lambda)+4C_Fb_2
\left(\omega+\frac{\partial}{\partial Y}\right)
\frac{\alphas}{2\pi}{\cal D}^-(\omega,Y,\lambda).
\label{eq:negativetraj}
\end{equation}
The leading contribution to ${\cal D}^-$ after setting $b_2=0$ in Eq.~(\ref{eq:negativetraj}) reads:
\begin{equation}\label{eq:Dmoinslambda}
{\cal D}^-(\omega, Y,\lambda)\approx\left(\frac{\lambda}{Y+\lambda}\right)^{\frac{b_1}
{4N_c\beta_0}}{\cal D}^-(\omega,\lambda).
\end{equation}
%Following from (\ref{eq:Dmoinslambda}), since 
%The exponent $\frac{b_1}{4N_c\beta_0}=\cO{10^{-2}\sqrt\alphas}$ induces a very small (non-Gaussian) correction, 
The exponent $b_1/(4N_c\beta_0)=\cO{10^{-2}\sqrt\alphas}$ induces a very small (non-Gaussian) correction, 
%it does not provide a Gaussian shape to the distribution
%and can therefore 
which can be neglected asymptotically, for $Y+\lambda\gg\lambda$. 
Thus, the (+) trajectory (\ref{eq:postraj}) 
provides the main contribution to the single inclusive distribution $D(\xi,Y)=xD(x,Y)$ at small 
$x\ll1$, after applying the inverse Mellin transform (\ref{eq:inversemellin}). Hard corrections 
proportional to $a_1$ and $a_2$ account for the energy balance in the hard fragmentation region
and are of relative order $\cO{\sqrt{\alphas}}$ and $\cO\alphas$ respectively with respect to the
$\cO1$ DLA contribution. 
The NLO expression (\ref{eq:twoloop}) results in corrections $\propto\beta_0$ at MLLA, and
$\propto\beta_0,\beta_1$ at NMLLA which provide a more accurate consideration of running coupling effects at
small $x\ll1$~\cite{Dremin:2000ep}.
In Ref.~\cite{Dremin:2000ep}, the mean multiplicities, multiplicity correlators in gluon and quark jets, and
the ratio of gluon and quark jet multiplicities were also studied at NMLLA, where corrections $\propto\beta_1$
were accordingly included. Here, we extend the NMLLA analysis to all moments of the fragmentation function.\\

The solution of Eq.~(\ref{eq:gluonDg}) can be written in the compact form:
\begin{equation}\label{eq:Dgomega}
{\cal D}^+(\omega,Y,\lambda)=E_+(\omega,\alphas(Y+\lambda)){\cal D}^+(\omega,\lambda),
\end{equation}
with the evolution ``Hamiltonian":
\begin{equation}\label{eq:Egg}
E_+(\omega,\alphas(Y+\lambda))=\exp\left[\int_{0}^{Y}dy\,\gamma(\omega, \alphas(y+\lambda))
\right].
\end{equation}
that describes the parton jet evolution from its initial 
virtuality $Q$ to the lowest possible energy scale $Q_0$, at which the parton-to-hadron
transition occurs. 
In Eq.~(\ref{eq:Egg}), $\gamma(\omega, \alphas(y))$ is the anomalous dimension that
mixes $g\to gg$ and $g\to q\bar q$ splittings and is mainly dominated by soft
gluon bremsstrahlung ($g\to gg$). Introducing the shorthand notation
$\gamma_\omega=\gamma(\omega,\alphas(Y))$, the MLLA anomalous dimension has been determined 
in the past~\cite{Fong:1990nt,Dokshitzer:1991wu}, setting $a_2=0$ and $\beta_1=0$ in Eq.~(\ref{eq:gluonDg}), and is given by 
\begin{eqnarray}
\gamma_\omega^{_{\rm MLLA}}\!&\!=\!&\!\frac12\left(-\omega+\sqrt{\omega^2+4\gamma_0^2}\right)\cr
\!&\!+\!&\!\frac{\alphas}{2\pi}\left[-\frac12a_1\left(1+\frac{\omega}{\sqrt{\omega^2
+4\gamma_0^2}}\right)+\beta_0\frac{\gamma_0^2}{\omega^2+4\gamma_0^2}\right]+{\cal O}(\alphas^{3/2}),
\label{eq:Pgg}
\end{eqnarray}
where $\gamma_0^2$ is the DLA anomalous dimension amounting to 
\begin{equation}
\gamma_0^2=\frac{4N_c\alphas}{2\pi}=\frac{4N_c}{\beta_0(Y+\lambda)}.
\end{equation}
The first term of Eq.~(\ref{eq:Pgg}) is the DLA main contribution, of order ${\cal O}(\sqrt{\alphas})$, which
physically accounts for soft gluon multiplication, the second and third terms are 
SL corrections ${\cal O}(\alphas)$ accounting for the energy balance ($\propto a_1$) and
running coupling effects ($\propto\beta_0$). %The last sentence may be somewhat misleading
It is important to make the difference between orders and relative orders mentioned above. 
Indeed, if one looks at the l.h.s. of the evolution equation (\ref{eq:gluonDg}) for ${\cal D}^+$, 
$(\omega+\partial/\partial Y)\partial {\cal D^{+}}/\partial Y=\cO\alphas$, the first term in the 
r.h.s is $\cO\alphas$, the second one proportional to $a_1$ is $\cO{\alphas^{3/2}}$, and the third one,
proportional to $a_2$, is $\cO{\alphas^{2}}$ such that after factorising the whole equation by $\cO\alphas$ one is 
left with the relative orders of magnitude in $\sqrt\alphas$. Setting Eq.~(\ref{eq:Dgomega}) in (\ref{eq:gluonDg})
leads to the perturbative differential equation
\begin{equation}\label{eq:eqgammanmlla}
(\omega+\gamma_\omega)\gamma_\omega-\frac{2N_c\alphas}{\pi}=-\beta(\alphas)\frac{d\gamma_\omega}{d\alphas}
-a_1(\omega+\gamma_\omega)\frac{\alphas}{2\pi}-\frac{a_1}{2\pi}\beta(\alphas)
+a_2(\omega^2+2\omega\gamma_\omega+\gamma_\omega^2)\frac{\alphas}{2\pi},
\end{equation}
which will be solved after inserting the two-loop coupling (\ref{eq:twoloop}) in order to include 
corrections $\propto\beta_1$ as well. 
%Here, 
%$$
%\beta(\alphas)=-\beta_0\frac{\alphas^2}{2\pi}-\beta_1\frac{\alphas^3}{4\pi^2}+{\cal O}(\alphas^4)
%$$ 
%is the QCD $\beta(\alphas)$ function~\cite{Caswell:1974gg} which keeps trace of running coupling effects
%in the present framework. 
The equation can be solved iteratively (perturbatively) by setting the MLLA
anomalous dimension written in Eq.~(\ref{eq:Pgg}) in the main and subleading contributions of 
Eq.~(\ref{eq:eqgammanmlla}), to find: 
%Neglecting contributions that go beyond ${\cal O}(\alphas)$, it is straightforward to find
\begin{eqnarray}
\gamma_\omega^{_{\rm NMLLA+NLO^*}}\!\!&\!\!=\!\!&\!\!\gamma_\omega^{_{\rm MLLA}}+\frac{\gamma_0^4}{16N_c^2}
\left\{a_1^2\frac{\gamma_0^2}{(\omega^2+4\gamma_0^2)^{3/2}}+\frac{a_1\beta_0}{2}\left(\frac1{\sqrt{\omega^2+4\gamma_0^2}}-\frac{\omega^3}{(\omega^2+4\gamma_0^2)^2}\right)
\right.\cr
\!\!&\!\!+\!\!&\!\!\left.\beta_0^2\left(\frac{2\gamma_0^2}
{(\omega^2+4\gamma_0^2)^{3/2}}-\frac{5\gamma_0^4}{(\omega^2+4\gamma_0^2)^{5/2}}\right)
-4N_c\frac{\beta_1}{\beta_0}\frac{\ln2(Y+\lambda)}{\sqrt{\omega^2+4\gamma_0^2}}
\right\}\cr
\!\!&\!\!+\!\!&\!\!\frac{1}{4}a_2\gamma_0^2\left[\frac{\omega}{(\omega^2+4\gamma_0^2)^{1/4}}
+(\omega^2+4\gamma_0^2)^{1/4}\right]^2+{\cal O}(\gamma_0^4),
\label{eq:nmllagamma}
\end{eqnarray}
which is the main theoretical result of this paper. Terms proportional to $a_1^2$, $a_1\beta_0$ and $\beta_0^2$
are of order ${\cal O}(\alpha_s^{3/2})$, and
were previously calculated in the (N)MLLA+LO scheme  described in~\cite{Albino:2004yg}. Those 
proportional to $\beta_1$ and $a_2$  are computed for the first time 
in our NMLLA+NLO* framework. Indeed, the single correction $\propto\beta_1$ is
obtained replacing Eq.~(\ref{eq:twoloop}) in the l.h.s.
of (\ref{eq:eqgammanmlla}), which leads to the equation,
$$
\gamma_\omega^2+\omega\gamma_\omega-\gamma_0^2+\frac{\beta_1}
{4N_c\beta_0}\gamma_0^4\ln2(Y+\lambda)+\ldots\!=\!0
\Rightarrow \gamma_\omega\!=\!\gamma_\omega^{_{\rm DLA}}-\frac{\gamma_0^4}{4N_c}
\left\{\frac{\beta_1}{\beta_0}\frac{\ln2(Y+\lambda)}{\sqrt{\omega^2+4\gamma_0^2}}+\ldots\right\}
$$
with $\gamma_\omega^{_{\rm DLA}}=\frac12\left(-\omega+\sqrt{\omega^2+4\gamma_0^2}\right)$.
Since $\ln(Y+\lambda)={\cal O}(1)$ and $\omega={\cal O}(\sqrt{\alphas})$,
and following $\alphas$ power counting, this correction has naturally the same
order of magnitude ${\cal O}(\alphas^{3/2})$ as the other terms
and should not be neglected. The other new correction $\propto a_2\gamma_0^2\propto\alpha_s$  
adds those NMLLA contributions arising from the $\propto\omega$ terms in 
the LO splitting functions (\ref{eq:expomega1})--(\ref{eq:expomega2}), known to better account
for energy conservation. Since this correction is multiplied by a term $[\ldots]^2={\cal O}(\sqrt{\alpha_s})$,
the overall result is ${\cal O}(\alpha_s^{3/2})$ and, thus, of the same order of magnitude as the previous
terms such that, the full resummed result is ${\cal O}(\alpha_s^{3/2})$.

%%%%%%%%%%%%%%%%%%%%%%%%%%%%%%%%%%%%%%%%%%%%%%%%%%%%%%%%%%%%%%%%%%%%%%%%%%%%%%%%%%%%%%%%

%\section{Evolution of parton fragmentation functions at NMLLA +NLO$^*$ (distorted-Gaussian parametrisation)}
\subsection{Distorted Gaussian (DG) parametrisation for the fragmentation function}
%\subsection{Method of the distorted Gaussian}
\label{subsec:moftheDG}

%The single inclusive distribution which leads to the hump-backed-plateau was found after exactly solving the 
%Eq.~(\ref{eq:gluonDg}) with $a_2=0$ and $\beta_1=0$. The differential Eq.~(\ref{eq:gluonDg}) 
%was solved and the solution was expressed in terms of confluent hypergeometric functions that were then reduced
%to an integral representation of the Bessel series of the second kind~\cite{Azimov:1984np}.
%The approach of the DG which makes use of the expanded MLLA anomalous dimension Eq.~(\ref{eq:Pgg}) will be 
%extended to the NMLLA$+$NLO$^*$ anomalous dimension Eq.~(\ref{eq:nmllagamma}) in this framework. 
The distorted Gaussian (DG) parametrisation of the single inclusive distribution of hadrons in jets
at small $x$ (or $\omega\to0$) was introduced by Fong and Webber in 1991~\cite{Fong:1990nt}, and   
in $x$-space it reads:
\begin{equation}\label{eq:xDg}
D^+(\xi,Y,\lambda) = \frac{{\cal N}}{\sigma\sqrt{2\pi}}\exp\left[\frac18k-\frac12s\delta-
\frac14(2+k)\delta^2+\frac16s\delta^3+\frac1{24}k\delta^4\right]\,,
\end{equation} 
where, $\delta=(\xi-\bar\xi)/\sigma$, ${\cal N}$ is the asymptotic average multiplicity inside a jet, and
$\bar\xi$, $\sigma$, $s$, and $k$ are respectively the mean peak position, the dispersion, the skewness, and
kurtosis of the distribution.  
The distribution should be displayed in the interval $0\leq \xi\leq Y$ which depends on the jet energy, and 
the values of $Q_0$ and $\lqcd$. The three scales of the process are organised in the form 
$Q\gg Q_0\geq\lqcd$. The formula (\ref{eq:xDg}) reduces %is indeed universal, it corresponds 
to a Gaussian for $s=k=0$ and its generic expression does not depend on the approach 
or level of accuracy used for the computation of its evolution.\\ %of the hadron distribution.
%However, the extension from the MLLA to
%the NMLLA$+$NLO$^*$ scheme means that higher-order corrections to 
%The evolution of its different components, ${\cal N}$, $\sigma$, $\bar\xi$, $s$ and $k$ will be computed 
%through the NMLLA$+$NLO$^*$ anomalous dimension, Eq.~(\ref{eq:nmllagamma}).\\

\begin{figure}[htbp]
\begin{center}
\epsfig{file=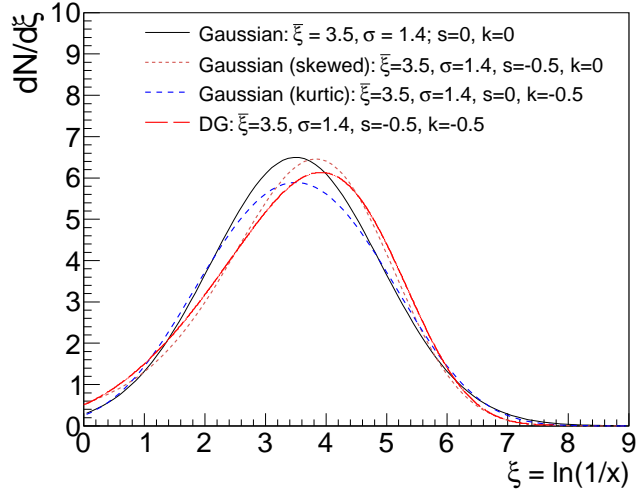, width=8.5truecm}
\caption{\label{fig:DG} Comparison of various Gaussian-like hadron distributions in jets sharing the same mean $\xi$
  position and width ($\bar\xi=3.5$ and $\sigma=1.4$)
 but with different third and fourth moments: (i) symmetric Gaussian, (ii) skewness $s=-0.5$, 
(iii) negative kurtosis $k=-0.5$, and (iv) full distorted Gaussian with $s = k =-0.5$.}
%of the peak position and $\sigma$, for $s\ne0$ and $k\ne0$ independently.} 
\end{center}
\end{figure}

As an example of the effects of non-zero skewness and kurtosis,  we compare in Fig.~\ref{fig:DG} the shapes of
four different single-inclusive hadron distributions of width $\sigma=1.4$ and mean position at $\bar\xi=3.5$ in the
interval $0\leq\xi\lesssim 7$ typical of jets at LEP-1 energies: (i) an exact Gaussian,
%(ii) a shifted Gaussian with $\bar\xi=3.5$, $\sigma=1.4$, skewness $s=0$, kurtosis 
%$k=0$, (iii) a second Gaussian with $\xi=3.5$, $\sigma=2$, $s=0$, $k=0$, 
(ii) a skewed Gaussian with %$\bar\xi=3.5$, $\sigma=1.4$, 
$s=-0.5$, $k=0$,  
(iii) a kurtic Gaussian with %$\bar\xi=3.5$, $\sigma=1.4$, 
$s=0$, $k=-0.5$,
and (iv) a DG %obtained by the combination of 
including both ``distorting'' $s,k$ components above. %the skewness and kurtosis.
%Figure~\ref{fig:DG} (right) displays the Gaussian compared with the DG given by superposition of all
%components. 
As can be seen, the shape of the DG differs from that of the pure Gaussian, mainly away from
the hump region. A negative skewness displaces the peak of the Gaussian to higher $\xi$ values while adding a
longer tail to low $\xi$, and a negative kurtosis tends to make ``fatter'' its width.\\

In order to obtain the evolution of the different DG components, we will proceed by following the same steps as
in~\cite{Fong:1990nt} but making use instead of the expanded NMLLA$+$NLO$^*$ anomalous dimension,
Eq.~(\ref{eq:nmllagamma}), computed here. Defining $K_n$ as the $n$-th moment of the single inclusive distribution:
\begin{equation}\label{eq:Kn}
K_n(Y,\lambda)=\left(-\frac{d}{d\omega}\right)^n\ln\left[{\cal D}^+(\omega,Y,\lambda)\right]_{\omega=0},
\end{equation}
the different components (normalised moments) of the DG are given by\footnote{We list also $k_5$ which is
  needed to obtain the maximum peak position $\ximax$ from $\bar\xi$, as discussed below.}:
\begin{equation}\label{eq:K12345}
{\cal N}= K_0,\quad \bar\xi= K_1,\quad \sigma=\sqrt{K_2},\quad s=\frac{K_3}{\sigma^3},\quad k=\frac{K_4}{\sigma^4},\quad k_5=\frac{K_5}{\sigma^5};
\end{equation}
such that after plugging Eq.~(\ref{eq:Egg}) into (\ref{eq:Dgomega}) and what results from it into (\ref{eq:Kn}),
one is left with
\begin{equation}\label{eq:moments}
K_{n\geq0}=\int_0^Ydy\left(-\frac{\partial}
{\partial\omega}\right)^n\gamma_\omega(\alphas(y+\lambda))\bigg|_{\omega=0},
\end{equation}
which is more suitable for analytical calculations since it directly involves the anomalous
dimension expression~(\ref{eq:nmllagamma}). 

\paragraph{Multiplicity.}
The multiplicity is obtained from the zeroth moment, i.e. the integral, of the single-particle distribution.
Setting $\omega=0$ in Eq.~(\ref{eq:nmllagamma}), one obtains
\begin{eqnarray}\label{eq:gammavsmult}
\gamma_{\omega}(0,\alphas)\!&\!=\!&\!\gamma_0-\frac1{8N_c}\left(a_1-\frac{\beta_0}2\right)\gamma_0^2\\
\!&\!+\!&\!\frac12\left[a_2+\frac1{32N_c^2}\left(\frac{a_1^2}{2}+a_1\beta_0+\frac{3\beta_0^2}{8}\right)
-\frac{\beta_1\ln2(Y+\lambda)}{4N_c\beta_0}\right]\gamma_0^3\,,\notag
\end{eqnarray}
from which the mean multiplicity ${\cal N}(Y,\lambda)$ can be straightforwardly derived 
%from Eq.~(\ref{eq:gammavsmult}) 
by integrating over $y$:
%The NMLLA$+$NLO$^*$ expression for the mean multiplicity 
%accounting for the two-loop coupling constant reads,
\begin{equation}\label{eq:meanmultip}
{\cal N}(Y,\lambda)={\cal N}_0\exp\left[f_{{\cal N}}(Y,\lambda)-f_{{\cal N}}(0,\lambda)\right]
\end{equation}
where
\begin{eqnarray}\label{eq:meanmultiprate}
f_{{\cal N}}(y,\lambda)\!\!&\!\!=\!\!&\!\!\sqrt{\frac{16N_c}{\beta_0}(y+\lambda)}-
\left(\frac{a_1}{\beta_0}-\frac12\right)\ln\sqrt{(y+\lambda)}-\frac{2N_c}{\beta_0}
\left[a_2+\frac1{4}\left(\frac{a_1}{4N_c}\right)^2
+\frac{a_1\beta_0}{32N_c^2}\right.\cr
\!\!&\!\!+\!\!&\!\!\left.\frac3{16}\left(\frac{\beta_0}{4N_c}\right)^2
-\frac{\beta_1}{4N_c\beta_0}(\ln2(y+\lambda)+2)\right]\sqrt{\frac{16N_c}{\beta_0(y+\lambda)}}.
\end{eqnarray} 
As expected, the mean multiplicity (\ref{eq:meanmultip}) including the two-loop $\alphas$ exactly coincides
with the expression obtained in~\cite{Dremin:2000ep}. This cross-check 
supports the validity of our ``master'' NMLLA$+$NLO$^*$ formula~(\ref{eq:nmllagamma}) 
for the anomalous dimension at small $\omega$, which is not surprising as the gluon jet evolution equation
solved in~\cite{Dremin:2000ep} for the mean multiplicity coincides with Eq.~(\ref{eq:gluonDg})  
after setting $\omega=0$ and ${\cal N}(Y,\lambda)={\cal D}^+(0,Y,\lambda)$. The first term
in Eq.~(\ref{eq:meanmultiprate}) is the DLA rate of multiparticle production, the second 
and third terms provide negative corrections that account for energy conservation 
and decrease the multiplicity.
%, while the $\propto\beta_1$ term provides a positive 
%correction that can drastically increase the multiplicity at low energy scales. 
However, the third term, proportional to $\beta_1$, is positive and can be large
since it accounts for NLO coupling corrections. Though, due to energy conservation, one may
expect the multiplicity to decrease in the present scheme  running coupling effects
take over and can drastically increase the multiplicity as well as single inclusive cross-sections at the 
energy scales probed so-far at $\epem$ colliders. Only at asymptotically high-energy scales, that is for
$Q_0\gg\lqcd$, the energy conservation becomes dominant over running coupling effects, thus inverting these trends.
The ratio of multiplicities in quark and gluon jets are discussed in Sect.~\ref{subsec:CFlambda} and compared
with the calculations of~\cite{Dremin:2000ep}. 
%was also performed within the same scheme of resummation and the results were compared with 
%$\epem$-annihilation data.
Performing the numerical evaluation for $n_f=5$ quark flavours\footnote{As will be seen below the 
dependence in $n_f$ is very weak and will not affect the final normalisation of the distribution.} we obtain
the final expression for the multiplicity:
\begin{eqnarray}
{\cal N}(Y)\!\!&\!\!\propto\!\!&\!\!\exp\left[2.50217\left(\sqrt{Y+\lambda}-\sqrt\lambda\right)
-0.491546\ln\frac{Y+\lambda}{\lambda}\right.\cr
\!\!&\!\!-\!\!&\!\!\left.\left(0.06889-0.41151\ln(Y+\lambda)\right)\frac1{\sqrt{Y+\lambda}}
+\left(0.06889-0.41151\ln\lambda\right)\frac1{\sqrt{\lambda}}\right].
\end{eqnarray}

\paragraph{Peak position.}
The energy evolution dependence of the mean peak position is obtained plugging Eq.~(\ref{eq:nmllagamma}) into
(\ref{eq:Egg}), and the latter into Eq.~(\ref{eq:Dgomega}) in order to get the $K_n$ moments of the
distribution from Eq.~(\ref{eq:Kn}). Thus, for $n=1$ one obtains
\begin{equation}\label{eq:sxipmlla} 
\bar\xi=\frac{Y}2+\frac{a_1}{\sqrt{16N_c\beta_0}}\left(\sqrt{Y+\lambda}-\sqrt{\lambda}\right)
-2N_c\frac{a_2}{\beta_0}(\ln (Y+\lambda)-\ln\lambda),
\end{equation}
%which we have written in terms of $Y$ for sufficiently high values of $\lambda$, since
%the truncated asymptotic expansion (\ref{eq:nmllagamma}) relies upon $\alphas(Q_0)\ll1$. 
The smallness of the constant in front of the NMLLA correction proportional to $(\ln (Y+\lambda)-\ln\lambda)$ 
should not drastically modify the MLLA peak position and should only affect it at small energy scales.\\

The position of the mean peak is related to the corresponding maximum and median values of the DG distribution by
the expressions~\cite{Dokshitzer:1991za}:
\begin{equation}\label{eq:diffxi}
\ximax-\bar\xi=-\frac12\sigma s\left(1-\frac14\frac{k_5}{s}+\frac56k\right),\quad
\xi_{\rm m}-\bar\xi=-\frac16\sigma s\left(1-\frac3{20}\frac{k_5}{s}+\frac12k\right)\,,
\end{equation}
for which we need the fifth moment of the DG, $k_5$, which reads:
\begin{eqnarray}
k_5(Y,\lambda)\!\!&\!\!=\!\!&\!\!\frac9{16}a_1\left(\frac3{Y+\lambda}\right)^{3/2}
\left[\frac{\beta_0(Y+\lambda)}{16N_c}\right]^{1/4}
\frac{1-\left(\frac{\lambda}{Y+\lambda}\right)^{5/2}}{\left[1-\left(\frac{\lambda}
{Y+\lambda}\right)^{3/2}\right]^{5/2}}
\left[1+5\left(\frac{f_1(Y,\lambda)}{64}\right.\right.\cr
\!\!&\!\!+\!\!&\!\!\left.\left.\frac{f_4(Y,\lambda)}{72}\right)\beta_0
\sqrt{\frac{16N_c}{\beta_0(Y+\lambda)}}\right].\label{eq:k5}
\end{eqnarray}
%In the NMLLA$+$NLO$^*$ context of the distorted Gaussian, 
%making it possible to calculate also these quantities separately.
%After inserting Eqs.~(\ref{eq:sigmalim}), (\ref{eq:skewnesslim}), 
%(\ref{eq:kurtosislim}) and (\ref{eq:k5lim}) into (\ref{eq:diffxi}).
%\begin{eqnarray}\label{eq:meanmedell}
%\ximax-\bar\xi=\frac1{32}a_1\left(1+\frac5{64}\beta_0\sqrt{\frac{16N_c}{\beta_0Y}}\right),\quad
%\xi_{m}-\bar\xi=\frac1{96}a_1\left(1+\frac{19}{320}\beta_0\sqrt{\frac{16N_c}{\beta_0Y}}\right)
%\end{eqnarray}
%Therefore, in the NMLLA$+$NLO$^*$ scheme by making use of (\ref{eq:diffxi}) and (\ref{eq:meanellbis}), 
%The maximum value of the distribution is reached at,
%\begin{equation}
%\ximax\approx\frac{Y}2+\sqrt{\frac{a_1^2}{16N_c\beta_0}Y}
%-2N_c\frac{a_2}{\beta_0}\ln Y-\frac{a_1^2}{16N_c\beta_0}.
%\label{eq:ximax}
%\end{equation}
%Asymptotically ($Y\to\infty$) and factorising by $Y$, one recovers the maximum of the peak
%position for the DLA spectrum given by Eq.~(\ref{eq:dlapeak}). 
%The ratio as was given in~\cite{Dokshitzer:1991za} is also an interesting quantity 
%\begin{equation}
%\label{eq:rationmeans}
%\frac{\ximax-\bar\xi}{\xi_{\rm m}-\bar\xi}=3\left(1+\frac3{160}\beta_0\sqrt{\frac{16N_c}{\beta_0Y}}\right),
%\end{equation}
%that can be fitted to the $\epem$ data.

The final numerical expressions for the mean and maximum peak positions, evaluated for $n_f=5$ quark flavours, read:
\begin{eqnarray}
\bar{\xi}(Y)\!\!&\!\!=\!\!&\!\!0.5Y
+0.592722\left(\sqrt{Y+\lambda}-\sqrt\lambda\right)+0.002\ln\frac{Y+\lambda}{\lambda},\\
\ximax(Y)\!\!&\!\!=\!\!&\!\!0.5Y+0.592722\left(\sqrt{Y+\lambda}-\sqrt\lambda\right)-\frac{1}{2}\sigma\,s+
0.002\ln\frac{Y+\lambda}{\lambda}\,.
\end{eqnarray}
%where we have made the difference between the mean $\xi$ and $\ximax$ considered in Appendix~\ref{section:hocorrtomoments}.

\paragraph{Width.}
The DG distribution dispersion $\sigma$ follows from its definition in Eq.~(\ref{eq:moments}) for $n$~=~2. 
The full expression for the second moment $K_2(Y,\lambda)$ can be found in Appendix~\ref{app:moments},
Eq.~(\ref{eq:K2}), from which taking the squared root, followed by the Taylor expansion in
$(1/\sqrt{y+\lambda}$ or $\sqrt{\alphas})$ and keeping trace of all terms in $(1/(y+\lambda)$ or $\alphas$),
the NMLLA+NLO$^{*}$ expression for the width is obtained:
\begin{eqnarray}
\sigma(Y,\lambda)\!\!&\!\!=\!\!&\!\!\left(\frac{\beta_0}{144N_c}\right)^{1/4}
\sqrt{(Y+\lambda)^{3/2}-\lambda^{3/2}}
\left\{1-\frac{\beta_0}{64}f_1(Y,\lambda)\sqrt{\frac{16N_c}{\beta_0(Y+\lambda)}}
\right.\cr
\!\!&\!\!+\!\!&\!\!\left.\left[\frac9{16}a_2f_2(Y,\lambda)-\frac{3}{64}\left(\frac{3a_1^2}{16N_c^2}f_2(Y,\lambda)
+\frac{a_1\beta_0}{8N_c^2}f_2(Y,\lambda)-\frac{\beta_0^2}{64N_c^2}f_2(Y,\lambda)\right.\right.\right.\cr
\!\!&\!\!+\!\!&\!\!\left.\left.\left.\frac{3\beta_0^2}{128N_c^2}f_1^2(Y,\lambda)\right)
+\frac{\beta_1}{64\beta_0}(\ln2(Y+\lambda)-2)f_3(Y,\lambda)\right]\!\frac{16N_c}{\beta_0(Y+\lambda)}\right\},
\label{eq:sigma}
\end{eqnarray}
where the functions $f_i$ are also defined in Appendix~\ref{app:moments}. The new correction term,
proportional to $(1/(Y+\lambda))$, is of order $\cO\alphas$ and decreases the width of the distribution and so
does $\lambda$ for the truncated cascade with $Q_0>\lqcd$.
The numerical expression for the width (for $n_f=5$ quark flavours) reads:
\begin{eqnarray}
\sigma(Y)\!\!&\!\!=\!\!&\!\!0.36499\sqrt{(Y+\lambda)^{3/2}-\lambda^{3/2}}
\left\{1-0.299739f_1(Y,\lambda)\frac1{\sqrt{Y+\lambda}}-\left[1.12479f_2(Y,\lambda)\right.\right.\cr
\!\!&\!\!+\!\!&\!\!\left.\left. 0.0449219f_1^2(Y,\lambda)
+\left(0.32239-0.246692\ln(Y+\lambda)\right)f_3(Y,\lambda)\right]\frac1{Y+\lambda}\right\}\,.
\end{eqnarray}

\paragraph{Skewness.}
The NMLLA term of the third DG moment, $K_3$, turns out to vanish like the leading order
one~\cite{Dokshitzer:1991za}. According to the definition in Eq.~(\ref{eq:moments}), the skewness $s=K_3\sigma^{-3}$ 
presents an extra subleading term which in this resummation scheme comes from the expansion of the second 
contribution to $\sigma^{-3}$, proportional to $1/\sqrt{(Y+\lambda)}$, as written in
Eq.~(\ref{eq:sigmaminus3}) of Appendix~\ref{app:moments}, such that
\begin{equation}
\label{eq:skewness}
s(Y,\lambda)=-\frac{a_1}{16}
\frac{\left(\frac{144N_c}{\beta_0}\right)^{1/4}}
{\sqrt{(Y+\lambda)^{3/2}-\lambda^{3/2}}}\left[1-
\frac{\beta_0}{64}f_1(Y,\lambda)\sqrt{\frac{16N_c}{\beta_0(Y+\lambda)}}\right].
\end{equation}
In~\cite{Fong:1990nt}, only the first term of this expression was provided, the subleading contribution
given here is thus new. This NMLLA+NLO$^{*}$ correction to Eq.~(\ref{eq:skewness})
increases the skewness of the distribution, while for increasing $\lambda$ it 
should decrease again, thus revealing two competing effects. The net result is a displacement of the tails 
of the HBP distribution downwards to the left and upwards to the right from the peak
position and depending on the sign given by both effects (Fig.~\ref{fig:DG}). 
The final numerical expression for the skewness (for $n_f=5$ quark flavours) reads:
\begin{eqnarray}
s(Y)\!\!&\!\!=\!\!&\!\!-\frac{1.94704}{\sqrt{(Y+\lambda)^{3/2}-\lambda^{3/2}}}
\left[1-0.299739f_1(Y,\lambda)\frac1{\sqrt{Y+\lambda}}\right].
\end{eqnarray}

\paragraph{Kurtosis.}
The evolution of the kurtosis follows from the expressions for the fourth DG moment, given in Eqs.~(\ref{eq:K4})
and (\ref{eq:sigmaminus4}) of Appendix~\ref{app:moments}. As shown in the same appendix, the proper Taylor
expansion in powers of  $(1/\sqrt{Y+\lambda})$ which keeps trace of higher-order corrections and leads to:
\begin{eqnarray}
k(Y,\lambda)\!\!&\!\!=\!\!&\!\!-\frac{27}{5(Y+\lambda)}\sqrt{\frac{\beta_0(Y+\lambda)}{16N_c}}
\frac{1-\left(\frac{\lambda}{Y+\lambda}\right)^{5/2}}
{\left[1-\left(\frac{\lambda}{Y+\lambda}\right)^{3/2}\right]^2}
\left\{1+\frac{\beta_0}{16}(f_1(Y,\lambda)-\frac53f_4(Y,\lambda))\sqrt{\frac{16N_c}{\beta_0(Y+\lambda)}}\right.\cr
\!\!&\!\!+\!\!&\!\!\left.\left[\left(\frac{25}{24}f_5(Y,\lambda)
-\frac94f_2(Y,\lambda)\right)a_2+\frac{a_1^2}{256N_c^2}
\left(9f_2(Y,\lambda)-\frac{25}{2}f_5(Y,\lambda)\right)\right.\right.\cr
\!\!&\!\!+\!\!&\!\!\left.\left.\frac{a_1\beta_0}{256N_c^2}(6f_2(Y,\lambda)-5f_5(Y,\lambda))
+\frac{\beta_0^2}{256N_c^2}\left(-\frac{3}4f_2(Y,\lambda)+\frac{54}8f_1^2(Y,\lambda)
+\frac{275}{24}f_5(Y,\lambda)\right.\right.\right.\cr
\!\!&\!\!-\!\!&\!\!\left.\left.\left.15f_1(Y,\lambda)f_4(Y,\lambda)\right)+\frac{5\beta_1}{96\beta_0}
\left(\ln2(Y+\lambda)-\frac{2}{3}\right)f_6(Y,\lambda)\right.\right.\cr
\!\!&\!\!-\!\!&\!\!\left.\left.\frac{\beta_1}{16\beta_0}(\ln2(Y+\lambda)-2)f_3(Y,\lambda)\right]
\frac{16N_c}{\beta_0(Y+\lambda)}\right\},
\label{eq:kurtosis}
\end{eqnarray}
where the functions $f_i$ can be again found in Appendix~\ref{app:moments}. 
The new NMLLA+NLO$^{*}$ correction for the kurtosis
%, evaluated for $n_f=5$ (see appendix \ref{eq:numcompevaluation}) 
affects the distribution by making it smoother in the tails and wider in the hump region.
The final numerical expression for the kurtosis (for $n_f=5$ quark flavours) reads:
\begin{eqnarray}
k(Y)\!\!&\!\!=\!\!&\!\!-\frac{2.15812}{\sqrt{Y+\lambda}}\frac{1-\left(\frac{\lambda}{Y+\lambda}\right)^{5/2}}
{\left[1-\left(\frac{\lambda}{Y+\lambda}\right)^{3/2}\right]^2}\left\{1+\left[1.19896f_1(Y,\lambda)-
1.99826f_4(Y,\lambda)\right]\frac1{\sqrt{Y+\lambda}}\right.\cr
\!\!&\!\!+\!\!&\!\!\left.\left[1.07813f_1^2(Y,\lambda)+4.49915f_2(Y,\lambda)+1.28956f_3(Y,\lambda)
-2.39583f_1(Y,\lambda)f_4(Y,\lambda)\right.\right.\cr
\!\!&\!\!-\!\!&\!\!\left.\left.3.76231f_5(Y,\lambda)+0.0217751f_6(Y,\lambda)\right.\right.\cr
\!\!&\!\!-\!\!&\!\!\left.\left.(0.986767f_3(Y,\lambda)
-0.822306f_6(Y,\lambda))\ln(Y+\lambda)\right]\frac1{Y+\lambda}\right\}.
\end{eqnarray}

\begin{figure}[htbp]
\begin{center}
\epsfig{file=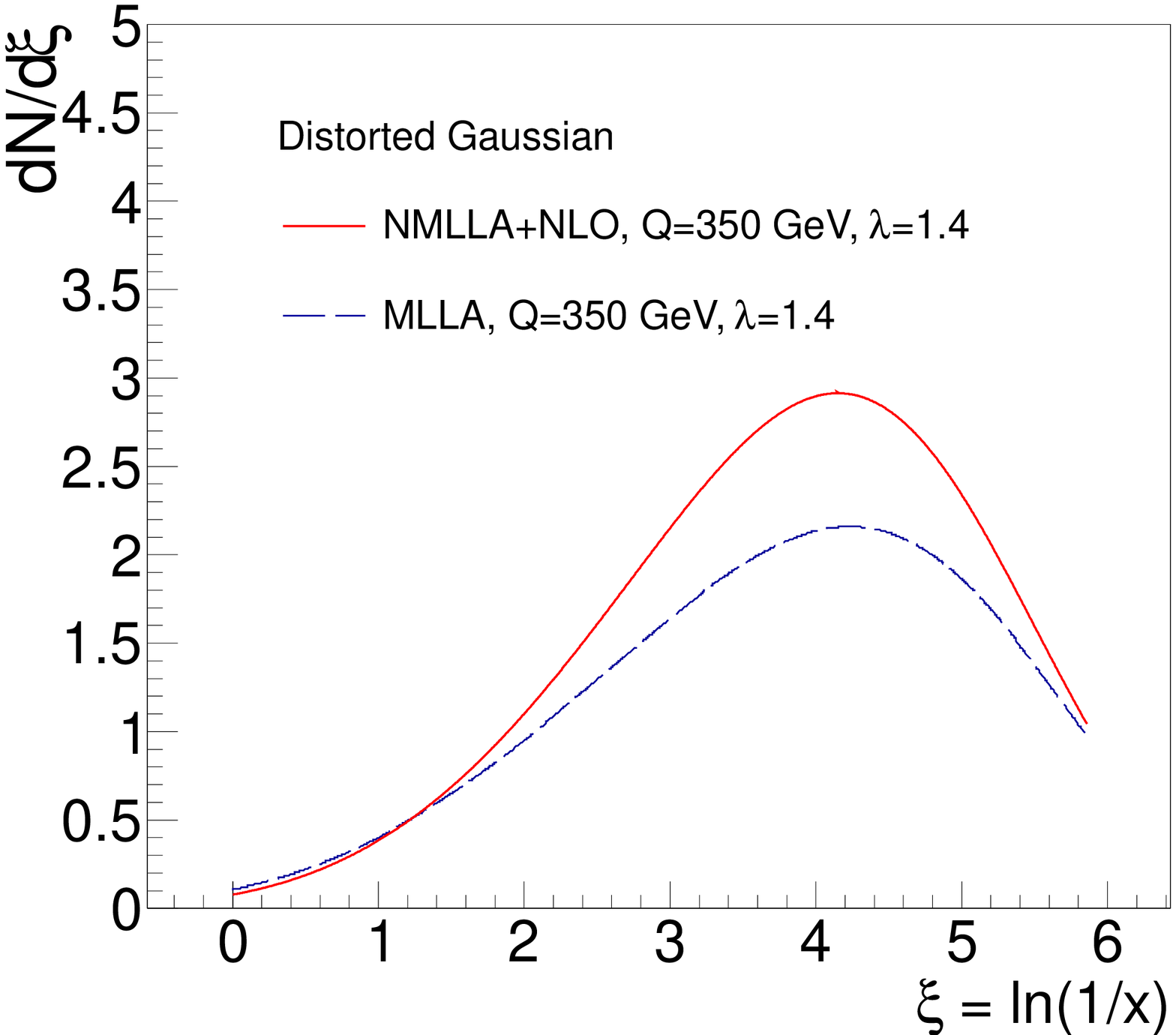,width=8.truecm}
\epsfig{file=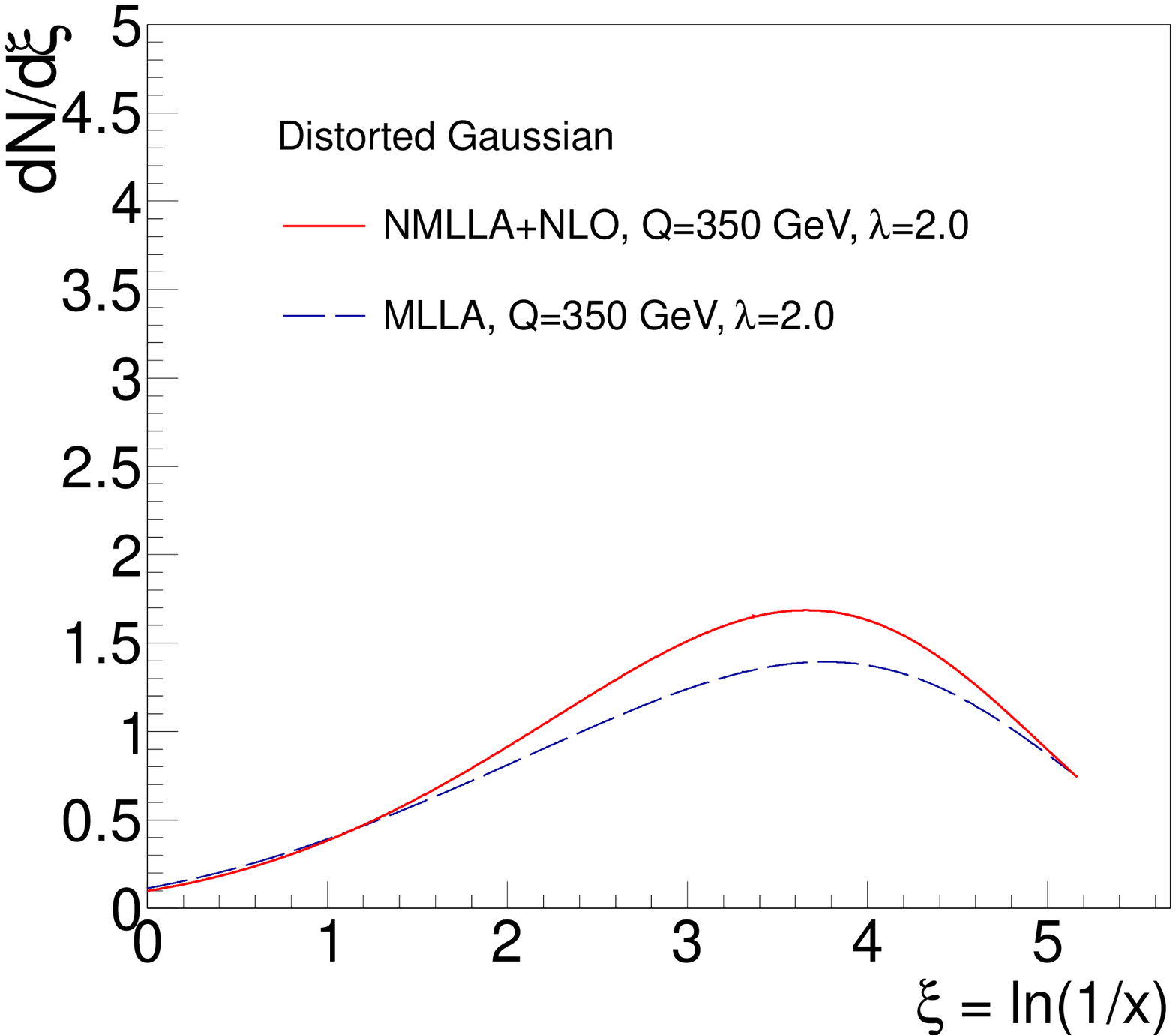, width=8.truecm}
\caption{\label{fig:DGmllavsnmlla}Comparison of the distorted Gaussian hadron distributions obtained for a jet of
virtuality $Q~=~350$~GeV evolved using NMLLA$+$NLO$^*$ (solid curve) and MLLA (dashed curve) equations, for two
hadronisation parameters: $\lambda=1.4$ (left) and $\lambda=2.0$ (right).} 
\end{center}
\end{figure}

\paragraph{Final DG expression.}
The final expression of the DG parametrisation of the single inclusive distribution of soft hadrons inside
gluon and quark jets, Eq.~(\ref{eq:xDg}), can be obtained summing all its individually-derived 
%These expressions are used to display the figures of the paper that are compared with previous MLLA predictions.
%The above 
NMLLA$+$NLO$^*$-resummed components: the mean multiplicity ${\cal N}(Y,\lambda)$ Eq.~(\ref{eq:meanmultip}), the mean
peak position $\bar\xi(Y,\lambda)$ Eq.~(\ref{eq:sxipmlla}), the dispersion $\sigma(Y,\lambda)$
Eq.~(\ref{eq:sigma}), the skewness Eq.~$s(Y,\lambda)$ (\ref{eq:skewness}),
and kurtosis $k(Y,\lambda)$ Eq.~(\ref{eq:kurtosis}).
%should be replaced in (\ref{eq:xDg}) as an intermediate step 
%in order to get the single inclusive distribution of soft hadrons inside gluon and quark jets.
%The fifth moment $k_5$ will not be added to the distorted Gaussian 
%but it will be considered for the estimation of the mean, median and maximum of the distribution in subsection 
%\ref{subsec:meanmedian}. 
In Fig.~\ref{fig:DGmllavsnmlla}, we display the resulting DG %after inserting the NMLLA$+$NLO$^*$ resummed
%ontributions (\ref{eq:sxipmlla}), (\ref{eq:meanmultip}), (\ref{eq:sigma}), (\ref{eq:skewness})and (\ref{eq:kurtosis}) into (\ref{eq:xDg}) 
for two different values of the hadronisation parameter $\lambda=1.4$ ($Y=5.8$, $Q_0=1$~GeV, $\lqcd=0.25$~GeV)
and $\lambda=2.0$ ($Y=5.2$, $Q_0=1$~GeV, $\lqcd=0.25$~GeV) for a jet of virtuality $Q=350$~GeV and
reconstructed jet energy $E=500$~GeV inside a radius cone $\theta=0.7$. The distribution is compared to the
corresponding MLLA predictions with the Fong-Webber results from~\cite{Fong:1990nt} after setting to zero all
terms proportional to $1/Y$ in the same expressions.\\

The contributions from the set of NMLLA$+$NLO$^*$ corrections to the MLLA DG appear to be quite 
substantial and decrease for increasing $\lambda$, since $\lambda$ guarantees the convergence of the
perturbative series for $Q_0\gg\lqcd$. Physically, for higher values of the shower energy cut-off
$Q_0$, the strength of the coupling constant decreases and the probability for the emission of soft gluon 
bremsstrahlung decreases accordingly, making the multiplicity distribution and the peak position smaller. The 
difference between the MLLA and NMLLA$+$NLO$^*$ resummed distributions is, as mentioned above, mainly due to 
running-coupling effects, proportional to $\beta_1$, at large $\xi$ (small $x$) which is not unexpected because
in this region they are more pronounced due to the $\ln(xE\theta)$ dependence in the
denominator of the strong coupling. On the other hand, energy conservation plays a more important 
role in the hard fragmentation region $x\sim1$ ($\xi\sim0$), where the NMLLA$+$NLO$^*$ DG is somewhat 
suppressed compared with the MLLA DG. 

%\paragraph{Numerical formul{\ae}:}
%%\section{Beyond limiting spectrum for $n_f=5$}
%%\label{eq:numcompevaluation}
%We take the expressions (\ref{eq:sxipmlla}), (\ref{eq:meanmultip}) (\ref{eq:skewness}), (\ref{eq:kurtosis}) 
%and (\ref{eq:k5}) and perform the numerical evaluation for $n_f=5$ quark flavours. As will be seen below the 
%dependence in $n_f$ is very weak and will not affect the final normalisation of the distribution.
%The expressions are:

%%%%%%%%%%%%%%%%%%%%%%%%%%%%%%%%%%%%%%%%%%%%%%%%%%%%%%%%%%%%%%%%%%%%%%%%%%%%%%%%%%%%%%%%

%
%%%%%%%%%%%%%%%%%%%%%%%%%%%%%%%%%%%%%%%%%%%%%%%%%%%%%%%%%%%%%%%%%%%%%%%%%%%%%%%%%%%%%%%%
%
\subsection{Multiplicities for the single inclusive $\boldsymbol{D_g}$ and $\boldsymbol{D_q}$ distributions}
\label{subsec:CFlambda}

\begin{figure}[htbp]
\begin{center}
\epsfig{file=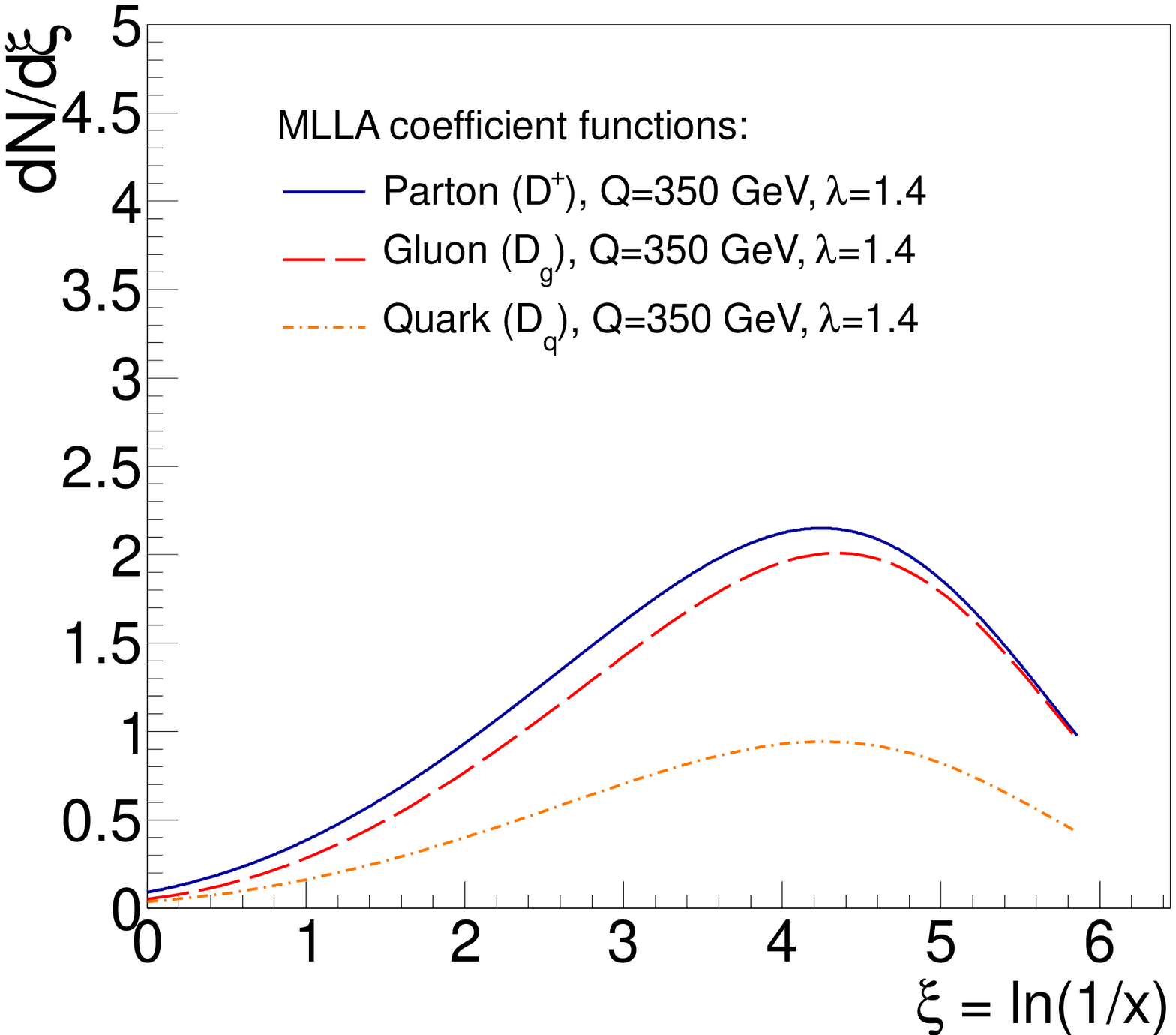,width=8.truecm}
\epsfig{file=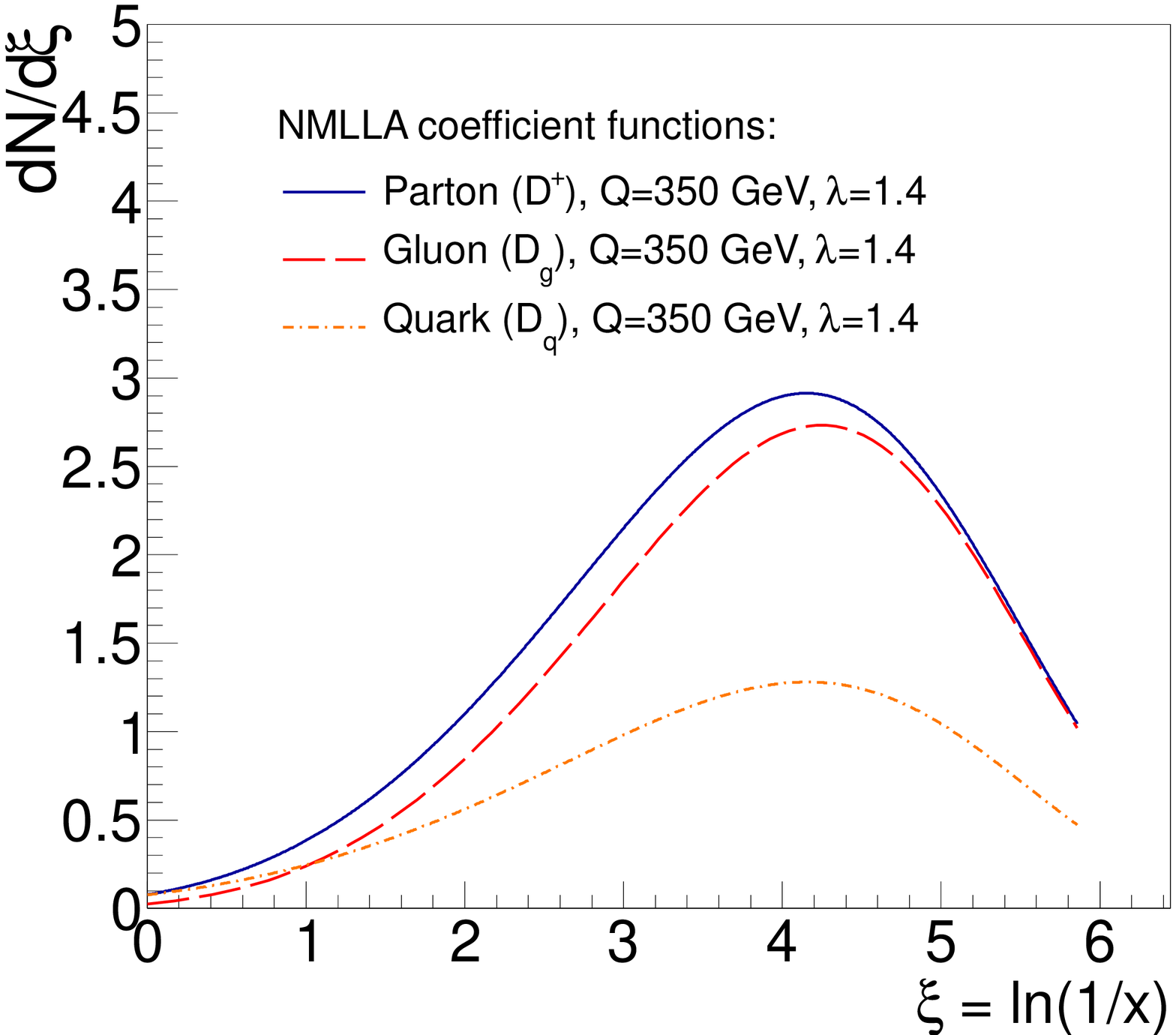,width=8.truecm}
\caption{\label{fig:CF}Comparison of the quark Eq.~(\ref{eq:Dqh}), gluon Eq.~(\ref{eq:Dgh}), and 
parton ${\cal D}^+(\xi,Y)$ Eq.~(\ref{eq:xDg}), distributions of hadrons %computed numerically
  for a jet of virtuality $Q=350$~GeV and hadronisation parameter $\lambda=1.4$ evolved using MLLA (left) and
  NMLLA$+$NLO$^*$ (right) equations.}
\end{center}
\end{figure}

In this section we determine the coefficient function involved in Eq.~(\ref{eq:DqDgdistr})
that provide higher-order corrections to the quark/gluon multiplicity ratio.
As shown through Eq.~(\ref{eq:Dmoinslambda}), the ${\cal D}^-(\omega,\lambda)$ component 
is negligible and thus the solutions for the gluon and quark single inclusive 
distributions can be directly obtained from ${\cal D}^+$ in the form
\begin{subequations}
\begin{eqnarray}\label{eq:DqDgdistrbis}
{\cal D}_q(\omega,Y,\lambda)\!&\!\approx\!&\!C_q^g(\Omega)
{\cal D}^+(\omega,Y,\lambda),\qquad C_q^g(\Omega)=\frac{P_{qg}(\Omega)}{P_{++}(\Omega)-P_{--}(\Omega)},\\
{\cal D}_g(\omega,Y,\lambda)\!&\!\approx\!&\!C_g^g(\Omega){\cal D}^+(\omega,Y,\lambda),
\qquad C_g^g(\Omega)=\frac{P_{++}(\Omega)-P_{qq}(\Omega)}{P_{++}(\Omega)-P_{--}(\Omega)}.
\end{eqnarray} 
\end{subequations}
Making use of the expressions (\ref{eq:expomega1})--(\ref{eq:expomega2}) and (\ref{eq:postraj})--(\ref{eq:postraj1}),
and expanding in $\omega$ results in
\begin{eqnarray}\label{eq:nmllacf}
C_q^g(\Omega)\approx\frac{C_F}{N_c}\left[1+c_q^{(0)}\Omega+c_q^{(1)}\Omega^2+\cO{\Omega^3}\right],\;
C_g^g(\Omega)\approx1+c_g^{(0)}\Omega+c_g^{(1)}\Omega^2+\cO{\Omega^3},
\end{eqnarray}
where the numerical values of the constants, for $n_f=5$ quark flavours, read
\begin{eqnarray}
c_q^{(0)}\!&\!=\!&\!\frac{a_1-b_1}{4N_c}-\frac34\stackrel{n_f=5}{=}-0.049,\\
c_q^{(1)}\!&\!=\!&\!\frac78+\frac{a_1-b_1}{16N_c}\left(\frac{a_1-b_1}{N_c}-3\right)
+\frac{C_F}{N_c}b_2-a_2\stackrel{n_f=5}{=}0.608,\\
c_g^{(0)}\!&\!=\!&\!-\frac{b_1}{4N_c}\stackrel{n_f=5}{=}-0.247,\\ 
c_g^{(1)}\!&\!=\!&\!\frac{b_1}{16N_c^2}(b_1-a_1)+\frac{C_F}{N_c}
\left(b_2-\frac58+\frac{\pi^2}6\right)\stackrel{n_f=5}{=}0.045.
\end{eqnarray}
The $c_i^{(0)}$ numerical constants in Eq.~(\ref{eq:nmllacf}) were obtained in~\cite{Dokshitzer:1991wu}.
Performing the inverse Mellin-transform back to the $x$-space, or making the equivalent replacement
$\Omega\to\frac{\partial}{\partial\xi}+\frac{\partial}{\partial Y}$, one has
\begin{subequations}
\begin{eqnarray}\label{eq:Dqh}
D_q(\xi,Y,\lambda)\!&\!\approx\!&\!\frac{C_F}{N_c}\left[1+c_q^{(0)}\left(\frac{\partial}{\partial\xi}+
\frac{\partial}{\partial Y}\right)+c_q^{(1)}\left(\frac{\partial}{\partial\xi}+
\frac{\partial}{\partial Y}\right)^2\right]
D^+(\xi,Y,\lambda),\\
D_g(\xi,Y,\lambda)\!&\!\approx\!&\!\left[1+c_g^{(0)}\left(\frac{\partial}{\partial\xi}+
\frac{\partial}{\partial Y}\right)+c_g^{(1)}\left(\frac{\partial}{\partial\xi}+
\frac{\partial}{\partial Y}\right)^2\right]D^+(\xi,Y,\lambda),
\label{eq:Dgh}
\end{eqnarray}
\end{subequations}
which in a more compact form can be rewritten as
\begin{eqnarray}
D_{a}(\xi,Y,\lambda)\!&\!\approx\!&\!\frac{C_A}{N_c}\left[D^+(\xi,Y,\lambda)
+c_A^{(0)}\left(\frac{\partial D^+(\xi,Y,\lambda)}{\partial\xi}+
\frac{\partial D^+(\xi,Y,\lambda)}{\partial Y}\right)\right.\cr
\!&\!+\!&\!\left.c_A^{(1)}\left(\frac{\partial^2D^+(\xi,Y,\lambda)}
{\partial\xi^2}+2\frac{\partial^2D^+(\xi,Y,\lambda)}
{\partial\xi\partial Y}
+\frac{\partial^2D^+(\xi,Y,\lambda)}{\partial Y^2}\right)\right]\label{eq:DAplusnum}
\end{eqnarray}
for numerical considerations. 
The first and second derivatives in Eqs.~(\ref{eq:Dqh}) and (\ref{eq:Dgh}) can be evaluated numerically. 
They provide corrections which are suppressed for the first and second terms of orders
${\cal O}(\sqrt{\alphas})$ and ${\cal O}(\alphas)$ respectively. In Fig.~\ref{fig:CF}, 
we compare the quark ($D_q$), gluon ($D_g$) and parton ($D^+$) hadron spectra obtained in the MLLA
(left) and NMLLA-NLO$^{*}$ (right) schemes for a jet of virtuality $Q=350$~GeV and %(perturbative) 
hadronisation parameter $\lambda=1.4$. The NMLLA-NLO$^{*}$ distributions are obtained from the above Eqs.~(\ref{eq:Dqh}),
(\ref{eq:Dgh}) and (\ref{eq:xDg}), while the MLLA are obtained setting to zero $c_q^{(1)}$ and $c_g^{(1)}$ in
Eqs.~(\ref{eq:Dqh}) and (\ref{eq:Dgh}) respectively and removing the $\cO\alphas$ 
corrections in (\ref{eq:xDg}) for $D^+(\xi,Y)$.\\

A clear difference is observed in the quark and gluon jet initiated distributions given
by the colour factor $C_F/N_c=4/9$ and the role of higher-order corrections which prove  
more sizable for the NMLLA$+$NLO$^*$ scheme over the whole phase space $0\leq\xi\leq Y$, 
as observed in the right panel of Fig.~\ref{fig:CF}. 
In~\cite{Dokshitzer:1991wu} however, the role of $\cO{\sqrt\alphas}$ corrections, 
proportional to $c_q^{(0)}$ and $c_g^{(0)}$ in Eqs.~(\ref{eq:Dqh}) and (\ref{eq:Dgh}), 
was reabsorbed into the inclusive spectrum $D^+(\xi,Y)$ through a shift to a slightly different jet energy
$E_A=E\exp\left(c_A^{(0)}\right)$, which allowed for a direct comparison between the MLLA $D^+(\xi,Y)$ and the hadronic
energy-momentum spectrum  (for a complete review see~\cite{Khoze:1996dn}). 
Asymptotically ($Q\to\infty$), the solution of the original Eq.~(\ref{eq:DAplusnum}) 
has a Gaussian shape near its maximum:
\begin{equation}
D_{a}(\xi,Q^2)\approx\frac{C_A}{N_c}\frac{{\cal N}}{\sigma\sqrt{2\pi}}
\exp\left[-\frac1{2\sigma^2}(\xi-\bar\xi)^2\right],
\end{equation}
normalised by the inverse asymptotic value of the mean multiplicity ratio $r^{-1}=C_F/N_c$
in a quark jet.
The ratio of gluon and quark multiplicities can be recovered by replacing 
$\omega=0$ ($\frac{\partial}{\partial\xi}=0$) in Eqs.~(\ref{eq:Dqh}) and (\ref{eq:Dgh}),
such that, after expanding the result in powers of $\sqrt{\alphas}$, one is left with
\begin{equation}\label{eq:ratiogq}
r=\frac{{\cal N}_g}{{\cal N}_q}=\frac{N_c}{C_F}\left(1-r_1\gamma_0-r_2\gamma_0^2\right),
\end{equation}
where, as a result of the expansion,
\begin{eqnarray}
r_1\!\!&\!\!=\!\!&\!\!c_q^{(0)}-c_g^{(0)}=\frac{a_1}{4N_c}-\frac34,\\
r_2\!\!&\!\!=\!\!&\!\!c_q^{(1)}-c_g^{(1)}-r_1c_q^{(0)}-\frac{r_1}{8N_c}
\left(a_1-\frac{\beta_0}2\right)=\tilde a_2-a_2+r_1\left(\frac34-
\frac{a_1}{8N_c}+\frac{\beta_0}{16N_c}\right),
\end{eqnarray}
with
$$
\tilde a_2=\frac78+\frac{C_F}{N_c}\left(\frac58-\frac{\pi^2}{6}\right).
$$
Notice that up to the order $\cO\alphas$, the multiplicity ratio does not involve corrections
proportional to $\beta_1$, which only appear beyond this level of accuracy~\cite{Dremin:2000ep}. 
Up to the NMLLA order in $\cO\alphas$, Eq.~(\ref{eq:ratiogq}) coincides with the expression 
found in~\cite{Dremin:1994bj}, which gives further support to the calculations carried out in our work.
A more updated evaluation of the mean multiplicity ratio, including two-loop splitting functions, was
given recently in~\cite{Bolzoni:2013rsa}.

%%%%%%%%%%%%%%%%%%%%%%%%%%%%%%%%%%%%%%%%%%%%%%%%%%%%%%%%%%%%%%%%%%%%%%%%%%%%%%%%%%%%%%%%
%
%\section{Evolution of the parton fragmentation functions at NMLLA +NLO$^*$: Limiting spectrum for the DG parametrisation}
\subsection{Limiting spectrum for the DG parametrisation}
%\subsection{Local hadron parton duality and limiting spectrum for the NMLLA$+$NLO$^*$ scheme}
\label{subsec:lphdandlimspec}

The so-called limiting spectrum, $\lambda\to0$, implies pushing the validity of the partonic evolution
equations down to (non-perturbative) hadronisation scales, $Q_0\approx\lqcd$~\cite{Dokshitzer:1982ia}. Such an approach
provides a minimal (and successful) approach with predictive power for the measured experimental
distributions. We derive here the evolution of the distorted Gaussian moments for this limit which involves
formul{\ae} depending only on $\lqcd$ as a single parameter.

\paragraph{Multiplicity.}
Among the various moments of the DG parametrisation, only its integral (representing the total hadron multiplicity)
needs an extra free parameter to fit the data. 
%Perturbative QCD is incomplete in that it cannot describe the physics of hadrons entirely. 
%An intuitive solution to this problem is provided by the LPHD, which states that 
The ``local parton hadron duality'' (LPHD) hypothesis is a powerful assumption which states that the
distribution of partons in inclusive processes is identical to that of the final hadrons, up to 
an overall normalization factor,
%the number of particles produced in the jet, which is normalised by 
%a constant ${\cal K}^{\rm ch}$, 
i.e. that the mean multiplicity of the measured charged
hadrons is proportional to the partonic one through a constant ${\cal K}^{\rm ch}$,
$$
{\cal N}^{\rm ch}(Y)={\cal K}^{\rm ch}{\cal N}(Y)\,.
$$
%for the limit where $Q_0\approx\Lambda_{\text{QCD}}$, the so-called limiting spectrum. 
Thus, in the limiting spectrum the mean multiplicity reads
\begin{eqnarray}\label{eq:multlim}
{\cal N}^{\rm ch}(Y)\!\!&\!\!=\!\!&\!\!{\cal K}^{\rm ch}\exp\left\{\sqrt{\frac{16N_c}{\beta_0}Y}-
\left(\frac{a_1}{\beta_0}-\frac12\right)\ln\sqrt{Y}-\frac{2N_c}{\beta_0}
\left[a_2+\frac1{4}\left(\frac{a_1}{4N_c}\right)^2
+\frac12\frac{a_1\beta_0}{16N_c^2}\right.\right.\cr
\!\!&\!\!+\!\!&\!\!\left.\left.\frac3{16}\left(\frac{\beta_0}{4N_c}\right)^2
-\frac{\beta_1}{4N_c\beta_0}(\ln2Y+2)\right]\sqrt{\frac{16N_c}{\beta_0Y}}\right\},
\end{eqnarray}
which is in agreement with the mean multiplicity first found in~\cite{Dremin:2000ep}, supported by the
improved solution of the evolution equations accounting for the same set of corrections.

\paragraph{Peak position.}
For the limiting spectrum, the mean peak position Eq.~(\ref{eq:sxipmlla}) can be approximated as follows:
\begin{equation}\label{eq:meanellter}
\bar\xi=\frac{Y}2+\frac{a_1}{16N_c}\sqrt{\frac{16N_c}{\beta_0}Y}-2N_c\frac{a_2}{\beta_0}\ln Y
\end{equation}
thanks to the fortuitous smallness ${\cal O}(10^{-3})$ of the NMLLA correction to $\bar\xi$
at high-energy where $Y+\lambda\gg\lambda$.
Notice that, as shown in~\cite{Fong:1990nt}, the MLLA version of Eq.~(\ref{eq:meanellter}) up to the 
second order is finite. The origin of the third $\propto\ln Y$ correction in this resummation
framework comes from the truncated expansion of the anomalous dimension Eq.~(\ref{eq:nmllagamma}) 
in ${\cal O}(\alphas)$, which is proportional to $1/Y$ by making $(-\partial\gamma_\omega/\partial\omega$)
at $\omega=0$, and hence yields the $\propto\ln Y$ term after integrating over $Y$. 
Therefore, we assume that Eq.~(\ref{eq:meanellter}) is valid for $Q\gg Q_0\approx\lqcd$.\\

The maximum of the peak position for the limiting spectrum DG can be obtained via Eq.~(\ref{eq:diffxi}) which
involves the mean peak position as well as the other higher-order moments.
In a generic form, the moments of the distorted Gaussian associated with the dispersion (\ref{eq:sigma}),
skewness (\ref{eq:skewness}), kurtosis (\ref{eq:kurtosis}), and $k_5$ (\ref{eq:k5}), are finite for $n\geq2$
for the limiting spectrum and can be written as
\begin{eqnarray}\label{eq:symbKn}
K_n(\alphas(Y+\lambda),\alphas(\lambda))\!&\!\!\simeq\!\!&\!\!\alphas(Y+\lambda)^{-(n+1)/2}\left[{\cal K}_n^{(0)}
+{\cal K}_n^{(0)}\sqrt{\alphas(Y+\lambda)}+{\cal K}_n^{(0)}\alphas(Y+\lambda)\right.\cr
\!&\!\!-\!\!&\!\!\left.\left\{\alphas(Y+\lambda)\Leftrightarrow\alphas(\lambda)\right\}\right]\,,
\end{eqnarray} 
where the constants ${\cal K}_n^{(0)}$ and the functions $f_i(\lambda\to0)\to1$ are written in Appendix~\ref{app:moments}. 
In other words, the second $\lambda$-dependent part of $K_n$ in Eq.~(\ref{eq:symbKn}) can be 
dropped as $\lambda\to0$ for sufficiently high energy scales, $Y+\lambda\gg\lambda$, where 
$\alphas(Y+\lambda)\ll\alphas(\lambda)$ in the r.h.s. of Eq.~(\ref{eq:symbKn}).
Performing the same approximation in Eq.~(\ref{eq:symbKn}) as $\lambda\to0$, 
% for the dispersion, skewness, kurtosis and $k_5$ 
the expressions for the rest of moments of the fragmentation functions in the limiting spectrum are
derived below. Thus inserting Eqs.~(\ref{eq:sigmalim}), (\ref{eq:skewnesslim}), 
(\ref{eq:kurtosislim}) and (\ref{eq:k5lim}) into (\ref{eq:diffxi}), we obtain :
\begin{eqnarray}\label{eq:meanmedell}
\ximax-\bar\xi=\frac1{32}a_1\left(1+\frac5{64}\beta_0\sqrt{\frac{16N_c}{\beta_0Y}}\right),\quad
\xi_{m}-\bar\xi=\frac1{96}a_1\left(1+\frac{19}{320}\beta_0\sqrt{\frac{16N_c}{\beta_0Y}}\right)
\end{eqnarray}

\paragraph{Width.}  
The width of the DG distribution in the limiting spectrum is obtained from Eq.~(\ref{eq:sigma}):
\begin{subequations}
\begin{eqnarray}
\sigma(Y)\!\!&\!\!=\!\!&\!\!\sqrt{\frac13Y}\left(\frac{\beta_0Y}{16N_c}\right)^{1/4}
\left\{1-\frac{\beta_0}{64}\sqrt{\frac{16N_c}{\beta_0Y}}+
\left[\frac9{16}a_2-\frac{3}{64}\left(\frac3{16N_c^2}a_1^2
+\frac{a_1\beta_0}{8N_c^2}+\frac{\beta_0^2}{128N_c^2}\right)\right.\right.\cr
\!\!&\!\!+\!\!&\!\!\left.\left.\frac{\beta_1}{64\beta_0}(\ln2Y-2)\right]\!\frac{16N_c}{\beta_0Y}\right\}.
\label{eq:sigmalim}
\end{eqnarray}

\paragraph{Skewness.}
The skewness of the DG distribution in the limiting spectrum reads, from Eq.~(\ref{eq:skewness}),
\begin{equation}\label{eq:skewnesslim}
s(Y)=-\frac{a_1}{16}\sqrt{\frac3{Y}}
\left(\frac{16N_c}{\beta_0Y}\right)^{1/4}\left(1-\frac{\beta_0}{64}\sqrt{\frac{16N_c}{\beta_0Y}}\right)\,,
\end{equation}

\paragraph{Kurtosis.}
The kurtosis can be derived from Eq.~(\ref{eq:kurtosis}):
\begin{eqnarray}\label{eq:kurtosislim}
k(Y)\!\!&\!\!=\!\!&\!\!-\frac{27}{5Y}\sqrt{\frac{\beta_0Y}{16N_c}}\left\{1-\frac{\beta_0}{24}
\sqrt{\frac{16N_c}{\beta_0Y}}-\left[\frac{29}{24}a_2+\left(\frac{7}{512N_c^2}
a_1^2-\frac{a_1\beta_0}{256N_c^2}-\frac{59}{6144N_c^2}\beta_0^2
\right)\right.\right.\cr
\!\!&\!\!+\!\!&\!\!\left.\left.\frac{\beta_1}{96\beta_0}
\left(\ln2Y-\frac{26}{3}\right)\right]\frac{16N_c}{\beta_0Y}\right\}.
\end{eqnarray}
Accordingly, we give the last component, $k_5$, following from Eq.~(\ref{eq:k5}):
\begin{eqnarray}
k_5(Y)\!\!&\!\!=\!\!&\!\!\frac9{16}\,a_1\left(\frac3Y\right)^{3/2}\left(\frac{\beta_0Y}{16N_c}\right)^{1/4}
\left(1+\frac{85}{576}\beta_0\sqrt{\frac{16N_c}{\beta_0Y}}\right).
\label{eq:k5lim}
\end{eqnarray}
\end{subequations}

%\if 0
%Ordinarily, evolution at low scales is not possible in perturbation theory because the convergence of the
%perturbative series is spoiled and can even be singular. However, by using hypergeometric 
%functions it was possible to write the MLLA solution of the evolution equations (with $a_2=0,\beta_1=0$) 
%for $D^+(\xi,Y)$ and therefore the $\sqrts$ dependence of the spectrum as a collinear and infrared safe 
%(CIS) observable. The convergence of the solution in this approach as $\lambda\to0$ was however
%demonstrated to be accidental, reproducing by chance the energy-momentum spectrum of
%hadrons even for hard partons $x\sim1$. 
%In the end, by using the hypergeometric representation of the cross-section, it became 
%possible to match the MLLA exact solution with the MLLA distorted Gaussian.\\
%\fi

\begin{figure}[htbp]
\begin{center}
\epsfig{file=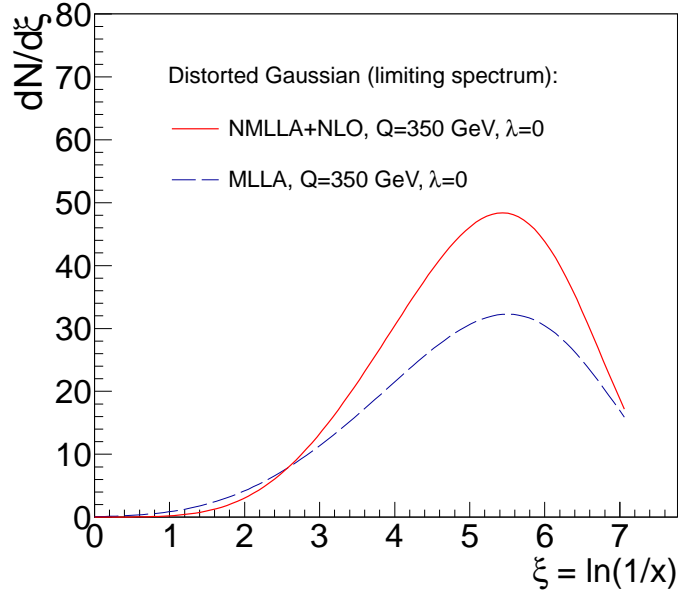, width=9.0truecm}
\caption{\label{fig:DGmllavsnmllals}Comparison of the distorted Gaussian hadron distributions obtained for a jet of
virtuality $Q~=~350$~GeV evolved using MLLA and NMLLA$+$NLO$^*$ equations, in the limiting spectrum 
(i.e. $Q_0=\lqcd$, hadronisation parameter $\lambda=0$).}
\end{center}
\end{figure}

\paragraph{Final DG  (limiting spectrum) expression.}
In order to get the DG in the limiting spectrum, one should replace
Eqs.~(\ref{eq:multlim})--(\ref{eq:kurtosislim}) into Eq.~(\ref{eq:xDg}). 
We note that in our NMLLA+NLO$^{*}$ framework, the ${\cal K}^{\rm ch}$ from the DG can be smaller than that found 
in~\cite{Dremin:2000ep} since it should fix the right normalisation enhanced by second-loop coupling 
constant effects. Notice also that setting subleading corrections to zero, we recover the results from 
\cite{Fong:1990nt} as expected.
In Fig.~\ref{fig:DGmllavsnmllals}, the MLLA and NMLLA$+$NLO$^*$ distorted Gaussians are displayed 
in the limiting spectrum approximation for a jet virtuality $Q=350$~GeV in the interval 
$0\leq\xi\leq Y$, for $Y=7.5$.\\

%As a consequence of the approximation, 
We can see a sizable difference between the MLLA $D^+(\xi,Y)$ and the NMLLA$+$NLO$^*$ $D^+(\xi,Y)$ evolutions,
which is mainly driven by the two-loop $\propto\beta_1$ correction in the mean multiplicity and other moments of
the DG, as mentioned above. The account of energy conservation can be observed at low $\xi$, i.e.
for harder partons. Similar effects have been discussed in~\cite{Sapeta:2008km} where an exact numerical
solution of the MLLA evolution equations was provided with one-loop coupling constant.  
Numerical solutions of exact MLLA equations provide a perfect account of energy conservation at every 
splitting vertex of the branching process in the shower. For this reason, accounting for higher-order corrections 
${\cal O}(\alphas^{n/2})$ to the truncated series of the single inclusive spectrum of hadrons
should follow similar features and trends to that provided by the numerical solutions of~\cite{Sapeta:2008km}
(see also \cite{Lupia:1997in}), although our NMLLA+NLO$^{*}$ solution incorporates in addition the two-loop
coupling constant.\\

\begin{figure}[htbp]
\begin{center}
\epsfig{file=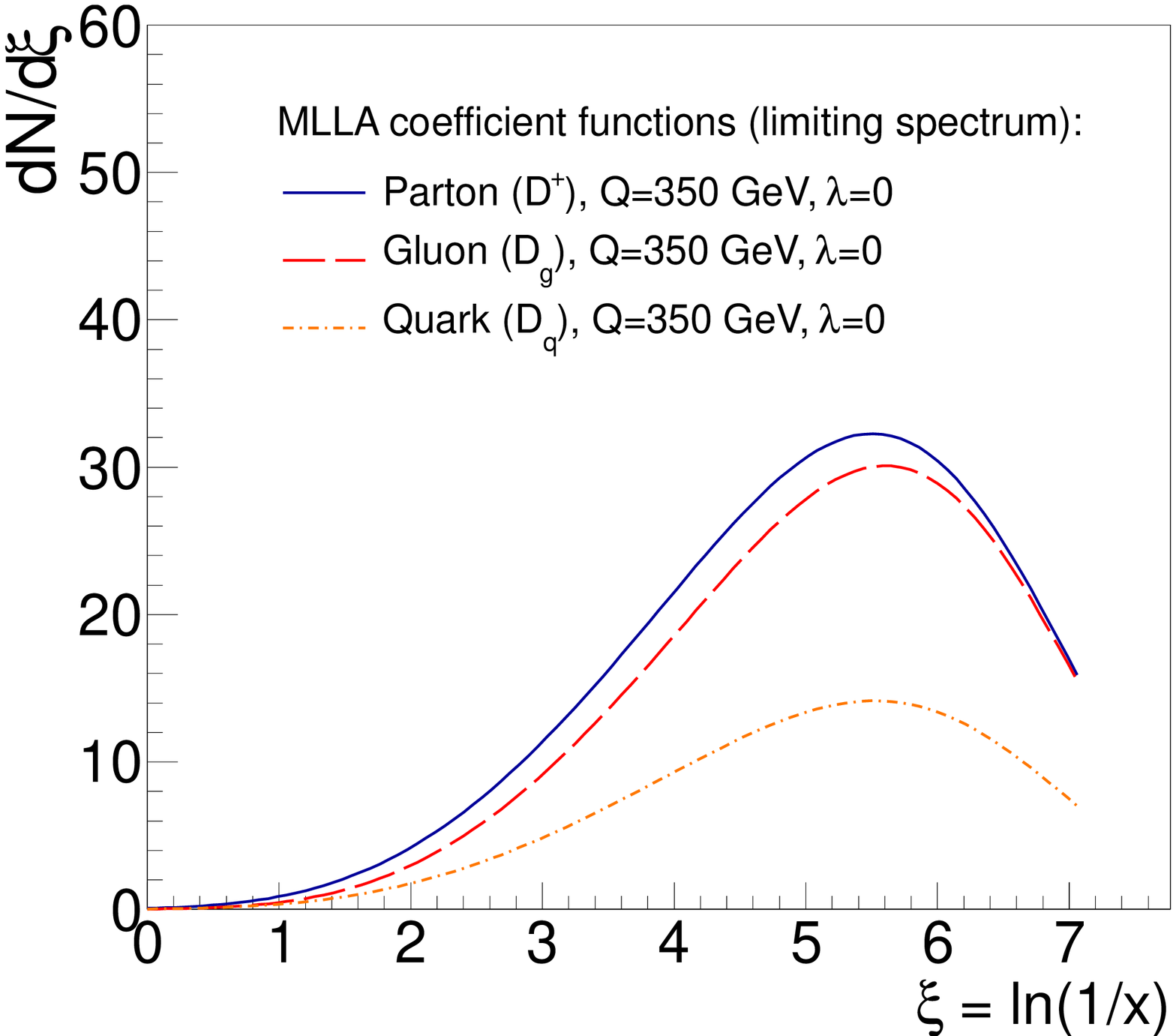, height=7.truecm,width=7.9truecm}
\epsfig{file=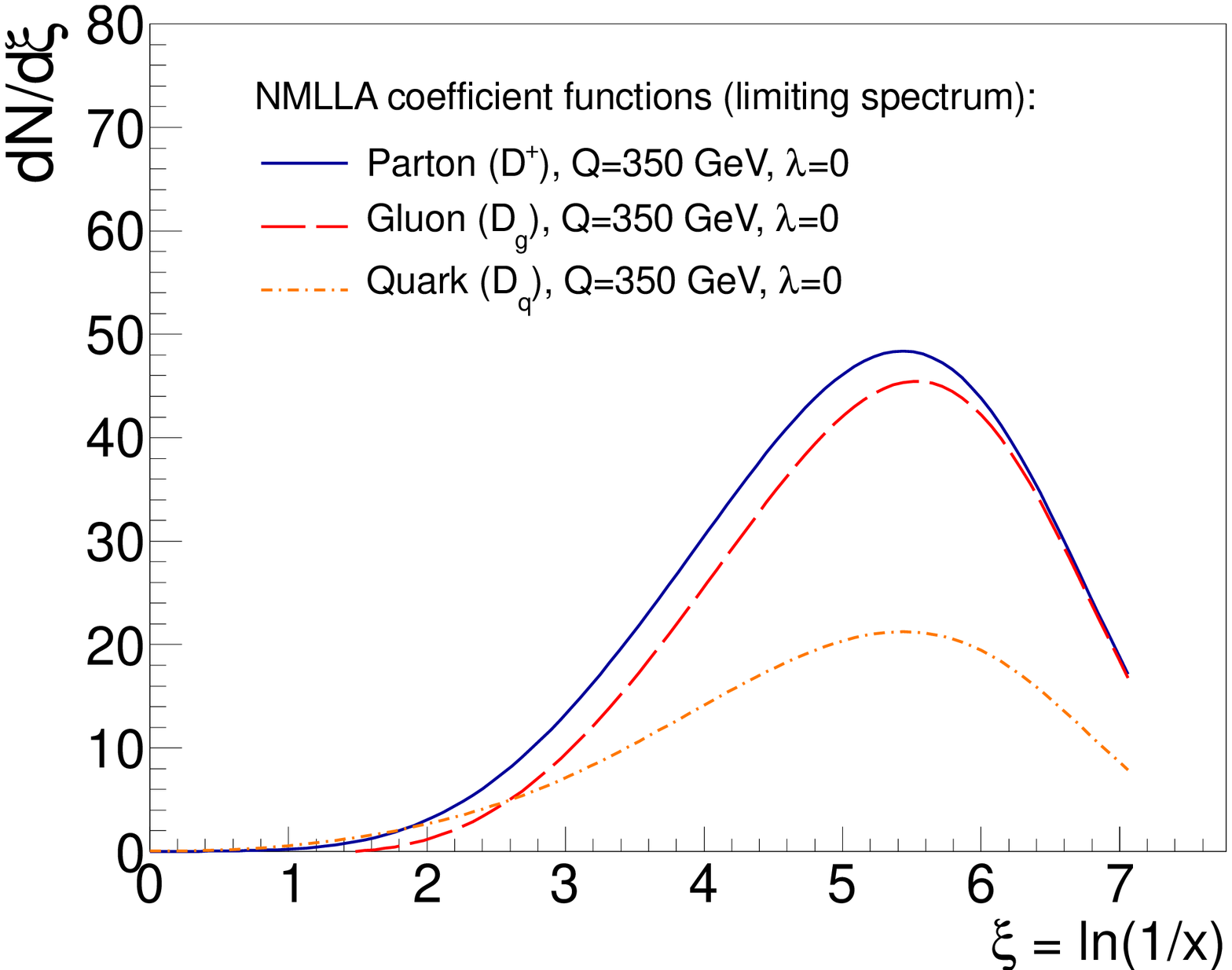, height=7.truecm,width=8.0truecm}
\caption{\label{fig:CFLS}Comparison of the distorted Gaussian hadron distributions obtained for a quark
  Eq.~(\ref{eq:Dqh}), gluon Eq.~(\ref{eq:Dgh}), and ${\cal D}^+(\xi,Y)$ Eq.~(\ref{eq:xDg}), for a
  jet of virtuality $Q=350$~GeV
  evolved using MLLA (left) and NMLLA$+$NLO$^*$ (right) equations, in the limiting spectrum  
  (i.e. $Q_0=\lqcd$, hadronisation parameter $\lambda=0$).}
\end{center}
\end{figure}

In Fig.~\ref{fig:CFLS} we display the same set of curves as in the Fig.~\ref{fig:CF} 
%of the previous paragraph \ref{subsec:CFlambda}
with the right normalisation given by the coefficient functions for quark and gluon jets.
The overall corrections provided by the coefficient functions slightly decrease the normalisation 
of the spectrum in a gluon jet as well as its width $\sigma$. In the quark jet, upon normalisation by the 
colour factor $C_F/N_c$, the normalisation is decreased while the width is slightly enlarged. In order to 
better visualise the less trivial enlargement for the width, we can for instance consider
$e^+e^-$-annihilation into hadrons at the LEP-2 centre of mass energy $\sqrts=196$~GeV for a quark jet of
virtuality $Q=\sqrts/2=98$~GeV with $Y=\ln(\sqrts/(2\lqcd))\approx6.0$ for $\lqcd=0.25$~GeV. If the
resulting distribution $D_q(\xi,Y)$ is refitted to a DG and compared with the $D^+(\xi,Y)$, the enlargement of 
the width compared with that given by ${\cal D}^+$ (\ref{eq:sigmalim}) can reach $10\%$. This latter effect is
mainly due to the positive $\cO\alphas$ correction to the coefficient function $C_q^g$ given by the larger
numerical coefficient $c_q^{(0)}=0.487$. Similar effects have been discussed in~\cite{Sapeta:2008km}.
In conclusion, we will directly fit the ${\cal D}^+(\xi,Y)$
distribution to the data of final state hadrons in the limiting spectrum approximation.

\subsection{Higher-order corrections for the DG limiting spectrum}
\label{sec:beyondNMLLA}

%\subsection{Limiting spectrum of the DG parameterisation}
%\label{subsec:higherordercorrecs}

The exact solution of the MLLA evolution equations with one-loop coupling constant entangles corrections
which go beyond $\cO{\sqrt\alphas}$, though the equations are originally obtained in this approximation 
\cite{Azimov:1985by}. 
%Though the evolution equations are approached equations within the MLLA scheme, the exact solution entangles
%higher-order corrections that go beyond the MLLA $\cO{\sqrt{\alphas}}$ scope. 
The exact solution resums fast convergent Bessel series in the 
limiting spectrum $\lambda\to0$. Using the DG parametrisation it is possible to match
the exact solution in the vicinity of the peak position $\delta\ll1$ after determining the DG moments: 
$\xi_1=\bar\xi$, $\xi_2=\mean{\xi^2}$, $\xi_3=\mean{\xi^3}$, $\xi_4=\mean{\xi^4}$, related to the 
dispersion, skewness and kurtosis through~\cite{Dokshitzer:1991ej}:
\begin{eqnarray}
\sigma^2\!&\!=\!&\!\xi_2-\bar\xi^2,\label{eq:sigmadeff}\\
s\!&\!=\!&\!\frac1{\sigma^3}(\xi_3-3\xi_2\bar\xi+2\bar\xi^3),\label{eq:skewdeff}\\
k\!&\!=\!&\!\frac1{\sigma^4}(\xi_4-4\xi_3\xi_1-3\xi_2^2+12\xi_2\bar\xi^2-6\bar\xi^4),
\label{eq:kurtdeff}
\end{eqnarray}
where $\xi_n$ is determined via
\begin{equation}\label{eq:exactmoments}
\xi_n=Y^n\cdot{\cal L}_n(B+1,B+2,z),\;B=\frac{a_1}{\beta_0},\;z=\sqrt{\frac{16N_c}{\beta_0}Y}
\end{equation} 
discussed in more detail in Appendix~\ref{section:hocorrtomoments}. Similarly, these extra corrections, which better
account for energy conservation and provide an improved description of the shape of the inclusive hadron
distribution in jets, will be computed and added hereafter to all the NMLLA+NLO$^{*}$ DG moments, as it was
done in~\cite{Dokshitzer:1991wu} for the particular case of the mean peak position, $\bar\xi$, but %$\xi_{\rm m}$, but
extended here also to all other components: Eqs.~(\ref{eq:meanellter}), (\ref{eq:sigmalim}),
(\ref{eq:skewnesslim}) and (\ref{eq:kurtosislim}).

%that we will consider in subsection \ref{subsec:meanmedian}.
\paragraph{Multiplicity.}
The extra ``hidden'' corrections discussed in Appendix~\ref{section:hocorrtomoments} result in one extra term
for the multiplicity in the DG limiting spectrum, which is inversely proportional to $Y$ and amounts to:
\begin{equation}\label{eq:extraNch}
\Delta{\cal N}=-\frac{0.168007}{Y},\quad \mbox{ for $n_f$~=~3, and }\quad \Delta{\cal N}=-\frac{0.23252}{Y},\quad \mbox{ for $n_f$~=~5}\,.
\end{equation}
However, we can use directly the full-NLO result obtained in~\cite{Dremin:2000ep} for the multiplicity.
In this case the extra correction amounts to:
\begin{eqnarray}
\Delta{\cal N}\!\!&=&\!\!-(0.08093 + 0.16539\ln Y)\frac1{Y},\mbox{ for $n_f$~=~3, and }\\
\,\Delta{\cal N}\!\!&=&\!\!-(0.00068 - 0.161658\ln Y)\frac1{Y},\, \mbox{ for $n_f$~=~5}\,.\label{eq:extraNchDremin}
\end{eqnarray}
%The previous choice was justified in subsection~\ref{subsec:CFlambda}. As concluded in~\cite{Dremin:2000ep},
although the terms $\propto\frac1{\sqrt{Y}}$ and $\propto\frac1Y$ %in Eq.~(\ref{eq:dreminmult_nf5}) 
are almost constant and practically compensate to each other at the currently accessible energies.
%Therefore, the MLLA gluon multiplicity (\ref{eq:mllafinalmult5}) would be a good approximation to the higher order result 
%in Eq.~(\ref{eq:dreminmult_nf5}). 

\paragraph{Peak position.}
The mean peak value of the DG distribution, $\bar\xi$, truncated as done in
Eq.~(\ref{eq:sxipmlla}) can be improved as discussed in~\cite{Dokshitzer:1991wu}.
The NMLLA correction proportional to $\ln Y$ is of relative order ${\cal O}(\sqrt{\alphas})$ 
and is very small ${\cal O}(10^{-3}\ln Y)$ compared to the second term. There is one extra correction
(numerical constant) to $\bar\xi$ coming from the exact solution of Eq.~(\ref{eq:gluonDg}) 
with $a_2=0$, written in terms of Bessel series in Appendix~\ref{section:hocorrtomoments}. 
%That is why,  in addition to the NMLLA correction to $\bar\xi$ found in this context,
%The expansion of the Bessel series in the exact solution provides an extra NMLLA term given by the 
%constant $-a_1(2a_1+3\beta_0)/(32N_c\beta_0)$. 
Indeed, substituting Eq.~(\ref{eq:besselratio}) into (\ref{eq:xi1}) (see
Appendix~\ref{section:hocorrtomoments} for a complete derivation), one obtains the extra NMLLA term to
$\bar\xi$: 
\begin{equation}\label{eq:extrabarxi}
\Delta\bar\xi=-\frac{\beta_0}{32N_c}B(2B+3),
\end{equation}
from the expansion of the Bessel series through the Eq.~(\ref{eq:xi1})
%like in~\cite{Dokshitzer:1991wu},
that should be added to Eq.~(\ref{eq:sxipmlla}).  
Therefore, the full resummed expression of the mean peak position reads
\begin{equation}\label{eq:meanell3}
\bar\xi=\frac{Y}2+\frac{a_1}{16N_c}\sqrt{\frac{16N_c}{\beta_0}Y}-2N_c\frac{a_2}{\beta_0}\ln Y-\frac{a_1(2a_1+3\beta_0)}{32N_c\beta_0}
\end{equation}
in its complete NMLLA+NLO$^{*}$ form. 
The corresponding position of the maximum is related to the mean peak value by the 
expression~\cite{Dokshitzer:1991za}:
\begin{equation}\label{eq:diffxi2}
\ximax-\bar\xi=-\frac12\sigma s=\frac{3a_1}{32N_c}\,,
\end{equation}
such that
\begin{equation}
\ximax=\frac{Y}2+\sqrt{\frac{a_1^2}{16N_c\beta_0}Y}
-2N_c\frac{a_2}{\beta_0}\ln Y-\frac{a_1^2}{16N_c\beta_0}\,.
\label{eq:ximax}
\end{equation}
where the DLA width $\sigma$ and skewness $s$ are enough for the computation.
Asymptotically ($Y\to\infty$) and factorising by $Y$, one recovers the maximum of the peak
position for the DLA spectrum Eq.~(\ref{eq:dlapeak}). In the same approximation, since $s(Y)\to0$,
the expression of the mean peak position in Eq.~(\ref{eq:diffxi2}) coincides with that of the maximum of
the Gaussian distribution. Of course, the ensemble of NMLLA corrections written in Eq.~(\ref{eq:meanell3}) 
can be obtained from Eq.~(\ref{eq:gluonDg}), provided that one can determine the exact solution of the evolution 
equations. Notice that Eq.~(\ref{eq:ximax}) does not include any term $\propto\beta_1$, as this kind of
term appears when higher-order corrections are included in the evolution equations and their solutions.

\paragraph{Width.}
Similar extra corrections can be found for the dispersion by calculating $\xi_2$ through this
recursive procedure. By making use of Eq.~(\ref{eq:exactmoments}) and the full
derivation presented in Appendix~\ref{section:hocorrtomoments}, it was found in~\cite{Dokshitzer:1991ej}:
\begin{equation}
\frac{\xi_2}{Y^2}=\frac14+\frac{B(B+\frac13)}{z^2}+\frac{(B+\frac13)}{z^2}
\left(1-\frac{2B(B+2)}{z^2}\right)\frac{I_{B+2}(z)}{I_{B+1}(z)},
\end{equation}
such that, with $\sigma^2=\xi_2-\bar\xi^2$ given by Eq.~(\ref{eq:sigmadeff}),
one finds the extra correction (for $n_f=5$)
\begin{equation}\label{eq:extrasigma}
\frac{\Delta\sigma}{0.36499Y^{3/4}}=\frac{1.98667}{Y^{3/2}},
\end{equation}
which should be accordingly added to the r.h.s. of Eq.~(\ref{eq:sigmalim}).\\

\paragraph{Skewness.}
In the case of the skewness, the expression for $\xi_3$ reads
\begin{eqnarray}\label{eq:xi3sek}
\frac{\xi_3}{Y^3}\!&\!=\!&\!\frac18+\frac{3B(B+1)}{2z^2}\left(1-\frac{4B(B+3)}{3z^2}\right)
+\frac2z\left[\frac{3B+2}{8}-\frac{B(B+1)(B+3)}{z^2}\left(1\right.\right.\cr
\!&\!-\!&\!\left.\left.\frac{2B(B+2)}{z^4}\right)\right]\frac{I_{B+2}(z)}{I_{B+1}(z)}
\end{eqnarray}
such that, if one makes use of the expression~(\ref{eq:skewdeff}), 
the extra correction reads (for $n_f=5$)
\begin{equation}\label{eq:extraSkew}
\frac{\Delta s}{-1.94703/Y^{3/4}}=-\frac{1.64393}{Y},
\end{equation}
to be added to the r.h.s. of Eq.~(\ref{eq:skewnesslim}). Notice that Eq.~(\ref{eq:xi3sek}) was given
in~\cite{Dokshitzer:1991ej} without accounting for terms $\cO{z^{-4}}$ and $\cO{z^{-7}}$. Such terms cannot
be neglected when dealing with MLLA and NMLLA corrections.\\

\paragraph{Kurtosis.}
Finally, for the kurtosis, we obtain the formula for $\xi_4$:
\begin{eqnarray}
\frac{\xi_4}{Y^4}\!&\!=\!&\!\frac1{16}-\frac{(B+4)(15B^3+30B^2+5B-2)}{5z^2}\left(1-\frac{4B(B+3)}{3z^4}\right)+
\frac{9B^2+15B+2}{6z^2}\cr
\!&\!+\!&\!\frac1{z}\left[\frac{B+1}2+\frac{4(B+3)(B+4)(15B^3+30B^2+5B-2)}{15z^4}\left(1-\frac{2B(B+2)}{z^2}\right)
\right.\cr
\!&\!-\!&\!\left.\frac{5B^3+35B^2+50B+8}{5z^2}\right]\frac{I_{B+2}(z)}{I_{B+1}(z)},
\end{eqnarray}
which can be cast into Eq.~(\ref{eq:kurtdeff}) to obtain the corresponding correction which reads (for $n_f=5$):
\begin{equation}\label{eq:extraKurt}
\frac{\Delta k}{-2.15812/\sqrt{Y}}=-\frac{8.05771}{Y^{3/2}},
\end{equation}
to be also added to Eq.~(\ref{eq:kurtosislim}). 

\paragraph{Final numerical formul{\ae}.}
%\section{Numerical evaluation of the components of the DG in the limiting spectrum}
For easiness of comparison to the data, we provide here the final numerical expressions for the energy
evolution of the NMLLA+NLO$^{*}$ components of DG hadron 
distribution of jets in the limiting spectrum, evaluated from Eqs.~(\ref{eq:multlim}),
(\ref{eq:meanellter}), (\ref{eq:sigmalim}), (\ref{eq:skewnesslim}) and (\ref{eq:kurtosislim}) 
plus the higher-order corrections eqs. (\ref{eq:extraNchDremin}), (\ref{eq:extrabarxi}),
(\ref{eq:extrasigma}), (\ref{eq:extraSkew}) and (\ref{eq:extraKurt}). 
We include the expressions for $n_f=3,4,5$ active quark flavours, 
%in order to show the small variation of the numerical coefficients as a function of $n_f$, 
although only the cases $n_f=4,5$ are relevant for most phenomenological applications (jets are usually
measured with energies (well) above the charm and bottom-quark mass thresholds). 
% bottom-pair $\r, b\bar{\rm b}$ mass threshold).\\ 
%all the figures of the paper. 
For $n_f=3$ quark flavours, one finds
\begin{eqnarray}
{\cal N}(Y)&=&{\cal K}^{\rm ch}\exp\left[2.3094\sqrt{Y}-0.373457\ln Y
+\left(0.061654+0.456178\ln Y\right)\frac1{\sqrt{Y}}\right.\cr %-\frac{0.168007}{Y}\right],\\
&+&\left.(0.121834 - 0.14749\ln Y)\frac1{Y}\right]\label{eq:dreminmult_evol_nf3},\\
\bar{\xi}(Y)&=&0.5Y+0.539929\sqrt{Y}-0.05\ln Y,\label{eq:ximean_evol_nf3}\\
\ximax(Y)&=&0.5Y+0.539929\sqrt{Y}-0.291524-0.05\ln Y,\\
\sigma(Y)&=&0.379918Y^{3/4}\left[1-0.324759\frac1{\sqrt{Y}}
-\left(1.6206-0.296296\ln Y\right)\frac1{Y}+\frac{1.70797}{Y^{3/2}}\right],\\
s(Y)&=&-\frac{1.84616}{Y^{3/4}}\left[1-0.324759\frac1{\sqrt{Y}}-\frac{1.63978}{Y}\right],\\
k(Y)&=&-\frac{2.33827}{\sqrt{Y}}\left[1-0.866025\frac1{\sqrt{Y}}+\left(0.713767-0.197531\ln Y\right)\frac1Y
-\frac{6.99062}{Y^{3/2}}\right].
\end{eqnarray}
For $n_f=4$ quark flavours, relevant for jet analysis above the charm mass threshold ($m_{\rm
  c}\approx$~1.3~GeV) but below the bottom mass, one finds
\begin{eqnarray}
{\cal N}(Y)&=&{\cal K}^{\rm ch}\exp\left[2.4\sqrt{Y}-0.427778\ln Y
+\left(0.0214879+0.44352\ln Y\right)\frac1{\sqrt{Y}}\right.\cr
&+&\left.(0.0682865 - 0.158071\ln Y)\frac1{Y}\right]\label{eq:dreminmult_evol_nf4},\\
\bar{\xi}(Y)&=&0.5Y+0.564815\sqrt{Y}-0.0287888\ln Y,\label{eq:ximean_evol_nf4}\\
\ximax(Y)&=&0.5Y+0.564815\sqrt{Y}-0.319015-0.0287888\ln Y,\label{eq:ximax_evol_nf4}\\
\sigma(Y)&=&0.372678Y^{3/4}\left[1-0.312499\frac1{\sqrt{Y}}
-\left(1.31978-0.2772\ln Y\right)\frac1{Y}+\frac{1.83441}{Y^{3/2}}\right],\label{eq:sigma_evol_nf4}\\
s(Y)&=&-\frac{1.89445}{Y^{3/4}}\left[1-0.312499\frac1{\sqrt{Y}}-\frac{1.64009}{Y}\right],\label{eq:skew_evol_nf4}\\
k(Y)&=&-\frac{2.25}{\sqrt{Y}}\left[1-0.833333\frac1{\sqrt{Y}}+\left(0.740793-0.1848\ln Y\right)\frac1Y
-\frac{7.47314}{Y^{3/2}}\right];\label{eq:kurt_evol_nf4}
\end{eqnarray}
and for $n_f=5$ quark flavours relevant for jet analysis above the bottom mass threshold ($m_{\rm b}\approx$~4.2~GeV):
\begin{eqnarray}
{\cal N}(Y)&=&{\cal K}^{\rm ch}\exp\left[2.50217\sqrt{Y}
-0.491546\ln Y-\left(0.06889-0.41151\ln Y\right)\frac1{\sqrt{Y}}\right.\cr %-\frac{0.23252}{Y}\right], %\label{eq:N_evol_nf5}
&+&\left.(0.00068 - 0.161658\ln Y)\frac1{Y}\right]\label{eq:dreminmult_evol_nf5},\\
\bar{\xi}(Y)&=&0.5Y+0.592722\sqrt{Y}+0.002\ln Y,\label{eq:ximean_evol_nf5}\ \\
\ximax(Y)&=&0.5Y+0.592722\sqrt{Y}-0.351319+0.002\ln Y, \label{eq:ximax_evol_nf5}\\\
\sigma(Y)\!\!&\!\!=\!\!&\!\!0.36499Y^{3/4}\left[1-0.299739\frac1{\sqrt{Y}}
-\left(1.4921-0.246692\ln Y\right)\frac1{Y}+\frac{1.98667}{Y^{3/2}}\right],\label{eq:sigma_evol_nf5}\\
s(Y)\!\!&\!\!=\!\!&\!\!
-\frac{1.94704}{Y^{3/4}}\left[1-0.299739\frac1{\sqrt{Y}}-\frac{1.64393}{Y}\right],\label{eq:skew_evol_nf5}\\
k(Y)\!\!&\!\!=\!\!&\!\!-\frac{2.15812}{\sqrt{Y}}\left[1-0.799305\frac1{\sqrt{Y}}
+\left(0.730466-0.164461\ln Y\right)\frac1Y-\frac{8.05771}{Y^{3/2}}\right].\label{eq:kurt_evol_nf5}
\end{eqnarray}
The MLLA expressions first computed in~\cite{Fong:1990nt} can be naturally recovered from our results by 
keeping all terms up to $1/\sqrt{Y}$. 
%setting all corrections $\propto\frac1Y$ to zero. 
For $n_f=5$ quark flavours, they read:
\begin{eqnarray}
{\cal N}(Y)&=&{\cal K}^{\rm ch}\exp\left[2.50217\sqrt{Y}-0.491546\ln Y\right],\\
\bar{\xi}(Y)&=&0.5Y+0.592722\sqrt{Y},\\
\ximax(Y)&=&0.5Y+0.592722\sqrt{Y},\\
\sigma(Y)&=&0.36499Y^{3/4}\left[1-0.299739\frac1{\sqrt{Y}}\right],\\
s(Y)&=&-\frac{1.94704}{Y^{3/4}},\\
k(Y)&=&-\frac{2.15812}{\sqrt{Y}}\left[1-0.799305\frac1{\sqrt{Y}}\right],
\end{eqnarray}
which clearly highlight, by comparing to the corresponding full expressions above, the new NMLLA$+$NLO$^*$
terms computed in this work for the first time.

\subsection{Other corrections: finite mass, number of active flavours, power terms, and $\boldsymbol{\lqcd}$ rescaling}
\label{subsec:powcorrecs}

\begin{figure}[htbp]
\begin{center}
\epsfig{file=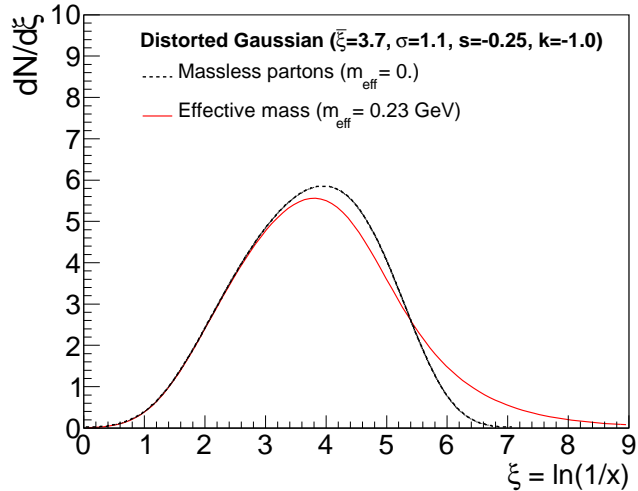, width=8.5truecm}
\caption{\label{fig:DG_vs_DGmeff} Comparison of the limiting-spectrum distorted Gaussian for jets typical of LEP-1
  energies (mean $\bar\xi=3.7$, width $\sigma=1.1$, skewness $s=k=-0.25$, and kurtosis $k=-1.$) with and without
  corrections for finite-mass effects ($\meff\approx\lqcd$) according to Eq.~(\ref{eq:Dmeff}).}
\end{center}
\end{figure}

\paragraph{Mass effects:}
In the approach discussed so far, the partons have been assumed massless and so their scaled energy and
momentum spectra are identical. Experimentally, the scaled momentum distribution 
$\xi_{\rm p}=\ln(\sqrts/(2\,p_h))$ is measured and, since the final-state hadrons are massive, the
equivalence of the theoretical and experimental spectra no longer exactly holds.  
%In~\cite{Khoze:1996py} the assumption was made that the cutoff $Q_0$ can be related to an effective mass $\meff\approx\lqcd$ %~0.32~GeV
%accounting for the typical mass-weighted fraction of pion, kaon and protons in a jet. This allows one
One can relate the  measured $\xi_p$ spectrum to the expected DG distribution (which depends on
$\xi\equiv\xi_{_{\rm E}}$) by performing the following change of variables~\cite{Khoze:1996py}: 
\begin{equation}\label{eq:Dmeff}
\frac1{\sigma_{\rm tot}}\frac{d\sigma^{\rm h}}{d\xi_p} \propto \frac{p_h}{E_h}D^+(\xi,Y)\,,\mbox{ with } 
\xi = \ln(1/x) = \ln\left({\frac{\sqrts/2}{\sqrt{(s/4)e^{-2\xi_p}+\meff^2}}}\right)\,,
\end{equation}
where the energy of a hadron with measured momentum $p_h=(\sqrts/2)\cdot\exp{-\xi_p}$ is
$E_h=\sqrt{p_h^2+\meff^2}$, and $\meff$ %\approx\lqcd$ %~0.32~GeV
is an effective mass of ${\cal O}(\lqcd)$ accounting for the typical mixture of pion, kaon and protons in a jet.
In Fig.~(\ref{fig:DG_vs_DGmeff}) we compare the DG distribution in the limiting-spectrum for
the typical HBP of LEP-1 jets with and without mass corrections,
using Eq.~(\ref{eq:Dmeff}) with $\meff~=0$ and $\meff~=~\lqcd \approx~0.23$~GeV. As expected, the net effect of the non-null mass of
the measured jet particles affects the tail of the distribution at high $\xi$ (i.e. at very low momenta) but
leaves otherwise relatively unaffected the rest of the distribution. In the analysis of experimental jet data
in the next Section, the rescaling given by Eq.~(\ref{eq:Dmeff}) will be applied to the theoretical DG
distribution for values of $\meff$~=~0--0.35~GeV to gauge the sensitivity of our results to finite-mass
effects. Since experimentally there are not many measurements in the large $\xi$ tail (i.e. very low
particle momenta) and here the distribution has larger uncertainties than in other ranges of the spectrum, 
the fits to the data turn out to be rather insensitive to $\meff$.
%The limiting momentum spectrum is based on massless partons. One can extend the formalism by adding 

\paragraph{Number of active flavours $n_f$: }
The available experimental $\epem$ data covers a range of jet energies $E_{\rm jet}\approx$~1--100~GeV
which, in its lowest range, crosses the charm ($m_{\rm c}\approx$~1.3~GeV) and bottom 
($m_{\rm b}\approx$~4.2~GeV) thresholds in the counting of the number of active quark flavours $n_f$ present
in the formul{\ae} for the energy-dependence of the DG moments. Although the differences are small, rather
than trying to interpolate the expressions for different values of $n_f$ in the heavy-quark crossing regions,
in what follows we will use the formula{\ae} for $n_f=5$ for the evolution of all moments and rescale the
obtained moments of the four lower-$\sqrts$ datasets from the BES experiment~\cite{Dunwoodie:2003xt} to account for
their lower effective value of $n_f$. The actual numerical differences between the evolutions of the DG
moments for $n_f=4$ and $n_f=5$ quark flavours -- given by
Eqs.~(\ref{eq:dreminmult_evol_nf4})--(\ref{eq:kurt_evol_nf4}) and 
(\ref{eq:dreminmult_evol_nf5})--(\ref{eq:kurt_evol_nf5}) respectively -- when evaluated for energies below the
bottom-quark threshold are quite small: 0--10\% for ${\cal N}(Y)$, below $1\%$ for $\ximax(Y)$, around 5\%
for the width $\sigma(Y)$, and 5--10\% for the skewness $s(Y)$ and kurtosis $k(Y)$. In this respect, the most
``robust'' ($n_f$-insensitive) observable is the peak position of the distribution.

\paragraph{Power-suppressed terms:}
Power corrections of order ${\cal O}(Q_0^n/Q^n)$ 
%have been considered for the extraction of the coupling constant $\alphas(Q^2)$ in high-energy processes. The origin of such 
%corrections in parton shower evolution at small $x$ is hidden in consistent parton kinematics
%considerations. They should 
appear if one sets more accurate integration bounds of the integro-differential evolution equations over $z$, 
such as $\frac{Q_0}{Q}\leq z\leq1-\frac{Q_0}{Q}$ instead of $0\leq z\leq1$, which actually leads to
Eq.~(\ref{eq:gluonDg}) after Mellin transformation with $Q_0\sim m_{\rm h}$, where $m_{\rm h}$ is the hadron
mass (for more details see review~\cite{Albino:2008aa,Albino:2008gy}). 
For the mean multiplicity, this type of corrections was considered in~\cite{Capella:1999ms}. They were proved
to be powered-suppressed and to provide small corrections at high-energy scales. Furthermore, they 
become even more suppressed in the limiting spectrum case where $Q_0$
% which is normally set to $1$~GeV for light quarks and gluons in high-energy processes, 
can be extended down to $\lqcd$ for infrared-safe observables like the hump-backed plateau. 
The MLLA computation of power corrections for differential observables is a numerical cumbersome task which,
for the hump-backed plateau, would add minor improvements in the very small $x$ domain 
$\ln(1/x)\to\ln(Q/\lqcd)$ away from the hump region of our interest, and thus they %such corrections 
would not introduce any significant shift to the main moments of the hadron distributions (in
particular its peak position $\ximax$, and width $\sigma$).

\paragraph{Rescaling of  the $\lqcd$ parameter:}
Technically, the $\lqcd$ parameter is a scheme-dependent integration constant of the QCD $\beta$-function.
Rescaling the QCD parameter by a constant, $\lqcd\to C\lqcd$, would give an equally acceptable definition.
In our formalism, such a variation would translate into a $\ln{C}$-shift of the constant term of the HBP peak,
Eq.~(\ref{eq:ximax})~\cite{Dokshitzer:1991wu}, which corresponds to higher-order contributions to the solution
of the evolution equations. 
%and new source of NMLLA$+$NLO$^*$ contributions to the solution of the evolution equations.
The approach adopted here is to connect $\lqcd$ to $\alphas$ in the $\overline{\rm MS}$ 
factorisation scheme through the two-loop Eq.~(\ref{eq:twoloop}) and, 
at this level of NLO accuracy, there is no ambiguity when comparing our extracted $\alphas$ 
results to other values obtained using the same definition.
\section{Extraction of $\boldsymbol{\alphas}$ from the %energy-dependence of the  
evolution of the distribution  of hadrons in jets in $\epem$ collisions}
%maximum peak position $\boldsymbol{\ximax(Q^2)}$}
\label{section:extracting}

In this last section, we confront our NMLLA+NLO$^*$ calculations with all the existing charged-hadron spectra
measured in jets produced in $\epem$ collisions in the range of energies $\sqrts\approx$~2--200~GeV.
The experimental distributions as a function of $\xi_{\rm p}=\ln(\sqrts/(2\,p_h))$ are fitted to the distorted
Gaussian parametrisation, Eq.~(\ref{eq:xDg}), and the corresponding DG components are derived for each dataset.
More concretely, we fit the experimental distributions to the expression:
%The final formula for the fits of the single inclusive distribution of charged hadrons in $\epem\to q\bar{q}$ 
%is given by:
\begin{equation}\label{eq:singleincldistfit}
\frac1{\sigma_{\rm tot}}\frac{d\sigma^{\rm h}}{d\xi}={\cal K}^{\rm ch}\frac{2C_F}{N_c}D^+(\xi,Y)\,,
\end{equation}
where $D^+(\xi,Y)$ is given by Eq.~(\ref{eq:Dmeff}) corrected to take into account the finite-mass effects of
the hadrons (for values of $\meff$~=~0--0.35~GeV, see below) with $Y=\ln[\sqrts/(2\,\lqcd)]$.
%with the components (\ref{eq:dreminmult_evol_nf5})--(\ref{eq:kurt_evol_nf5}). 
Each fit has five free parameters for the DG: maximum peak position, total multiplicity, width, skewness and
kurtosis. In total, we analyse 32 data-sets from the following experiments: 
BES at $\sqrts$~=~2--5~GeV~\cite{Dunwoodie:2003xt}; 
%BaBar\footnote{The individual distributions measured for 
%prompt pions, kaons and (anti)protons have been added up into a single charged-particle distribution.} 
%at $\sqrts$~=~10.54~GeV~\cite{Lees:2013rqd};
TASSO at $\sqrts$~=~14--44~GeV~\cite{Braunschweig:1988qm,Braunschweig:1990yd}; 
TPC at $\sqrts$~=~29~GeV~\cite{Aihara:1988su}; TOPAZ at $\sqrts$~=~58~GeV~\cite{Itoh:1994kb}; 
ALEPH~\cite{Barate:1996fi}, L3~\cite{Adeva:1991it} and OPAL~\cite{Akrawy:1990ha,Ackerstaff:1998hz} at $\sqrts$~=~91.2~GeV; 
ALEPH~\cite{Buskulic:1996tt,Heister:2003aj}, DELPHI~\cite{Abreu:1996mk} and OPAL~\cite{Alexander:1996kh} 
%and ALEPH~\cite{Heister:2003aj}
at $\sqrts$~=~133~GeV; and ALEPH~\cite{Heister:2003aj} and
OPAL~\cite{Ackerstaff:1997kk,Abbiendi:1999sx,Abbiendi:2002mj} in the range $\sqrts$~=~161--202~GeV. 
The total number of points is 1019 and the systematic and statistical uncertainties of the spectra are
added in quadrature.\\
%\footnote{Out of the 32 data-sets, the eight spectra of 
%Refs.~\cite{Braunschweig:1990yd,Itoh:1994kb,Barate:1996fi,Akrawy:1990ha,Heister:2003aj} quoted a too small  
%systematic uncertainty which has been increased in order to get a realistic goodness of fit, $\chi^2/{\rm ndf}\approx$~1,  
%for the DG parametrization.}.\\

\begin{figure}[htbp!]
\begin{center}
\epsfig{file=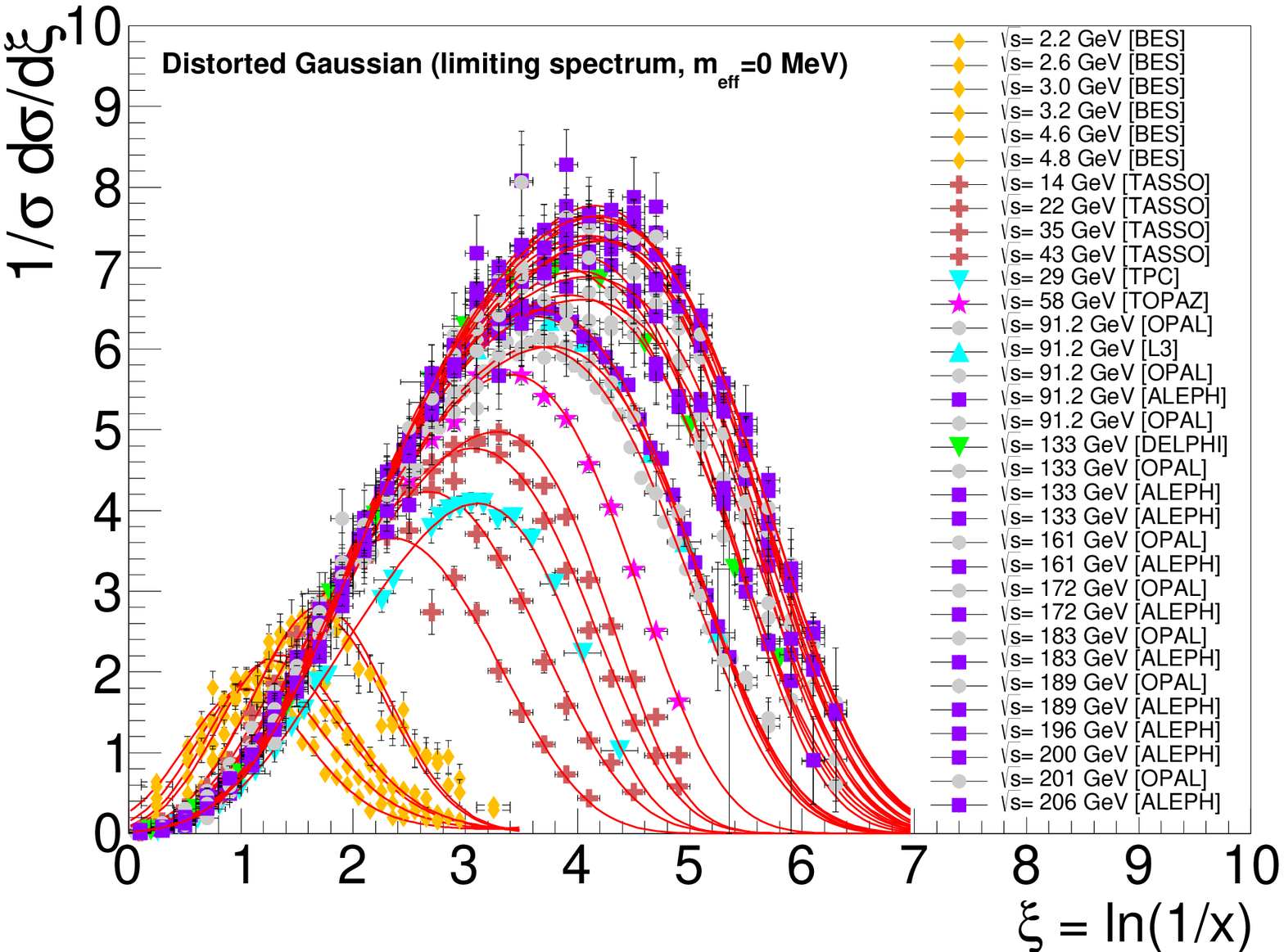,width=11.5truecm}\vspace{-1.0truecm}
\epsfig{file=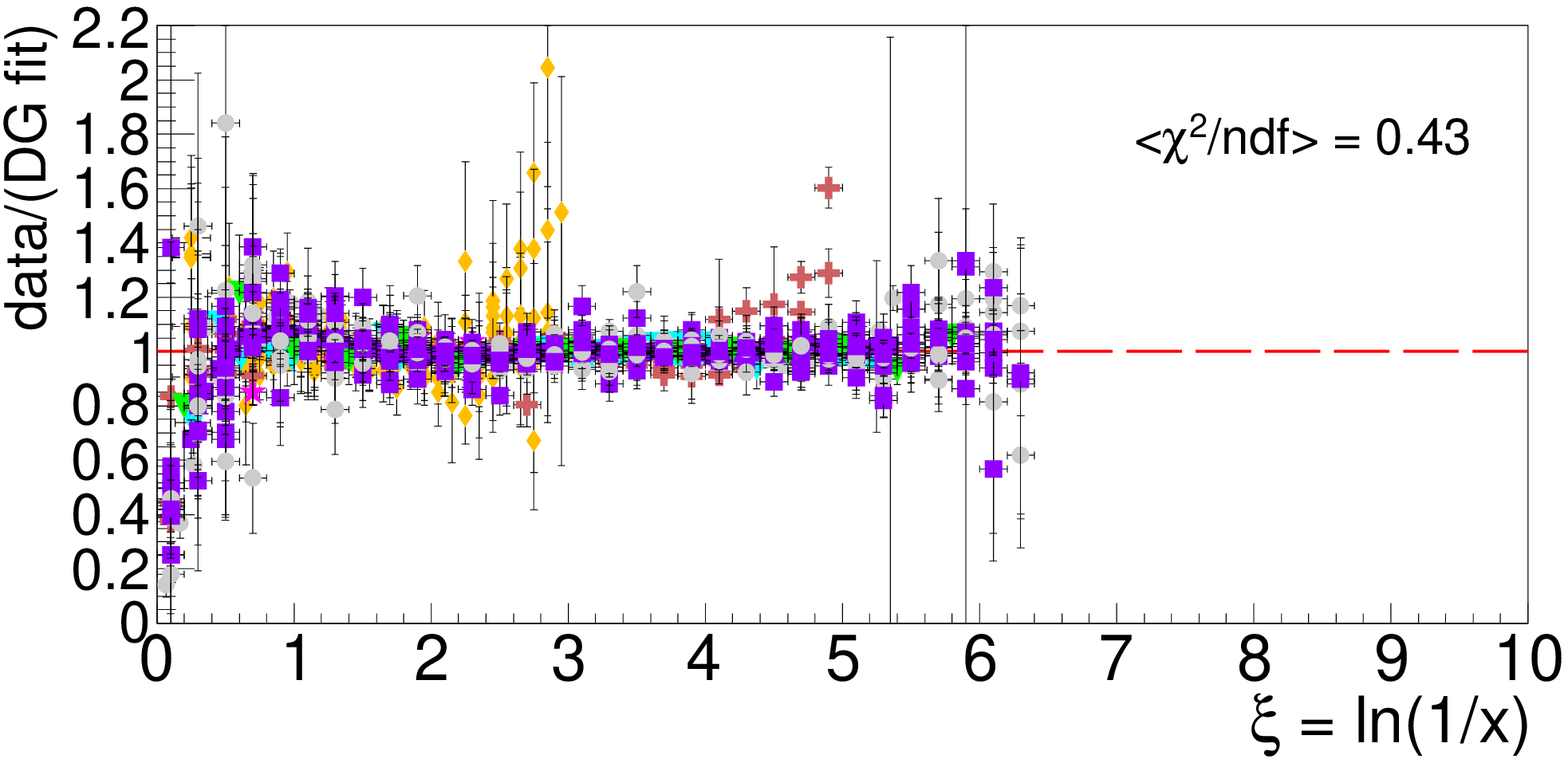,width=11.5truecm}
%\vspace{-0.5cm}
\caption{\label{fig:DGeedata_meff0}
Top: Single inclusive hadron distributions measured in jets in the world $\epem$ data at
$\sqrts\approx$~2--200~GeV as a function of $\xi=\ln(\sqrts/(2\,p_h))$ %. The curves are fits to the 
fitted to the distorted Gaussian Eq.~(\ref{eq:Dmeff}) with $\meff$~=~0. 
Bottom: Ratio of each set of data points to the corresponding DG fit. The value $\mean{\chi^{2}/{\rm ndf}}$
quoted is the average of all individual fits.}
\end{center}
\end{figure}

\begin{figure}[ht!]
\begin{center}
\epsfig{file=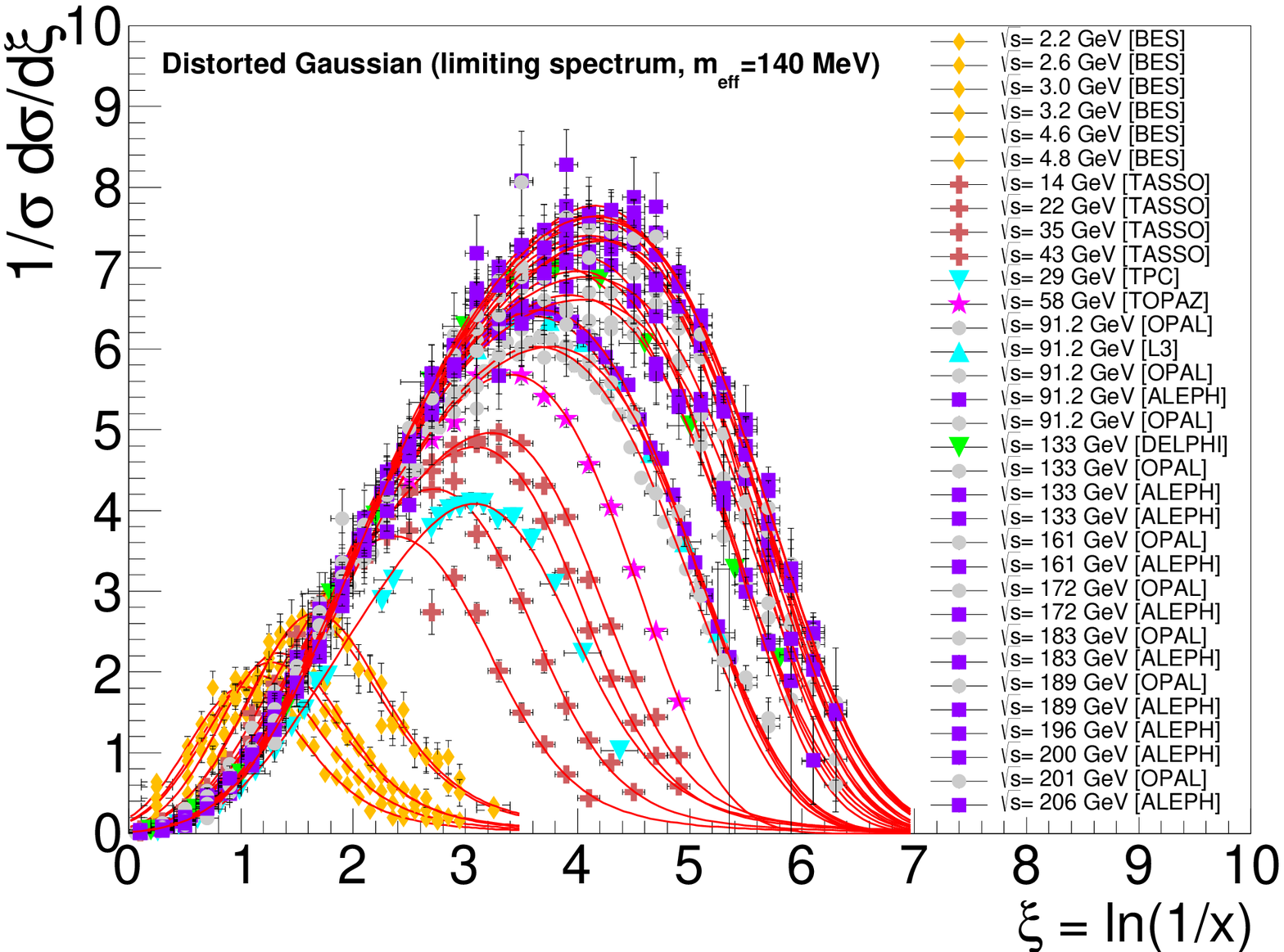,width=11.5truecm}\vspace{-1.0truecm}
\epsfig{file=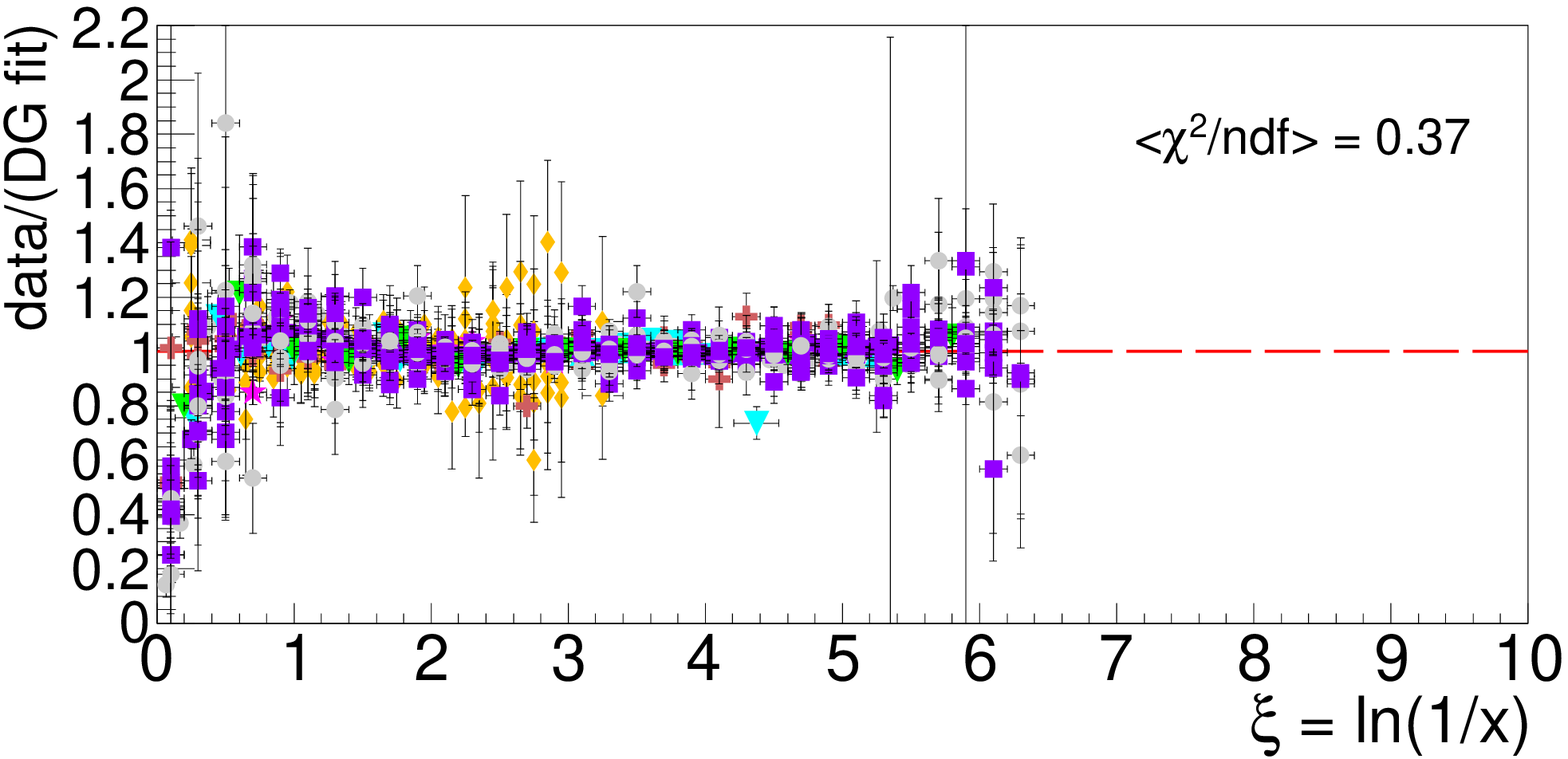,width=11.5truecm}
%\vspace{-0.5cm}
\caption{\label{fig:DGeedata_meff230}
Top: Single inclusive hadron distributions measured in jets in the world $\epem$ data at
$\sqrts\approx$~2--200~GeV as a function of $\xi=\ln(\sqrts/(2\,p_h))$ %. The curves are fits to the 
fitted to the distorted Gaussian Eq.~(\ref{eq:Dmeff}) with $\meff$~=~140~MeV
Bottom: Ratio of each set of data points to the corresponding DG fit. The value $\mean{\chi^{2}/{\rm ndf}}$
quoted is the average of all individual fits.}
\end{center}
\end{figure}

\begin{figure}[htbp!]
\begin{center}
\epsfig{file=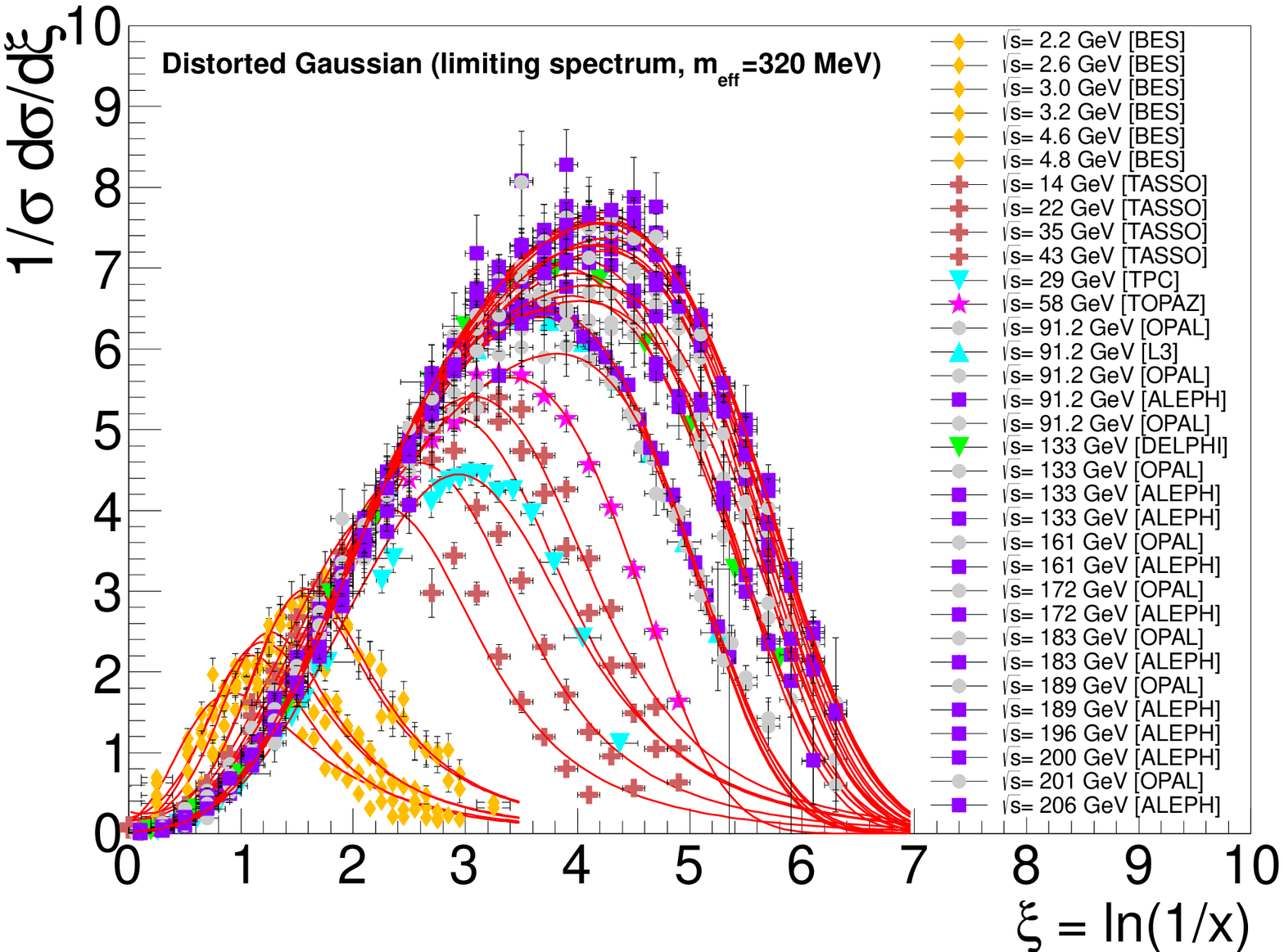,width=11.5truecm}\vspace{-0.975truecm}
\epsfig{file=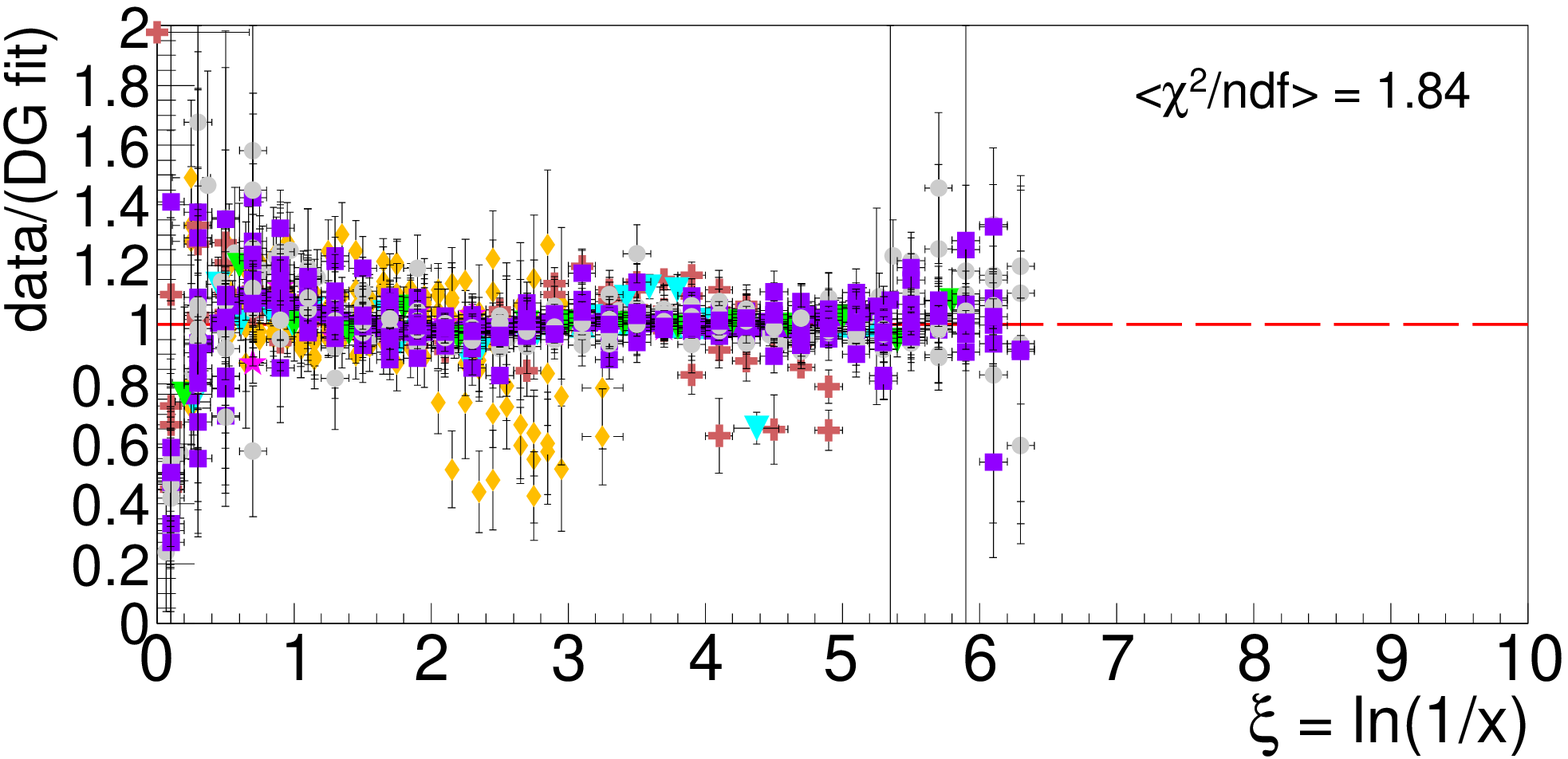,width=11.5truecm}
%\vspace{-0.5cm}
\caption{\label{fig:DGeedata_meff320}
Top: Single inclusive hadron distributions measured in jets in the world $\epem$ data at
$\sqrts\approx$~2--200~GeV as a function of $\xi=\ln(\sqrts/(2\,p_h))$ %. The curves are fits to the 
fitted to the distorted Gaussian Eq.~(\ref{eq:Dmeff}) with $\meff$~=~320~MeV. 
Bottom: Ratio of each set of data points to the corresponding DG fit. The value $\mean{\chi^{2}/{\rm ndf}}$
quoted is the average of all individual fits.}
\end{center}
\end{figure}

%\paragraph{Fits of the experimental fragmentation functions to the DG (limiting spectrum) :} %for $\meff$~=~0, 230,320~MeV:}
In order to assess the effect of finite-mass corrections discussed in the previous Section, we carry out the
DG fits of the data to Eq.~(\ref{eq:Dmeff}) for many values of $\meff$ in the range 0--320~MeV. The lower
value assumes that hadron and parton spectra are identical, the upper
%second one takes as effective mass a value equivalent to that of $\lqcd$ at NLO ($\MSbar$), and the third 
choice corresponds to an average of the pion, kaon and (anti)proton masses weighted by their corresponding
abundances (65\%, 30\% and 5\% approximately) in $\epem$ collisions. 
Representative fits of all the single-inclusive hadron distributions for $\meff$~=~0, 140, and 320~MeV are shown in
Figures~\ref{fig:DGeedata_meff0}--\ref{fig:DGeedata_meff320} respectively, %collects all the fits perfomed withthe data.   
%show all the single-inclusive hadron distributions fitted to the DG distribution for $\meff$~=~0, 120, and
%320~MeV respectively, %, Eq.~(\ref{eq:Dmeff}),
with the norm, peak, width, skewness, and kurtosis as free parameters. In all cases the individual data-model
agreement is very good, with goodness-of-fit per degree-of-freedom $\chi^2/{\rm ndf}\approx$~0.5--2.0, as
indicated in the data/fit ratios around unity in the bottom panels. 
The fits to all datasets with energies above $\sqrts$~=~50~GeV turn out to be completely insensitive to the
choice of $\meff$, i.e. the moments of the DG obtained are ``invariant'' with respect to the value of
$\meff$, whereas those at lower energies are more sensitive to it. The value of the effective mass that
provides an overall best agreement to the whole set of experimental distributions is $\meff\approx$~140~MeV,
which is consistent with a dominant pion composition of the inclusive charged hadron spectra.
%The finite-mass correction plays in general a small role in
%the fit results, although the lowest-$\sqrts$ data tend to prefer (somehow surprisingly) a zero $\meff$ value.\\

The general trends of the DG moments are already visible in these plots: 
as $\sqrts$ increases, the peak of the distribution shifts to larger values of $\xi$ (i.e. smaller relative
values of the charged-hadron momenta) and the spectrum broadens (i.e. its width $\sigma$ increases). In
the range of the current measurements, the peak moves from $\ximax\approx$~1 to $\ximax\approx$~4, and the
width increases from $\sigma\approx$~0.5 to 1.2.
The expected logarithmic-like energy dependence of the peak of the $\xi$
distribution, %Eq.~(\ref{eq:ximean_evol_nf5}) and 
given by Eq.~(\ref{eq:ximax_evol_nf5}),  due to soft gluon coherence (angular ordering), 
correctly reproduces the suppression of hadron production at small $x$ seen in
the data to the right of the distorted Gaussian peak. Although a decrease at large $\xi$ 
(very small $x$) is expected based on purely kinematic arguments, the peak position would vary twice as
rapidly with the energy in such a case in contradiction with the calculations and data.
%Though new ingredients have been taken into account such as the 
%second-loop correction to $\alphas$, the maximum of peak position (\ref{eq:ximax}) 
%does not include any $\propto\beta_1$ correction in the NMLLA$+$NLO$^*$ scheme, which would
%appear beyond this level of accuracy, and the new correction $\propto\ln(Y)$ is suppressed in 
%the energy range of our interest. 
The integral of the $\xi$ distribution gives the total charged-hadron multiplicity ${\cal N}^{\rm ch}$ which
increases exponentially as per Eq.~(\ref{eq:dreminmult_evol_nf5}).\\

%\paragraph{Energy-dependence of the DG moments:}

The $\sqrts$-dependence of each one of the individual DG moments is studied by fitting their evolution
to our NMLLA+NLO$^*$ limiting-spectrum predictions Eqs.~(\ref{eq:dreminmult_evol_nf5})--(\ref{eq:kurt_evol_nf5}) 
with $Y=\ln(\sqrts/(2\lqcd))$ %(\ref{eq:ximean_evol_nf5})--(\ref{eq:kurt_evol_nf5})
for $n_f$~=~5 quark flavours, with $\lqcd$ as the only free parameter. Before performing the combined
energy-dependence fit, the moments of the lowest-$\sqrts$ distribution from the BES experiment are corrected
to account for their different number of active flavours ($n_f$~=~3,4) as described in the previous Section.

\begin{figure}[htbp!]
\begin{center}
\epsfig{file=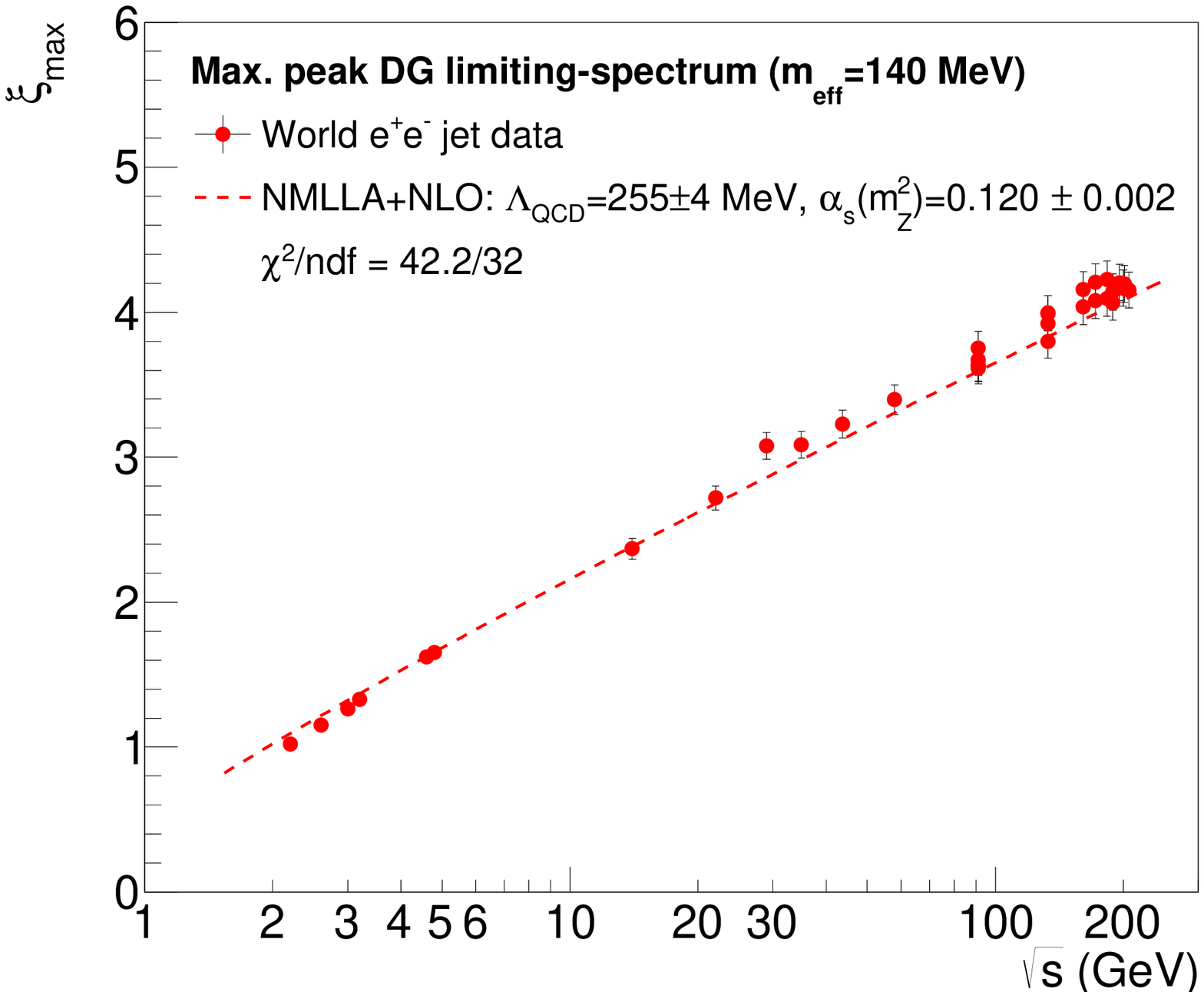,width=10.5truecm}
\caption{\label{fig:peak_evol} Energy evolution of the maximum peak position $\ximax$ of the
  spectrum of charged hadrons in jets measured in $\epem$ at collision energies
  $\sqrts\approx$~2--200~GeV, fitted to Eq.~(\ref{eq:ximax_evol_nf5}) with $Y~=~\ln(\sqrts/(2\lqcd))$,
  with finite-mass corrections ($\meff$~=~0.14~GeV). %(right) and without (left) finite-mass corrections.
  The extracted values of $\lqcd$ and equivalent NLO$_{_{\rm \MSbar}}$ $\alphasmZ$ and the
  goodness-of-fit per degree-of-freedom $\xi^2/$ndf, are quoted.}  
\end{center}
\end{figure}

%The result of the $\ximax(\sqrts)$ fit yields (Fig.~\ref{fig:peakevol})

\begin{figure}[htbp!]
\begin{center}
\epsfig{file=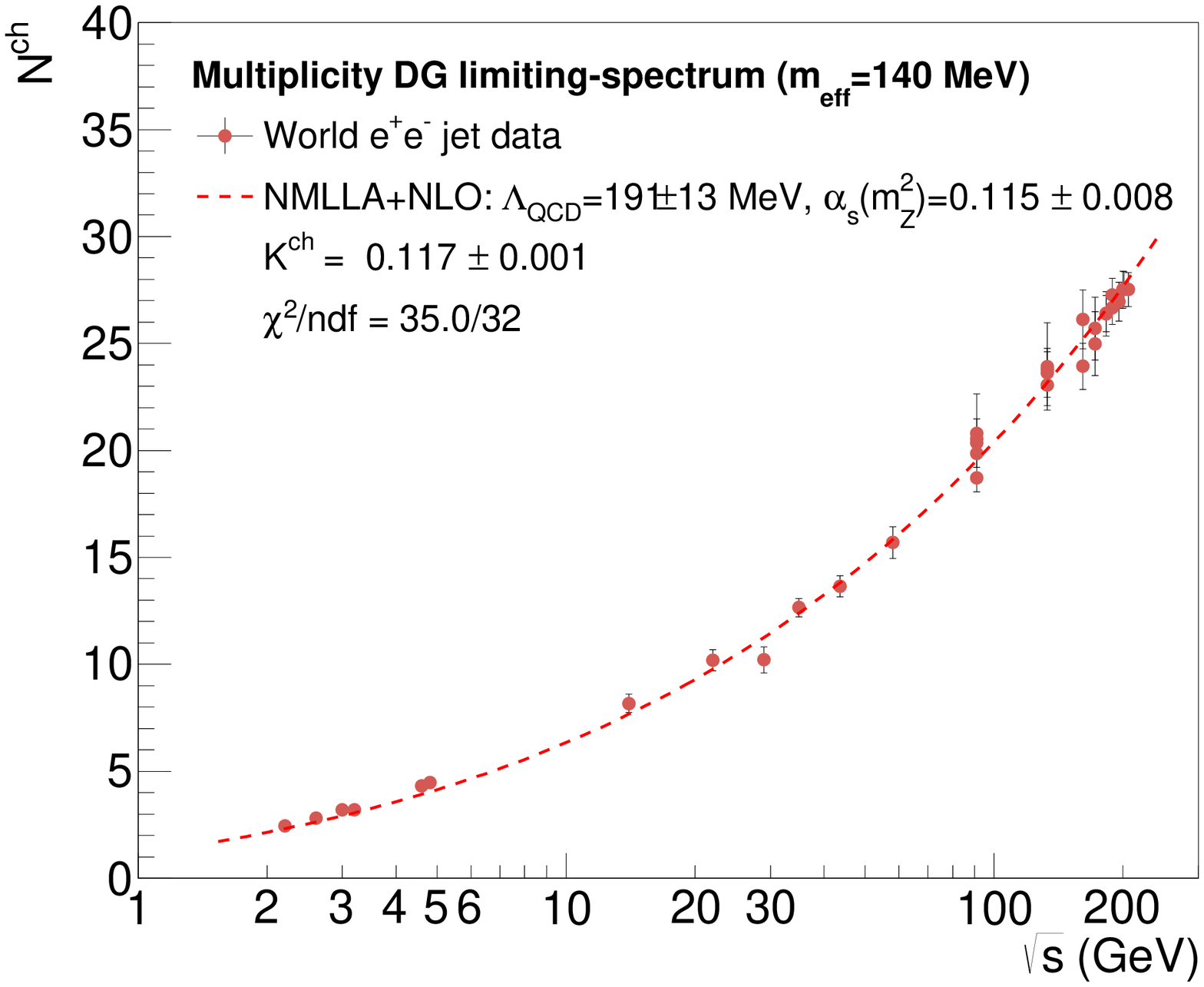,width=10.5truecm}
\caption{\label{fig:N_evol} Energy evolution of the total multiplicity ${\cal N}^{\rm ch}$ 
  spectrum of charged hadrons in jets measured in $\epem$ at collision energies
  $\sqrts\approx$~2--200~GeV, fitted to Eq.~(\ref{eq:dreminmult_evol_nf5}) with $Y=\ln(\sqrts/(2\lqcd))$,
  with finite-mass corrections ($\meff$~=~0.14~GeV). %with (right) and without (left) finite-mass corrections.
  The extracted values of the ${\cal K}^{\rm ch}$ normalization constant, $\lqcd$ and equivalent
  NLO$_{_{\rm \MSbar}}$ $\alphasmZ$, and the goodness-of-fit per degree-of-freedom $\xi^2/$ndf, are quoted.} 
\end{center}
\end{figure}

%Figure~\ref{fig:peak_evol} (left) shows the total 
%multiplicity ${\cal N}^{\rm ch}$ obtained from the integral of each one of the DGs plotted in
%Fig.~(\ref{fig:DGeedata}). The result Eq.~(\ref{eq:dreminmult_evol_nf5})

\begin{figure}[htbp]
\begin{center}
\epsfig{file=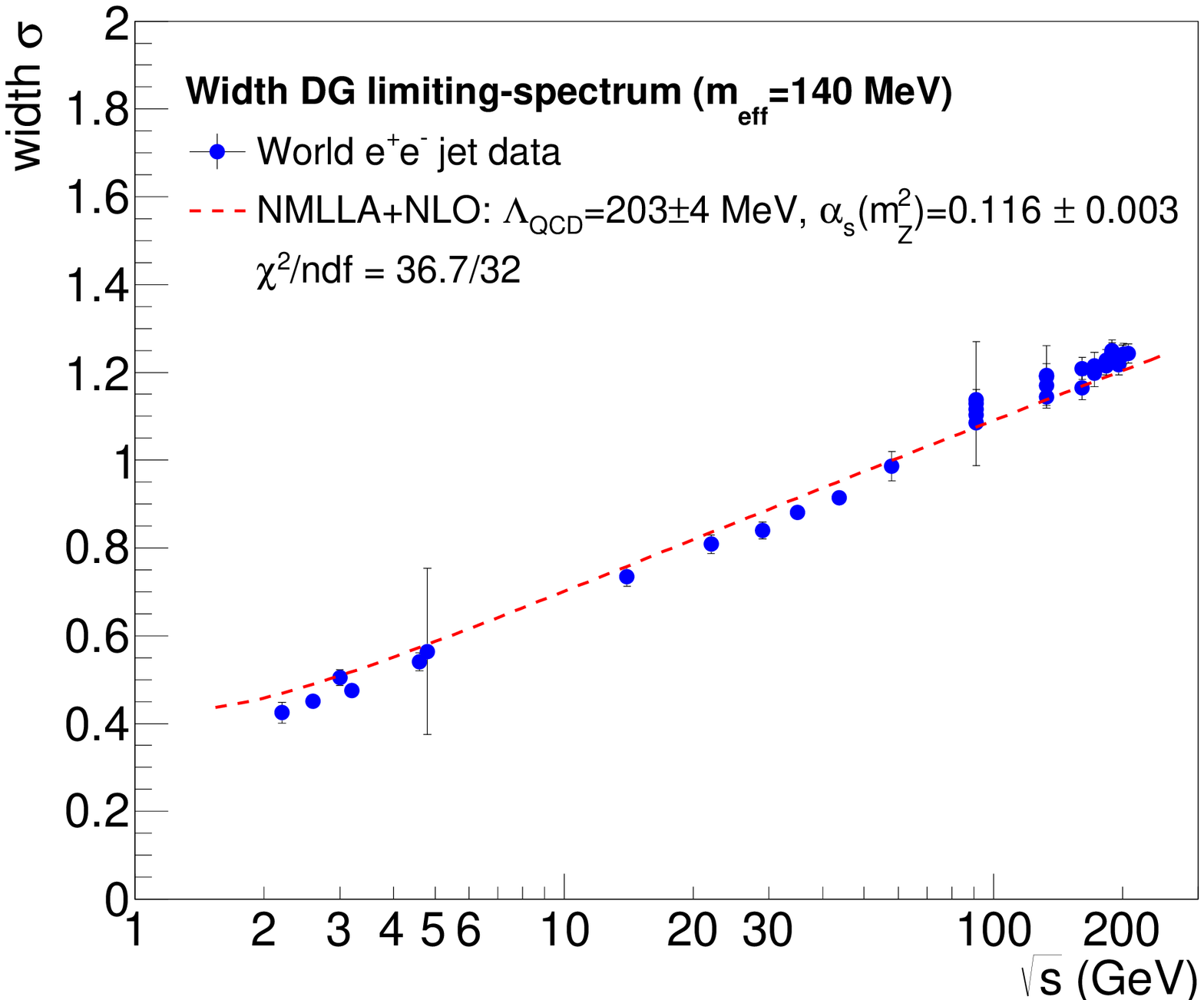,width=10.5truecm}
\caption{\label{fig:sigma_evol} Energy evolution of the width $\sigma$ 
  spectrum of charged hadrons in jets measured in $\epem$ at collision energies
  $\sqrts\approx$~2--200~GeV, fitted to Eq.~(\ref{eq:sigma_evol_nf5}) with $Y=\ln(\sqrts/(2\lqcd))$,
  with finite-mass corrections ($\meff$~=~0.14~GeV). %with (right) and without (left) finite-mass corrections.
  The extracted values of $\lqcd$ and equivalent NLO$_{_{\rm \MSbar}}$ $\alphasmZ$, and the
  goodness-of-fit per degree-of-freedom $\xi^2/$ndf, are quoted.}  
\end{center}
\end{figure}

% \begin{figure}[htbp]
% \begin{center}
% \epsfig{file=skewDG_vs_sqrts_meff0MeV.eps,width=8.15truecm}
% \epsfig{file=kurtosDG_vs_sqrts_meff0MeV.eps,width=8.15truecm}
% \caption{\label{fig:highermom_evol} Energy evolution of the skewness $x$ (left),and kurtosis $k$ (right) of the 
%   spectrum of charged hadrons in jets measured in $\epem$ at collision energies
%   $\sqrts\approx$~2--200~GeV, fitted to Eqs.~(\ref{eq:skew_evol_nf5}) and (\ref{eq:kurt_evol_nf5}), respectively,
% with $Y=\ln(\sqrts/(2\lqcd))$, with no finite-mass corrections ($\meff=0$).
%   The extracted values of $\lqcd$ and equivalent NLO$_{_{\rm \MSbar}}$ $\alphasmZ$ and the
%   goodness-of-fit per degree-of-freedom $\xi^2/$ndf, are quoted.}  
% \end{center}
% \end{figure}

\begin{figure}[htbp]
\begin{center}
\epsfig{file=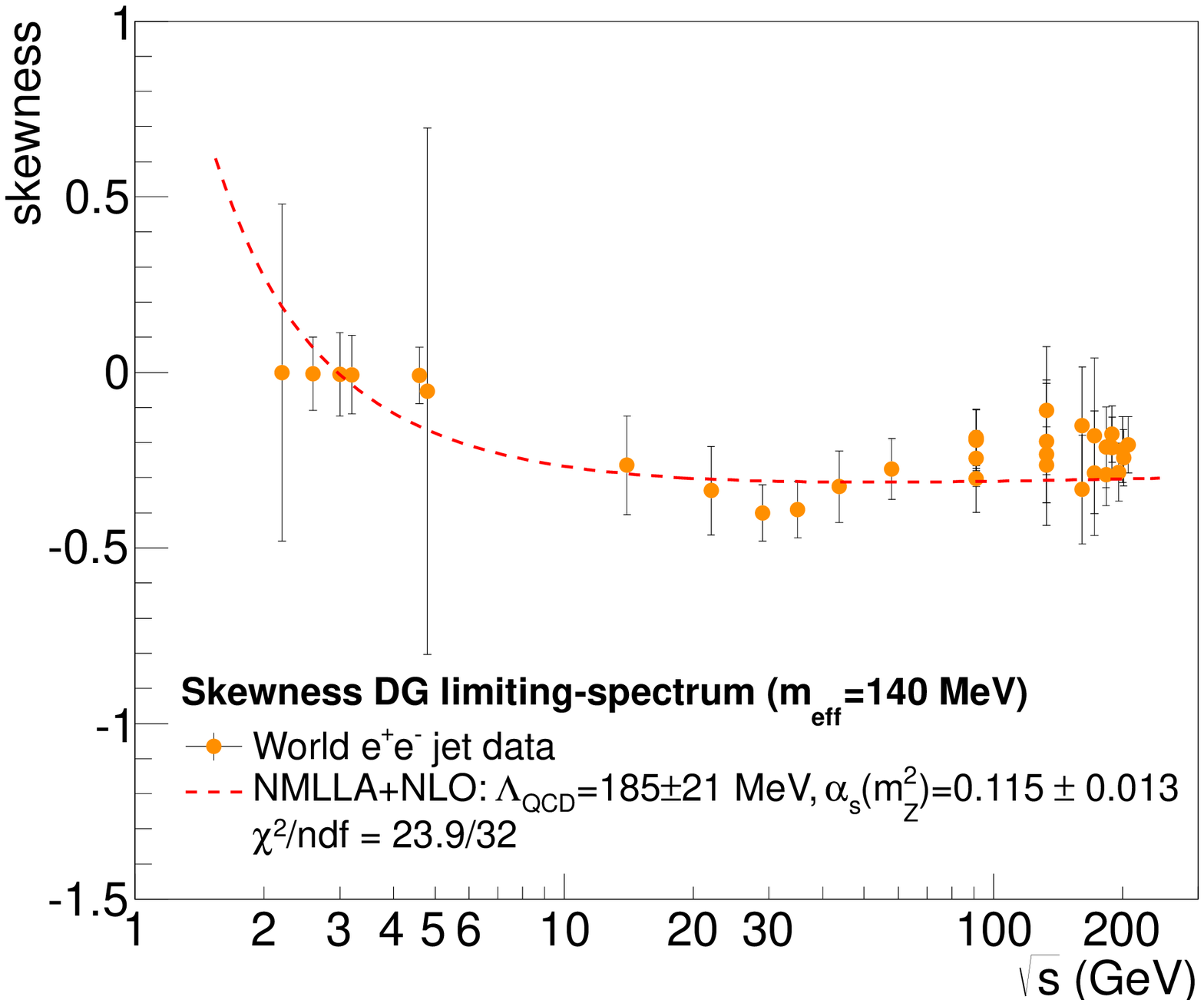,width=10.5truecm}
\caption{\label{fig:skewn_evol} Energy evolution of the skewness $s$ of the 
  spectrum of charged hadrons in jets measured in $\epem$ at collision energies
  $\sqrts\approx$~2--200~GeV, fitted to Eq.~(\ref{eq:skew_evol_nf5}) 
  with $Y=\ln(\sqrts/(2\lqcd))$, with finite-mass corrections ($\meff=0.14$~GeV).
  The extracted values of $\lqcd$ and equivalent NLO$_{_{\rm \MSbar}}$ $\alphasmZ$, and the
  goodness-of-fit per degree-of-freedom $\xi^2/$ndf, are quoted.}  
\end{center}
\end{figure}

\begin{figure}[htbp]
\begin{center}
\epsfig{file=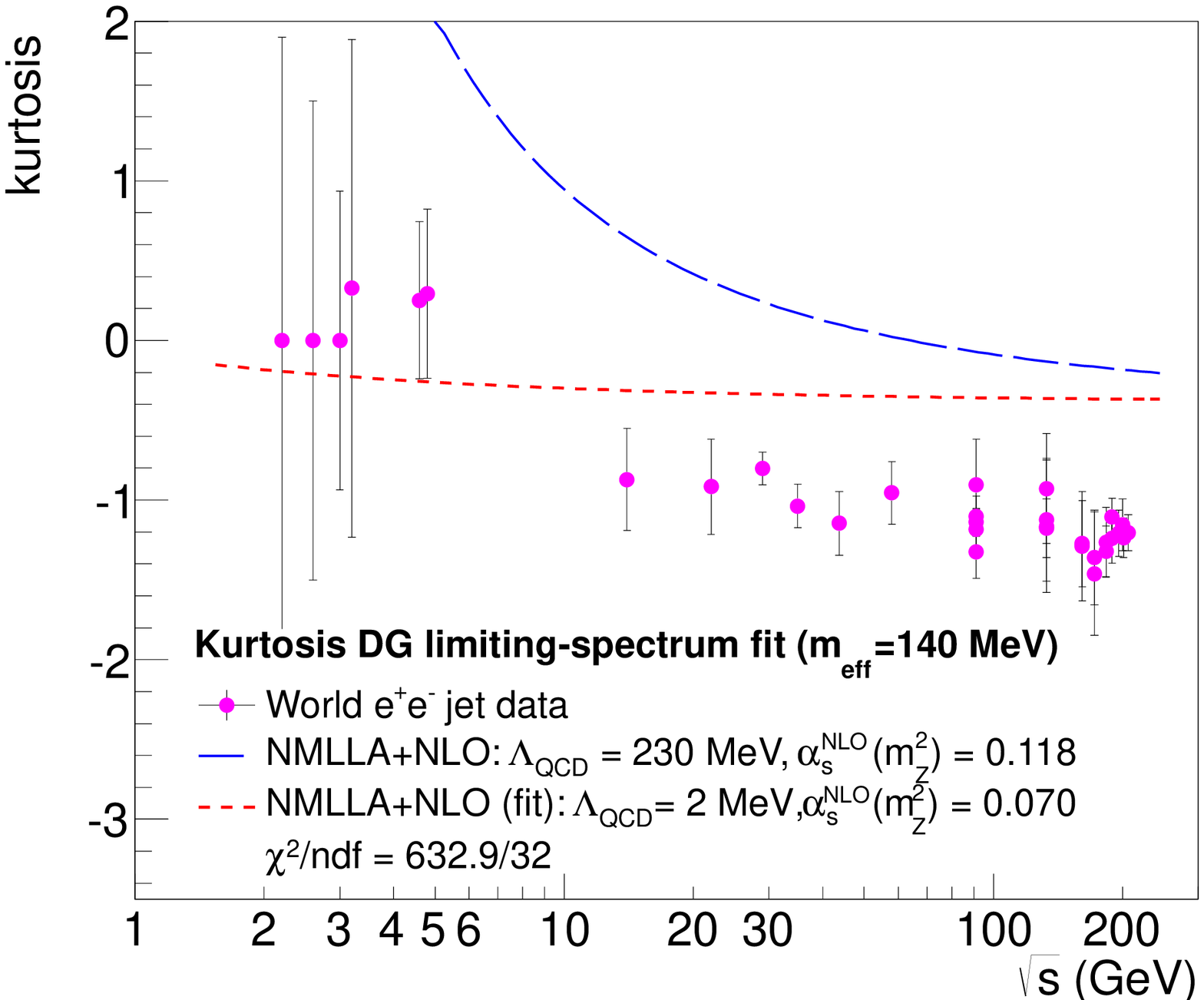,width=10.5truecm}
\caption{\label{fig:kurt_evol} Energy evolution of the kurtosis $k$ of the 
  spectrum of charged hadrons in jets measured in $\epem$ at collision energies
  $\sqrts\approx$~2--200~GeV, fitted to Eq.~(\ref{eq:kurt_evol_nf5}) with $Y=\ln(\sqrts/(2\lqcd))$, with
  finite-mass corrections ($\meff=0.14$~GeV). The resulting $\lqcd$, NLO$_{_{\rm \MSbar}}$ $\alphasmZ$, and
  goodness-of-fit per degree-of-freedom $\xi^2/$ndf are quoted. The long-dashed curve shows the expected
  theoretical dependence for $\lqcd$~=~230~MeV.}  
\end{center}
\end{figure}

The collision-energy dependencies of all the obtained DG components are plotted in
Figs.~\ref{fig:peak_evol}--\ref{fig:kurt_evol} %, \ref{fig:N_evol}, \ref{fig:sigma_evol}, \ref{fig:sigma_evol} and~\ref{fig:kurt_evol}
for $\meff=0.14$~GeV which, as aforementioned, provides the best individual fit to the DGs.
In any case, using alternative $\meff$ values results only in small changes in the derived values of
$\lqcd$, consistent with its quoted uncertainties. %derived from the $\sqrts$-evolution of all moments. 
Varying $\meff$ from zero to 0.32~GeV yields differences in the extracted $\lqcd$ parameter below $\pm$0.5\% for
the $\ximax$ fits and below $\pm$2\% for the other components, which indicate the robustness of our NMLLA+NLO$^*$
calculations for the limiting-spectrum DG with respect to finite-mass effects if a wide enough range of
charged-hadron and parent-parton (jet) energies are considered in the evolution fit.
%For all the components, except skewness and kurtosis, we plot the results for
%$\meff=0$ (left) and $\meff=0.32$~GeV (right) to estimate the influence of the finite-mass effects in the final global evolution.
%where they are individually fitted to our NMLLA+NLO$^*$ limiting-spectrum predictions. 
%iven by Eqs.~(\ref{eq:dreminmult_evol_nf5})-(\ref{eq:kurt_evol_nf5}) using $Y=\ln(\sqrts/(2\lqcd))$,
%.e. having $\lqcd$ as only free parameter.
The point-to-point uncertainties of the different moments, originally coming from the DG fit procedure alone,
have been enlarged so that their minimum values are at least 3\% for the peak position, and 5\% for the
multiplicity and width. Such minimum uncertainties are consistent with the spread of the DG moments obtained
for different experiments at the same collision-energies, and guarantee an acceptable global goodness-of-fit 
$\chi^2/{\rm ndf}\approx$~1 for their $\sqrts$-dependence. We note that not all measurements originally
corrected for feed-down contributions from weak decays of primary particles. This affects, in particular,
the multiplicities measured for the TASSO~\cite{Braunschweig:1988qm,Braunschweig:1990yd},
TPC~\cite{Aihara:1988su} and OPAL~\cite{Akrawy:1990ha} datasets which include charged particles from K$_{0}^{s}$ and $\Lambda$
decays. The effect on the peak position (and higher HBP moments) of including secondary particles from decays
is negligible ($<$0.5\%), but increases the total charged particles yields by 8\% according to experimental
data and Monte Carlo simulations~\cite{Sjostrand:2006za}. For these three data-sets, we have thus reduced
accordingly the value of ${\cal N}^{\rm ch}$.\\ 

The DG skewness and kurtosis are less well constrained by the individual fits to the measured fragmentation
functions and have much larger uncertainties than the rest of moments. As a matter of fact, in the case of the
kurtosis our NMLLA+NLO$^*$ prediction for its energy-evolution Eq.~(\ref{eq:kurt_evol_nf5}), fails to provide
a proper description of the data and seems to be above the data by a constant offset (Fig.~\ref{fig:kurt_evol}). Whether this fact is due to missing higher-order
contributions in our calculations or to other effects is not yet clear at this point. Apart from the kurtosis,
the QCD coupling value extracted 
%$\lqcd$ for all the other moments yields values around $\lqcd\approx$~220~MeV, or equivalently around %NLO$_{_{\rm \MSbar}}$
from all the other moments has values around $\alphasmZ=$~0.118, in striking agreement with the
current world-average obtained by other methods~\cite{Bethke:2012jm,Beringer:1900zz}.\\
%The final quantitative extraction of $\alphasmZ$ in our framework is discussed next.\\ 

\begin{table}[htbp!]
\caption{\label{tab:1} Values of $\lqcd$ and associated $\alphasmZ$ at NLO ($\overline{\rm MS}$ scheme, $n_f~=~5$ quark flavours)
  obtained from the fits of the $\sqrts$-dependence of the moments of the charged hadron distribution of jets
  in $\epem$ collisions obtained from their NMLLA+NLO$^{*}$ evolution. 
  %The errors of $\lqcd$ and $\alphasmZ$ include in quadrature global-fit and finite-mass uncertainties. 
  The last column provides the weighted-average
  of the individual measurements with its total propagated uncertainty.}  
%\vspace{-0.5cm}
\begin{center}
\begin{tabular}{lcccc|c}\hline
 DG moment:    & Peak position      & Multiplicity      & Width             & Skewness          & Combined \\\hline
 $\lqcd$ (MeV) &  255 $\pm$ 4       & 191 $\pm$ 13      & 203 $\pm$ 4       & 185 $\pm$ 21      & 249 $\pm$ 6\\
% $\lqcd$ (MeV) &  255 $\pm$ 4       & 153 $\pm$ 10      & 205 $\pm$ 6       & 185 $\pm$ 21      & 239 $\pm$ 6\\
 $\alphasmZ$   &  0.120 $\pm$ 0.002 & 0.115 $\pm$ 0.008 & 0.116 $\pm$ 0.003 & 0.115 $\pm$ 0.013 & 0.1195 $\pm$ 0.0022 \\\hline
\end{tabular}
\end{center}
\end{table}

%\paragraph{Final $\alphas$ extraction:}
Table~\ref{tab:1} lists each value of the $\lqcd$ parameter individually extracted from the energy evolutions of the four DG
components that are well described by our NMLLA+NLO$^*$ approach, and their associated values of %We can convert the extracted value of $\lqcd$ into 
$\alphasmZ$ obtained using the two-loop Eq.~(\ref{eq:twoloop}) for $n_f=5$ quark flavours. 
%The values are weighted-averages for $\meff$ 
%assuming that their uncertainties are fully correlated. 
Whereas the errors quoted for the different $\lqcd$ values include only uncertainties from
the fit procedure, the propagated $\alphasmZ$ uncertainties have been enlarged by a common factor such that
their final weighted average has a $\chi^2/$ndf close to unity. Such a ``$\chi^2$ averaging''
method~\cite{Beringer:1900zz} takes into account in a well defined manner any correlations between the
four extractions of $\alphas$, as well as underestimated systematic uncertainties.
The relative uncertainty of the $\alphasmZ$ determination from the DG moments evolution is about $\pm$1.5\% for the maximum
peak position, $\pm$3.5\% for the width, $\pm$7\% for the total multiplicity, and about $\pm$11\% for the skewness.
%. The skewness and kurtosis have much larger uncertainties and are not useful on this regard. 
The last column of Table~\ref{tab:1} lists the final values of $\lqcd$  and $\alphasmZ$ determined by taking
the weighted-average of the four individual measurements. 
We obtain a final value $\alphasmZ$~=~0.1195~$\pm$~0.0022 which is in excellent agreement with the
current world-average of the strong coupling at the Z mass~\cite{Bethke:2012jm,Beringer:1900zz}.
Our extraction of the QCD strong coupling has an uncertainty ($\pm$2\%) that is commensurate with that from other
$\epem$ observables such as jet-shapes ($\pm$1\%) and 3-jets rates
($\pm$2\%)~\cite{Bethke:2012jm,Beringer:1900zz}. 
%Since it has been obtained from a completely different theoretical approach than other existing methods
%The extracted value of the QCD mass scale from the fit of the energy evolution of the 
%peak position equals $\lqcd=0.245\pm0.008$. Although, it should not be identified with the
%$\Lambda_{\overline{\rm MS}}$ from the factorisation scheme, it can be taken in this 
%framework to determine the value of the coupling constant from Eq.~(\ref{eq:twoloop}) such 
%that $\alphasmZ=0.119$.
In a forthcoming work, we extend the extraction of the strong coupling via the
NMLLA+NLO$^*$ evolution of the moments of the hadron distribution in jet world-data measured not only in
$\epem$ but also including deep-inelastic $e^\pm\,p$ collisions~\cite{DdERedamy}.\\ 

\section{Conclusions and outlook}
\label{sec:summ}

We have computed analytically the energy evolution of the moments of the single-inclusive distribution of
hadrons inside QCD jets in the next-to-modified-leading-log approximation (NMLLA) including
next-to-leading-order (NLO) corrections to the $\alphas$ strong coupling. Using a distorted Gaussian
parametrization, we provide in a closed-form the numerical expressions for the energy-dependence of the
maximum peak position, total multiplicity, peak width, kurtosis and skewness of the limiting spectra where
the hadron distributions are evolved down to the $\lqcd$ scale.
Comparisons of all the existing jet data measured in $\epem$ collisions in the range
$\sqrts\approx$~2--200~GeV to the NMLLA$+$NLO$^*$ predictions for the moments of the hadron distributions allow
one to extract a value of the QCD parameter $\lqcd$ and associated two-loop coupling constant at the Z resonance,
$\alphasmZ$~=~0.1195~$\pm$~0.0022, in excellent agreement with the current world average obtained
with other methods. The NMLLA$+$NLO$^*$ approach presented here can be further extended to full NMLLA$+$NLO
through the inclusion of the two-loop splitting functions. Also, in a forthcoming phenomenological study we
plan to compare our approach not only to the world $\epem$ jet data but also to jet measurements in (the
current hemisphere of the Breit-frame of) deep-inelastic $e^\pm\,p$ collisions. The application of our
approach to the hadron distribution of TeV-jets 
produced in proton-proton collisions at LHC energies would further allow one to extract $\alphas$ from
parton-to-hadron FFs over a 
very wide kinematic range. The methodology presented here provides a new independent approach for the
determination of the QCD coupling constant complementary to other existing jet-based methods --relyiong on jet
shapes, and/or on ratios of N-jet production cross sections-- with a totally different set of experimental and
theoretical uncertainties. 

%%%%%%%%%%%%%%%%%%%%%%%%%%%%%%%%%%%%%%%%%%%%%%%%%%%%%%%%%%%%%%%%%%%%%%%%%%%%%%%%%%%%%%%%
%
\section*{Acknowledgments}

We are grateful to Yuri~Dokshitzer, Igor~Dremin, Kari~Eskola, Valery~A.~Khoze, Sven~Moch, Wolfgang~Ochs, and
Bryan~Webber for useful discussions and/or 
comments to a previous version of this manuscript. R.~P\'erez-Ramos acknowledges support from the Academy of
Finland, Projects No. 130472 and No.133005. 

%%%%%%%%%%%%%%%%%%%%%%%%%%%%%%%%%%%%%%%%%%%%%%%%%%%%%%%%%%%%%%%%%%%%%%%%%%%%%%%%%%%%%%%%
%

%\clearpage
\section*{Appendix}

\appendix

\section{Mellin-transformed splitting functions}
\label{sec:mellinsplittings}

The set of LO DGLAP splitting functions in Mellin space has been given in~\cite{Dokshitzer:1978hw}. 
It follows from Eqs.~(\ref{eq:pqgqq})--(\ref{eq:pgggq}) by making use of the Mellin transform given in
Eq.~(\ref{eq:mllaeveqgenter}) such that  
\begin{subequations}
\begin{eqnarray}\label{eq:expomega3}
P_{gg}(\Omega)\!&\!=\!&\!-4N_c\left[\psi(\Omega+1)+\gamma_E\right]+\frac{11N_c}{3}-
\frac{2n_f}{3}+\frac{8N_c(\Omega^2+\Omega+1)}{\Omega(\Omega^2-1)(\Omega+2)},\\
P_{gq}(\Omega)\!&\!=\!&\!\frac{\Omega^2+\Omega+2}{\Omega(\Omega+1)(\Omega+2)},\\
P_{qg}(\Omega)\!&\!=\!&\!2C_F\frac{\Omega^2+\Omega+2}{\Omega(\Omega^2-1)},\\
P_{qq}(\Omega)\!&\!=\!&\!-C_F\left[\psi(\Omega+1)+4\gamma_E-3-\frac2{\Omega(\Omega+1)}\right].
\label{eq:expomega4}
\end{eqnarray}
\end{subequations}
The expansion of the set of splitting functions (\ref{eq:expomega3})--(\ref{eq:expomega4}) 
in Mellin space is trivial and makes use of the Taylor expansion of the digamma function
as $\Omega\to0$:
$$
\psi(\Omega+1)=-\gamma_E+\frac{\pi^2}{6}\Omega+{\cal O}(\Omega^2),
$$
and $(1\pm\Omega)^{\alpha}\approx1\mp\alpha\Omega+\frac12\alpha(\alpha-1)\Omega^2+\ldots$, which leads 
to the formul{\ae} (\ref{eq:expomega1})--(\ref{eq:expomega2}).
%\section{Kinematics}
%Consider the splitting of one parton into two paratons $a(k)\to b(k_1)+c(k_2)$ such that
%$b$ and $c$ carry the energy-momentum fractions $z$ and $(1-z)$ of $a$ respectively. 
%Following the Sudakov parametrisation for $k_1$ and $k_2$, the four-momentums can be written as
%\begin{eqnarray}
%k_1^{\mu}\!\!&\!\!=\!\!&\!\!zk^{\mu}-k_\perp^{\mu}+\frac{\vec{k}^2+k_1^2}{z}\frac{n^{\mu}}{n\cdot k},\\
%k_2^{\mu}\!\!&\!\!=\!\!&\!\!(1-z)k^{\mu}+k_\perp^{\mu}+\frac{\vec{k}^2+k_2^2}{1-z}\frac{n^{\mu}}{2n\cdot k},
%\end{eqnarray}
%with light-like vectors $k_i^2=n^2=0$, and time-like transverse momentum $k_\perp$ such that, 
%$k\cdot k_\perp=n\cdot k_\perp=0$. Performing the scalar product $k_1\cdot k_2$, one has
%\begin{equation}\label{eq:k1k2}
%k_\perp^2=2z(1-z)k_1\cdot k_2.
%\end{equation}
%Let us write now $k=\left(E,\vec{k}\right)$, $k_1=\left(zE,\vec{k}_1\right)$, $k_2=\left((1-z)E,\vec{k}_2\right)$ such that, with 
%$k^2\approx0$ one has $\mid\!\vec{k}_1\!\mid=zE$, $\mid\!\vec{k}_1\!\mid=(1-z)E$. From energy-momentum conservation and the previous
%choices 
%\begin{equation}\label{eq:ksq}
%k^2=2k_1\cdot k_2=2z(1-z)E^2(1-\cos\theta)
%\end{equation}
%such that replacing (\ref{eq:ksq}) in (\ref{eq:k1k2}) one has:
%\begin{equation}\label{eq:kperp}
%k_\perp=2z(1-z)E\sin\frac{\theta}2.
%\end{equation}
%In the collinear limit however, one is left with $k_\perp\approx z(1-z)E\theta$.

\section{NMLLA$+$NLO$^*$ moments $K_n$ of the distorted Gaussian}
\label{app:moments}

We compute here the generic for the moments of the distorted Gaussian (DG) for $\lambda\ne0$ according to
Eq.~(\ref{eq:moments}) by introducing the following functions:
\begin{eqnarray}
f_1(Y,\lambda)\!\!&\!\!=\!\!&\!\!\frac{1-\frac{\lambda}{Y+\lambda}}{1-\left(\frac{\lambda}{Y+\lambda}\right)^{3/2}},\quad
f_4(Y,\lambda)=\frac{1-\left(\frac{\lambda}{Y+\lambda}\right)^2}{1-\left(\frac{\lambda}{Y+\lambda}\right)^{5/2}}\\
f_2(Y,\lambda)\!\!&\!\!=\!\!&\!\!\frac{1-\left(\frac{\lambda}{Y+\lambda}\right)^{1/2}}{1-\left(\frac{\lambda}{Y+\lambda}\right)^{3/2}},\quad
f_5(Y,\lambda)=\frac{1-\left(\frac{\lambda}{Y+\lambda}\right)^{3/2}}{1-\left(\frac{\lambda}{Y+\lambda}\right)^{5/2}},\\
f_3(Y,\lambda)\!\!&\!\!=\!\!&\!\!\frac{1-\left(\frac{\lambda}{Y+\lambda}\right)^{1/2}\left[\frac{\ln2\lambda-2}{\ln2(Y+\lambda)-2}\right]}
{1-\left(\frac{\lambda}{Y+\lambda}\right)^{3/2}},\quad
f_6(Y,\lambda)=\frac{1-\left(\frac{\lambda}{Y+\lambda}\right)^{3/2}\left[\frac{\ln2\lambda-2/3}{\ln2(Y+\lambda)-2/3}\right]}
{1-\left(\frac{\lambda}{Y+\lambda}\right)^{5/2}}.
\end{eqnarray}
Notice that $f_i(Y,\lambda=0)=1$. The expressions for $K_2$, $K_3$, $K_4$ and $K_5$ are then, respectively:
\begin{eqnarray}
K_2(Y,\lambda)\!\!&\!\!=\!\!&\!\!
\frac{Y+\lambda}3\sqrt{\frac{\beta_0(Y+\lambda)}{16N_c}}\left[1-\left(\frac{\lambda}{Y+\lambda}\right)^{3/2}\right]
\left\{1-\frac1{32}\beta_0f_1(Y,\lambda)\sqrt{\frac{16N_c}{\beta_0(Y+\lambda)}}
\right.\cr
\!\!&\!\!+\!\!&\!\!\left.\left[\frac9{8}a_2f_2(Y,\lambda)-\frac3{32}\left(\frac{3}{16N_c^2}a_1^2+\frac{a_1\beta_0}{8N_c^2}-
\frac{\beta_0^2}{64N_c^2}\right)f_2(Y,\lambda)\right.\right.\cr
\!\!&\!\!+\!\!&\!\!\left.\left.\frac{\beta_1}{32\beta_0}(\ln2(Y+\lambda)-2)f_3(Y,\lambda)\right]
\frac{16N_c}{\beta_0(Y+\lambda)}\right\}\label{eq:K2}\\
K_3(Y,\lambda)\!\!&\!\!=\!\!&\!\!-\frac{a_1}{64N_c}\sqrt{\frac{\beta_0}{N_c}}(Y+\lambda)^{3/2}
\left[1-\left(\frac{\lambda}{Y+\lambda}\right)^{3/2}\right]\left(1-\frac{\beta_0}{16}f_1(Y,\lambda)
\sqrt{\frac{16N_c}{\beta_0(Y+\lambda)}}\right)
\label{eq:K3}\\
\label{eq:K4}
K_4(Y,\lambda)\!\!&\!\!=\!\!&\!\!-\frac3{320}\left(\frac{\beta_0}{N_c}\right)^{3/2}(Y+\lambda)^{5/2}
\left[1-\left(\frac{\lambda}{Y+\lambda}\right)^{5/2}\right]\left\{1-\frac{5}{48}
\beta_0f_4(Y,\lambda)\sqrt{\frac{16N_c}{\beta_0(Y+\lambda)}}\right.\cr
\!\!&\!\!+\!\!&\!\!\left.\left[\frac{25}{24}a_2f_5(Y,\lambda)-\frac{5}{256}\left(\frac{5}{2N_c^2}
a_1^2+\frac{a_1\beta_0}{N_c^2}-\frac{55}{24N_c^2}\beta_0^2
\right)f_5(Y,\lambda)\right.\right.\cr
\!\!&\!\!+\!\!&\!\!\left.\left.\frac{5\beta_1}{96\beta_0}
\left(\ln2(Y+\lambda)-\frac{2}{3}\right)f_6(Y,\lambda)\right]\frac{16N_c}{\beta_0(Y+\lambda)}\right\}\\
K_5(Y,\lambda)\!\!&\!\!=\!\!&\!\!\frac{3a_1\beta_0^2(Y+\lambda)^2}
{4096N_c^3}\left(10+12\sqrt{\frac{N_c(Y+\lambda)}{\beta_0}}\right)
-\frac{3a_1\beta_0^2\lambda^2}
{4096N_c^3}\left(10+12\sqrt{\frac{N_c\lambda}{\beta_0}}\right).
\end{eqnarray}
Compared to MLLA, a new term appears in the expression (\ref{eq:K3}) of $K_3$. In order to determine the 
dispersion $\sigma$, the skewness $s$ and kurtosis of the distribution, we
need to normalise by the corresponding power of $\sigma$. After taking the $\sigma=\sqrt{K_2}$ and 
expanding the Taylor series in $1/\sqrt{Y}$, we find the following expressions:
\begin{eqnarray}\label{eq:sigmaminus3}
\sigma^{-3}(Y,\lambda)\!\!&\!\!=\!\!&\!\!\left(\frac3{Y+\lambda}\right)^{3/2}
\left(\frac{16N_c}{\beta_0(Y+\lambda)}\right)^{3/4}
\left[1-\left(\frac{\lambda}{Y+\lambda}\right)^{3/2}\right]^{-3/2}\left(1\right.\cr
\!\!&\!\!+\!\!&\!\!\left.
\frac{3\beta_0}{64}f_1(Y,\lambda)\sqrt{\frac{16N_c}{\beta_0(Y+\lambda)}}\right),\\
\label{eq:sigmaminus4}
\sigma^{-4}(Y,\lambda)\!\!&\!\!=\!\!&\!\!\left(\frac3{Y+\lambda}\right)^{2}
\frac{16N_c}{\beta_0(Y+\lambda)}\left[1-\left(\frac{\lambda}{Y+\lambda}\right)^{3/2}\right]^{-2}
\left\{1+\frac{\beta_0}{16}f_1(Y,\lambda)\sqrt{\frac{16N_c}{\beta_0(Y+\lambda)}}\right.\cr
\!\!&\!\!-\!\!&\!\!\left.\left[\frac9{4}a_2f_2(Y,\lambda)-
\frac{3}{16}\left(\frac{3a_1^2}{16N_c^2}f_2(Y,\lambda)
+\frac{a_1\beta_0}{8N_c^2}f_2(Y,\lambda)-\frac{\beta_0^2}{64N_c^2}f_2(Y,\lambda)\right.\right.\right.\cr
\!\!&\!\!+\!\!&\!\!\left.\left.\left.\frac{9\beta_0^2}{64N_c^2}f_1^2(Y,\lambda)\right)
+\frac{\beta_1}{16\beta_0}(\ln2(Y+\lambda)-2)f_3(Y,\lambda)\right]\!\frac{16N_c}{\beta_0(Y+\lambda)}\right\},\\
\sigma^{-5}(Y)\!\!&\!\!=\!\!&\!\!\left(\frac3{Y+\lambda}\right)^{5/2}
\left(\frac{16N_c}{\beta_0(Y+\lambda)}\right)^{5/4}\left[1-\left(\frac{\lambda}
{Y+\lambda}\right)^{3/2}\right]^{-5/2}\left(1\right.\cr
\!\!&\!\!+\!\!&\!\!\left.\frac{5\beta_0}{64}f_1(Y,\lambda)
\sqrt{\frac{16N_c}{\beta_0(Y+\lambda)}}\right).
\end{eqnarray}
Thus $\sigma^{-3}$, $\sigma^{-4}$ and $\sigma^{-5}$ expressions should be multiplied by $K_3$, $K_4$ and $K_5$ 
and the result re-expanded again in order to get the final results of Eqs.~(\ref{eq:skewness}), 
(\ref{eq:kurtosis}) and (\ref{eq:k5}) respectively.

\section{Higher-order corrections to the moments of the distorted Gaussian}
\label{section:hocorrtomoments}

We extract here some corrections to be incorporated into the perturbative expansion of 
the truncated series for the mean peak position, dispersion, skewness and kurtosis 
\cite{Dokshitzer:1991ej}. The presence of these corrections in the exact solution of the MLLA evolution 
equations is far from trivial and is thus detailed in this appendix. These corrections are indeed hidden 
in the exact solution of the MLLA evolution equations with one-loop coupling constant and can be extracted
after performing some algebraical calculations as described in~\cite{Dokshitzer:1991ej}
(see also~\cite{Dokshitzer:1991wu} and references therein). The exact solution was written in terms of 
confluent hypergeometric functions and then in terms of fast convergent Bessel series as follows~\cite{Dokshitzer:1991wu}:
\begin{equation}
{\cal D}^+(\xi,Y) = \frac{8N_c\,\Gamma(B)}{\beta_0}\
 \int_0^\frac{\pi}{2}\
  \frac{\dd\tau}{\pi}\,e^{-B\upalpha}
\  {\cal F}_{B}(\tau,Y,\xi),
\label{eq:ifD}
\end{equation}
where the integration
is performed with respect to $\tau$ defined by
$\displaystyle \upalpha = \frac{1}{2}\ln\frac{Y-\xi}{\xi}  + i\tau$ and with
\begin{eqnarray*}
{\cal F}_{B}(\tau,Y,\xi) &=& \left[ \frac{\cosh\upalpha
-\displaystyle{\frac{Y-2\xi}{Y}}
\sinh\upalpha} 
 {\displaystyle \frac{4N_cY}{\beta_0}\,\frac{\upalpha}{\sinh\upalpha}} \right]^{B/2}
  I_{B}(2\sqrt{Z(\tau,Y,\xi)}), \cr
&& \cr
&& \cr
Z(\tau,Y,\xi) &=&
\frac{4N_cY}{\beta_0}\,
\frac{\upalpha}{\sinh\upalpha}\,
 \left(\cosh\upalpha
-\frac{Y-2\xi}{Y}
\sinh\upalpha\right),
\label{eq:calFdef}
\end{eqnarray*}

$B=a_1/\beta_0$ and $I_B$ is the modified Bessel function of the first kind.
It was then possible to extract the moments of the DG from this more complicated approach also. In the 
end, the MLLA moments of the DG found in~\cite{Fong:1990nt} from the MLLA anomalous dimension 
allows one to cross check the MLLA expressions found in~\cite{Dokshitzer:1991wu}.
According to~\cite{Dokshitzer:1991ej},
\begin{equation}
\xi_n=Y^n\cdot{\cal L}_n(B+1,B+2,z),\;B=\frac{a_1}{\beta_0},\;z=\sqrt{\frac{16N_c}{\beta_0}Y},
\end{equation} 
where the function ${\cal L}_n$ was written in the form of the series,
$$
{\cal L}_n(B+1,B+2;z)=P_0^{(n)}(B+1,B+2;z)+\frac2z\frac{I_{B+2}(z)}{I_{B+1}(z)}\cdot P_1^{(n)}(B+1,B+2;z),
$$
with
\begin{equation}
P_0^{(n)}(B+1,B+2;z)=\sum_{k=0}^{n-1}\upalpha_{n-k}^{(n)}\left(\frac2z\right)^{2k},\;
P_1^{(n)}(B+1,B+2;z)=\sum_{k=0}^{n-1}\upbeta_{n-k}^{(n)}\left(\frac2z\right)^{2k}.
\end{equation}
The functions $I_{B+i}(z)$ correspond to the modified Bessel series of the second kind. 
The leading coefficients are defined as:
$$
\upalpha_n^{(n)}=2^{-n},\quad\upbeta_n^{(n)}=\frac{n}{2^n}\left(B+\frac{n-1}3\right)
$$
and the others $\upalpha_{n-k}^{(n)}$, $\upbeta_{n-k}^{(n)}$ for $k\ne0$ are the solutions of 
the triangular matrix
\begin{eqnarray}
&&
\begin{pmatrix}
1 & 0 & 0 & 0 & 0 & 0 \\ 
1 & B+2 & 0 & 0 & 0 & 0 \\
1 & 1 & -B-1 & 0 & 0 & 0 \\
1 & B+3 & B+3 & (B+2)(B+3) & 0 & 0 \\
1 & 2 & -2B & -2B & 2B(B+1) & 0 \\
1 & B+3 & B+4 & (B+3)(B+4) & (B+3)(B+4) & (B+2)(B+3)(B+4) \\
\end{pmatrix} 
\begin{pmatrix}
\upbeta_1^{(n)} \\
\upalpha_1^{(n)} \\
\upbeta_2^{(n)} \\
\upalpha_2^{(n)} \\
\upbeta_3^{(n)} \\
\upalpha_3^{(n)}
\end{pmatrix}\cr
&&=
\begin{pmatrix}
-\frac{\Phi^{(n)}_{-B-1}}{B+1} \\
0 \\
-\frac{\Phi^{(n)}_{-B}}{B} \\
0 \\
-\frac{\Phi^{(n)}_{-B+1}}{B-1} \\
0 
\end{pmatrix}.
\label{eq:matrix}
\end{eqnarray}
The functions $\Phi$ in the r.h.s. of Eq.~(\ref{eq:matrix}) are defined in the form
\begin{eqnarray}
\Phi^{(1)}_c\!&\!=\!&\!\frac12\left\{c\right\}_2+(B+1)\left\{c\right\}_1,\\
\Phi^{(2)}_c\!&\!=\!&\!\frac14\left\{c\right\}_4+\left(B+\frac53\right)\left\{c\right\}_3+(B+1)(B+2)\left\{n\right\}_2,\\
\Phi^{(3)}_c\!&\!=\!&\!\frac18\left\{c\right\}_6+\frac14\left(3B+7\right)\left\{c\right\}_5
+\frac12\left(3B^2+13B+13\right)\left\{c\right\}_4\cr
\!&\!+\!&\!(B+1)(B+2)(B+3)\left\{c\right\}_3,\\
\Phi^{(4)}_c\!&\!=\!&\!\frac1{16}\left\{c\right\}_8+\frac12\left(B+3\right)\left\{c\right\}_7+
\left(\frac32B^2+\frac{17}{2}B+\frac{34}3\right)\left\{c\right\}_6\cr
\!&\!+\!&\!\left[2(B+1)^3+10(B+1)^2+14(B+1)+\frac{24}5\right]\left\{c\right\}_5\cr
\!&\!+\!&\!(B+1)(B+2)(B+3)(B+4)\left\{c\right\}_4,
\end{eqnarray}
where the shorthand notation $\left\{c\right\}_p=c(c-1)\ldots(c-p+1)$ has been introduced for the sake of 
simplicity and $c=-B-1,-B,-B+1$ according to the r.h.s. of Eq.~(\ref{eq:matrix}). For instance, making
use of Eq.~(\ref{eq:exactmoments}), for $n=1$ one has,
$$
\xi_1=Y\cdot{\cal L}(B+1,B+2;z)=Y\left[P_0^{(1)}(B+1,B+2;z)+\frac2z\frac{I_{B+2}(z)}{I_{B+1}(z)}\cdot 
P_1^{(1)}(B+1,B+2;z)\right],
$$\label{eq:meanmedian}
where in this case:
$$
P_0^{(1)}(B+1,B+2;z)=\upalpha_1^{(1)}=\frac12,\quad P_1^{(1)}(B+1,B+2;z)=\upbeta_1^{(1)}=\frac12B,
$$
according to the recursive relations given above. Therefore,
\begin{equation}\label{eq:xi1}
\xi_1=\frac{Y}2\left[1+\frac2zB\frac{I_{B+2}(z)}{I_{B+1}(z)}\right].
\end{equation}
Expanding the ratio $I_{B+2}(z)/I_{B+1}(z)$ for large $z$ (large energy scale in $Y(E)$) and making
use of the asymptotic expansion for the Bessel functions, 
\begin{eqnarray}
I_\nu(z)\!&\!\approx\!&\!\frac{e^z}{\sqrt{2\pi z}}\left[1-\frac{1}{2z}\left(\nu^2-\frac14\right)
+\frac{1}{8z^2}\left(\nu^2-\frac94\right)\left(\nu^2-\frac14\right)\right.\cr
\!&\!-\!&\!\left.\frac{1}{48z^3}\left(\nu^2-\frac{25}4\right)
\left(\nu^2-\frac94\right)\left(\nu^2-\frac14\right)\right],
\end{eqnarray}
one has
\begin{equation}\label{eq:besselratio}
\frac{I_{B+2}(z)}{I_{B+1}(z)}=1-\frac{2B+3}{2z}+\frac{(2B+3)(2B+1)}{8z^2}+\frac{(2B+3)(2B+1)}{8z^3}+{\cal O}(z^{-4}).
\end{equation}

\clearpage

%bibliographystyle{plain}
\bibliographystyle{h-elsevier3}

%\bibliography{mybib}

%%%%%%%%%%%%%%%%%%%%%%%%%%%%%%%%%%%%%%%%%%%%%%%%%%%%%%%%%%%%%%%%%%%%%%%%%%%
%
\end{document}